\documentclass[traditabstract,a4]{aa} 		

\usepackage{graphicx}
\usepackage{txfonts}
\usepackage{natbib}
\usepackage{enumitem}
\usepackage{lscape}

\begin{document}

\title{Building the cosmic infrared background brick by brick with Herschel/PEP.
\thanks{Herschel is an ESA space observatory with science instruments provided by 
European-led Principal Investigator consortia and with important participation from NASA.}}

\author{S. Berta\inst{1}
        \and
        B. Magnelli\inst{1}
	\and
	R. Nordon\inst{1}
	\and
	D. Lutz\inst{1}
	\and 
	S. Wuyts\inst{1}
	\and 
	B. Altieri\inst{2}
	\and
	P. Andreani\inst{3}\fnmsep\inst{4}
	\and
	H. Aussel\inst{5}
	\and
	H. Casta{\~n}eda\inst{6}
	\and
	J. Cepa\inst{6}\fnmsep\inst{7}
	\and
	A. Cimatti\inst{8}
	\and
	E. Daddi\inst{5}
	\and
	D. Elbaz\inst{5}
	\and
	N.M. F{\"o}rster Schreiber\inst{1}
	\and
	R. Genzel\inst{1}
	\and
	E. Le Floc'h\inst{5}
	\and
	R. Maiolino\inst{9}
	\and
	I. P{\'e}rez-Fournon\inst{6} 
	\and
	A. Poglitsch\inst{1}
	\and
	P. Popesso\inst{1}
	\and
	F. Pozzi\inst{8}
	\and
	L. Riguccini\inst{5}
	\and
	G. Rodighiero\inst{10}
	\and
	M. Sanchez-Portal\inst{2}
	\and
	E. Sturm\inst{1}
	\and
	L.J. Tacconi\inst{1}
	\and 
	I. Valtchanov\inst{2}
	}

\offprints{Stefano Berta, \email{berta@mpe.mpg.de}}

\institute{Max-Planck-Institut f\"{u}r Extraterrestrische Physik (MPE),
Postfach 1312, 85741 Garching, Germany.
\and
Herschel Science Centre, ESAC, Villanueva de la Ca\~nada, 28691 Madrid, Spain.
\and
ESO, Karl-Schwarzschild-Str. 2, D-85748 Garching, Germany.
\and
INAF - Osservatorio Astronomico di Trieste, via Tiepolo 11, 34143 
Trieste, Italy.
\and
Laboratoire AIM, CEA/DSM-CNRS-Universit{\'e} Paris Diderot, IRFU/Service d'Astrophysique, 
B\^at.709, CEA-Saclay, 91191 Gif-sur-Yvette Cedex, France.
\and
Instituto de Astrof{\'\i}sica de Canarias, 38205 La Laguna, Spain. 
\and
Departamento de Astrof{\'\i}sica, Universidad de La Laguna, Spain.
\and
Dipartimento di Astronomia, Universit{\`a} di Bologna, Via Ranzani 1,
40127 Bologna, Italy.
\and
INAF - Osservatorio Astronomico di Roma, via di Frascati 33, 00040 Monte Porzio Catone, Italy.
\and
Dipartimento di Astronomia, Universit{\`a} di Padova, Vicolo dell'Osservatorio 3, 
35122 Padova, Italy.
}

\date{Received ...; accepted ...}

 
\abstract{
The cosmic infrared background (CIB) includes roughly half of the energy
radiated by all galaxies at all wavelengths across cosmic time, as observed at the present epoch. The {\em PACS Evolutionary
Probe} (PEP) survey is exploited here to study the CIB and its redshift differential, at 70, 100 and 160 $\mu$m, where the
background peaks. Combining PACS observations of the GOODS-S, GOODS-N, Lockman Hole and COSMOS areas, 
we define number counts spanning over more than two orders of magnitude in flux: from $\sim$1 mJy
to few hundreds mJy. Stacking of 24 $\mu$m sources and $P(D)$ statistics extend the analysis down to $\sim$0.2 mJy.
Taking advantage of the wealth of ancillary data in PEP fields, differential number counts $d^{\,2} N/dS/dz$ and CIB are studied up
to $z=5$. Based on these counts, we discuss the effects of confusion on PACS 
blank field observations and provide confusion limits for the three bands considered.
While most of the available backward evolution models predict  
the total PACS number counts with reasonable success, the consistency to redshift distributions and CIB 
derivatives can still be significantly improved. The new high-quality PEP data highlight 
the need to include redshift-dependent constraints in future modeling.
The total CIB surface brightness emitted above PEP 3$\sigma$ flux limits is 
$\nu I_\nu=4.52\pm1.18$, $8.35\pm0.95$ and $9.49\pm0.59$ $[$nW m$^{-2}$ sr$^{-1}]$
at 70, 100, and 160 $\mu$m, respectively. These values correspond to $58\pm7$\% and
$74\pm5$\% of the COBE/DIRBE CIB direct measurements at 100 and 160 $\mu$m. Employing the $P(D)$ analysis, these fractions increase to
$\sim$65\% and $\sim$89\%. More than half of the resolved CIB was emitted at redshift $z\le1$. 
The 50\%-light redshifts lie at $z=0.58$, 0.67 and 0.73 at the three PACS wavelengths.
The distribution moves towards earlier epochs at longer wavelengths: while the 70 $\mu$m CIB is mainly produced by 
$z\le1.0$ objects, the contribution of $z>1.0$ sources reaches 50\% at 160 $\mu$m. Most of 
the CIB resolved in the three PACS bands was emitted by galaxies with infrared luminosities in the range $10^{11}-10^{12}$ L$_\odot$.
}


\maketitle


\section{Introduction}\label{sect:intro}

With the exception of the cosmic microwave background, which represents the relic of the Big
Bang, emitted at the last scattering surface, the extragalactic background light (EBL) 
is the integral of the energy radiated by all galaxies, from $\gamma$-rays
to radio frequencies, across all cosmic epochs.
Its energy density distribution is characterized by two primary peaks: the first at $\lambda\sim1$ $\mu$m
produced by starlight, and a second peak at $\sim100-200$ $\mu$m mainly due to starlight absorbed and reprocessed
by dust in galaxies \citep[e.g.][]{hauser2001}. Contribution from other sources, such as active galactic nuclei, are
expected to be only 5-20\% to the total EBL in the mid- and far-IR \citep[e.g.][]{matute2006,jauzac2011,draper2011}.
The optical and infrared backgrounds dominate the EBL by
several orders of magnitude with respect to all other spectral domains. 

The cosmic infrared background (CIB) was detected and measured for the first 
time in the mid nineties, analyzing the data obtained with the DIRBE and FIRAS instruments 
aboard the {\em Cosmic Background Explorer} (COBE) satellite \citep{puget1996,
hauser1998,fixsen1998,lagache1999,lagache2000}. In the global budget, the CIB 
provides roughly half of the total EBL \citep[e.g.][]{dole2006,hauser2001}.
Since in the local Universe the infrared output of galaxies is only a third of the emission at optical wavelengths
\citep[e.g.][]{soifer1991}, this implies a strong evolution of infrared galaxy populations, towards an
enhanced far-IR output in the past, in order to account for the total measured CIB. 

The discovery of large numbers of distant sources emitting a substantial amount of their energy in the IR
\citep[e.g.][among many others]{smail1997,aussel1999,elbaz1999,papovich2004} demonstrated that, although locally rare,
powerful IR galaxies are indeed numerous at high redshift.
The deep extragalactic campaigns carried out in the nineties and 2000's with 
the {\em Infrared Space Observatory} (ISO, see 
Genzel \& Cesarsky \citeyear{genzel2000} for a summary), and the {\em Spitzer Space Telescope} 
(Werner et al. \citeyear{werner2004}; e.g. Papovich et al. \citeyear{papovich2004}) 
were very efficient in identifying and characterizing large samples of mid-IR sources. 
In contrast, at wavelengths near the CIB peak (100-200 $\mu$m), the performance of these telescopes was strongly
limited by their small apertures ($\le 85$ cm diameter), the prohibitive confusion limits, and
detectors' sensitivity.
Spitzer surveys at 70 and 160 $\mu$m produced limited samples of distant far-IR
objects \citep[e.g.][]{frayer2009}: in the 160 $\mu$m Spitzer/MIPS band only $\sim7$\% of
the CIB was resolved into individually detected objects  
(Dole et al., \citeyear{dole2004}, see Lagache et al. \citeyear{lagache2005} for a review). 
Performing stacking of 24 $\mu$m sources, \citet{dole2006} retrieved more than 70\% of the far-IR background
at 70 and 160 $\mu$m. \citet{bethermin2010a} extended this analysis by extrapolating to very faint flux
densities using a power-law fit to stacked number counts, and produced an estimate of the CIB surface
brightness in agreement with direct measurements.

Launched in May 2009, Herschel \citep{pilbratt2010} is providing stunning results: 
its large 3.5 m mirror,
and the high sensitivity {\em Photodetector Array Camera \& Spectrometer}
(PACS, performing imaging at 70, 100, 160 $\mu$m; Poglitsch et al., \citeyear{poglitsch2010})
are tailored to overtake the confusion and blending of sources that were hampering the detection of faint
far-IR sources in previous space missions.   
In \citet[][hereafter called ``Paper~I'']{berta2010}, we exploited data from the {\em PACS Evolutionary Probe}
(PEP) survey, covering the GOODS-N, Lockman Hole and COSMOS (to partial depth only) fields, and we resolved into individual sources
45\% and 52\% of the CIB at 100 and 160 $\mu$m. At longer wavelengths, in the spectral domain covered by the
{\em Spectral and Photometric Imaging Receiver} \citep[SPIRE,][]{griffin2010}, \citet{oliver2010} resolved
15\%, 10\% and 6\% of the CIB at 250, 350 and 500 $\mu$m. These fractions increased to 64\%, 60\% and 43\%,
respectively, using a $P(D)$ analysis \citep{glenn2010}. Finally, \citet{bethermin2010b} retrieved roughly 50\% of the
CIB applying stacking of 24 $\mu$m sources to BLAST 250, 350 and 500 $\mu$m maps.

The {\em PACS Evolutionary Probe} 
(PEP\footnote{http://www.mpe.mpg.de/ir/Research/PEP/}) 
is one of the major Herschel Guaranteed Time (GT) extragalactic projects.
It is structured as a ``wedding cake'' survey, based on four different layers
in order to cover wide shallow areas and deep pencil-beam fields. 
PEP includes the most popular and widely studied extragalactic blank fields:
COSMOS (2 deg$^2$), Lockman Hole, EGS and ECDFS (450-700 arcmin$^2$),  
GOODS-N and GOODS-S ($\sim$200 arcmin$^2$). In addition, the fourth tier of the
``cake'' consists of ten nearby lensing clusters, offering the chance 
to break the PACS confusion limit thanks to gravitational lensing. 
An in depth description of PEP fields and survey properties is presented by 
Lutz et al. (\citeyear{lutz2011}, in prep.).

Here we extend the analysis carried out in Paper~I to the deepest field observed by PEP, GOODS-S, reaching a
3 $\sigma$ depth of 1.2 mJy at 100 $\mu$m, and including also the 70 $\mu$m PACS bandpass.
The surface brightness of the CIB and its spectral energy distribution (SED) has become known
with increasing detail, from its initial discovery with COBE to the Spitzer era. 
The real conundrum is now represented by how the CIB evolves as a function of cosmic time.
This information is not only important to constrain models of galaxy evolution, but also
to shed light on the intrinsic spectra of TeV sources, such as BLAZARs, whose $\gamma$-ray photons interact
with the CIB \citep[e.g.][]{dominguez2011,franceschini2008,mazin2007}. Thanks to Herschel capabilities, in Paper~I we have initiated the study of the redshift
build-up of the local far-IR CIB, by associating to each object detected by PACS in GOODS-N a complete UV-to-FIR
SED and splitting number counts and CIB into redshift slices. \citet{jauzac2011} performed a similar study based on stacking on
70 and 160 $\mu$m Spitzer/MIPS maps of COSMOS, and \citet{lefloch2009} performed a similar analysis at 24 $\mu$m in the same
field. Here we take advantage of deeper PACS observations in GOODS-S and full coverage in COSMOS to further explore 
the history of CIB, with the aim to understand when it was primarily emitted. 
 
In this paper, the cosmic far-IR background is reconstructed ``brick by brick'' in three main ``storeys''. In Section
\ref{sect:detected_counts} the resolved component of number counts is presented, including slicing in redshift
and a comparison to available evolutionary models. Section \ref{sect:stacking_counts} deals with stacking of 24
$\mu$m sources onto PACS maps. The third step (Sect. \ref{sect:pdd_counts}) extends the analysis to even fainter fluxes
using the ``probability of deflection'', $P(D)$, technique. After a brief digression about source confusion in
Sect. \ref{sect:confusion}, we discuss the observed properties of CIB in Sect. \ref{sect:CIB}.
Finally, Sect. \ref{sect:summary} draws a summary of our findings.


\section{Level 1: detected sources}\label{sect:detected_counts}

In Paper~I we studied number counts based on {\em Science Demonstration Phase}
(SDP) and early routine phase data, i.e. based on the GOODS-N, Lockman Hole
and COSMOS fields. At the time of SDP, COSMOS was available only to 2/3 of its nominal depth.
\citet{altieri2010} exploited gravitational
lensing in the Abell 2218 cluster of galaxies to estimate 
the lens amplification of the fluxes of background galaxies. In this way, they
were able to push the analysis down to 1 mJy at 100 and 160 $\mu$m, thus
breaking the confusion limit for blank field observations 
(see Sect. \ref{sect:confusion}).

Here we complete the Herschel/PACS view of far-IR number counts and CIB, making
use of the deepest PEP field, GOODS-S, and the
full coverage of COSMOS, in addition to Paper I results. 
Data reduction, catalogs construction and simulations, aimed at deriving 
completeness, fraction of spurious sources and photometric reliability, are 
described in Lutz et al. (in prep.) and \citet{berta2010}. 
Table \ref{tab:fields} summarizes the main properties of the fields taken into
account here.

\begin{figure*}[!ht]
\centering
\rotatebox{-90}{
\includegraphics[height=0.45\textwidth]{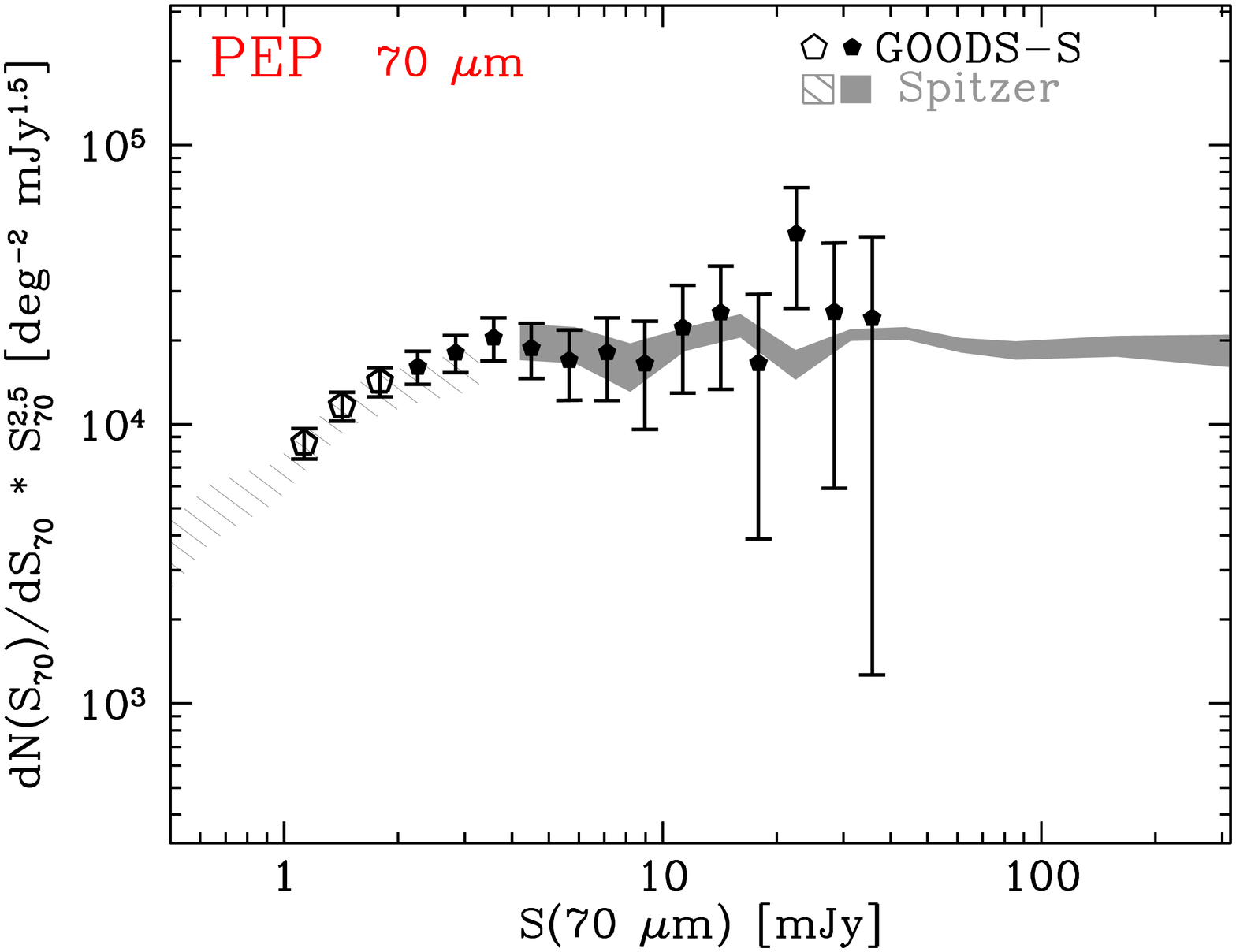}
}
\rotatebox{-90}{
\includegraphics[height=0.45\textwidth]{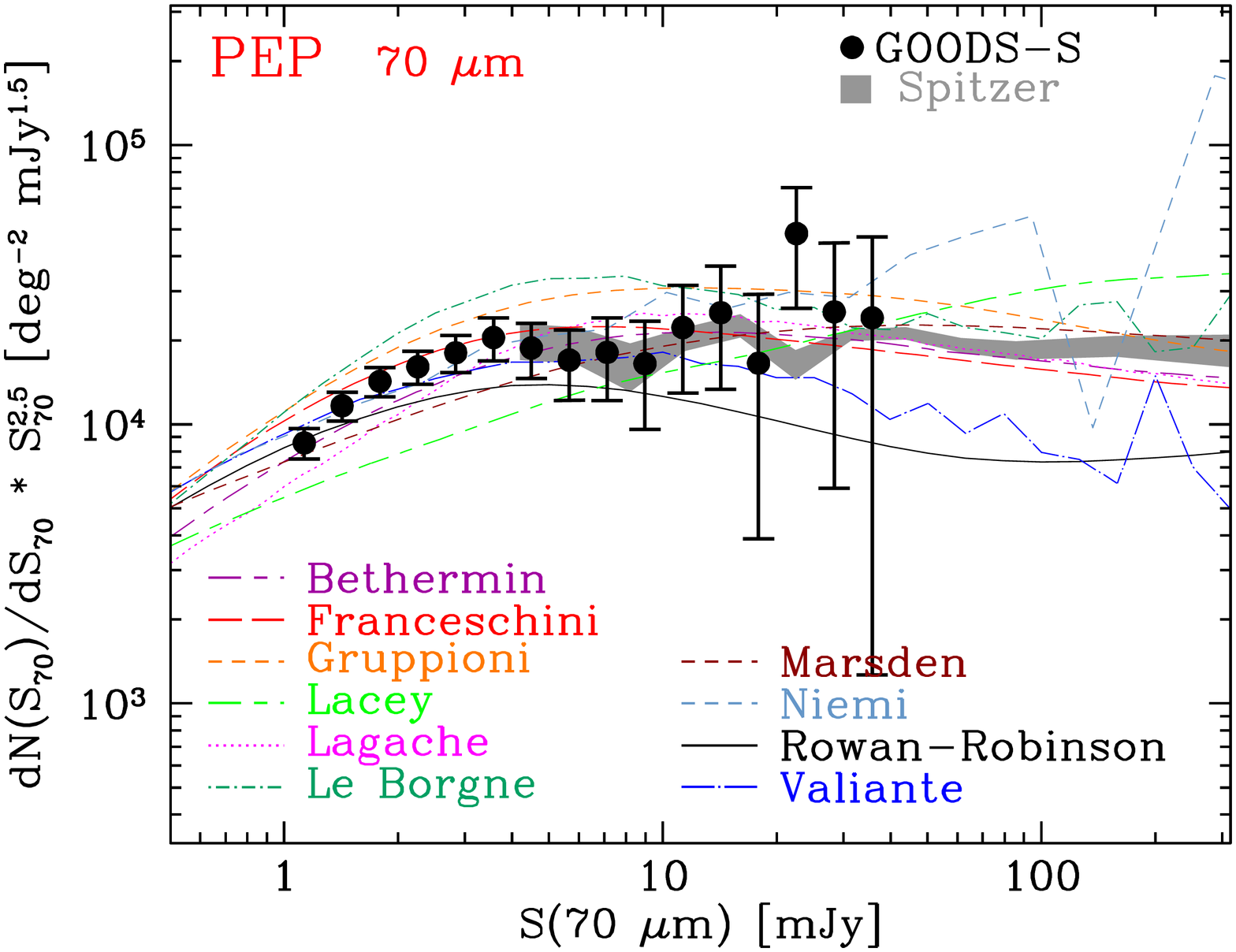}
}
\rotatebox{-90}{
\includegraphics[height=0.45\textwidth]{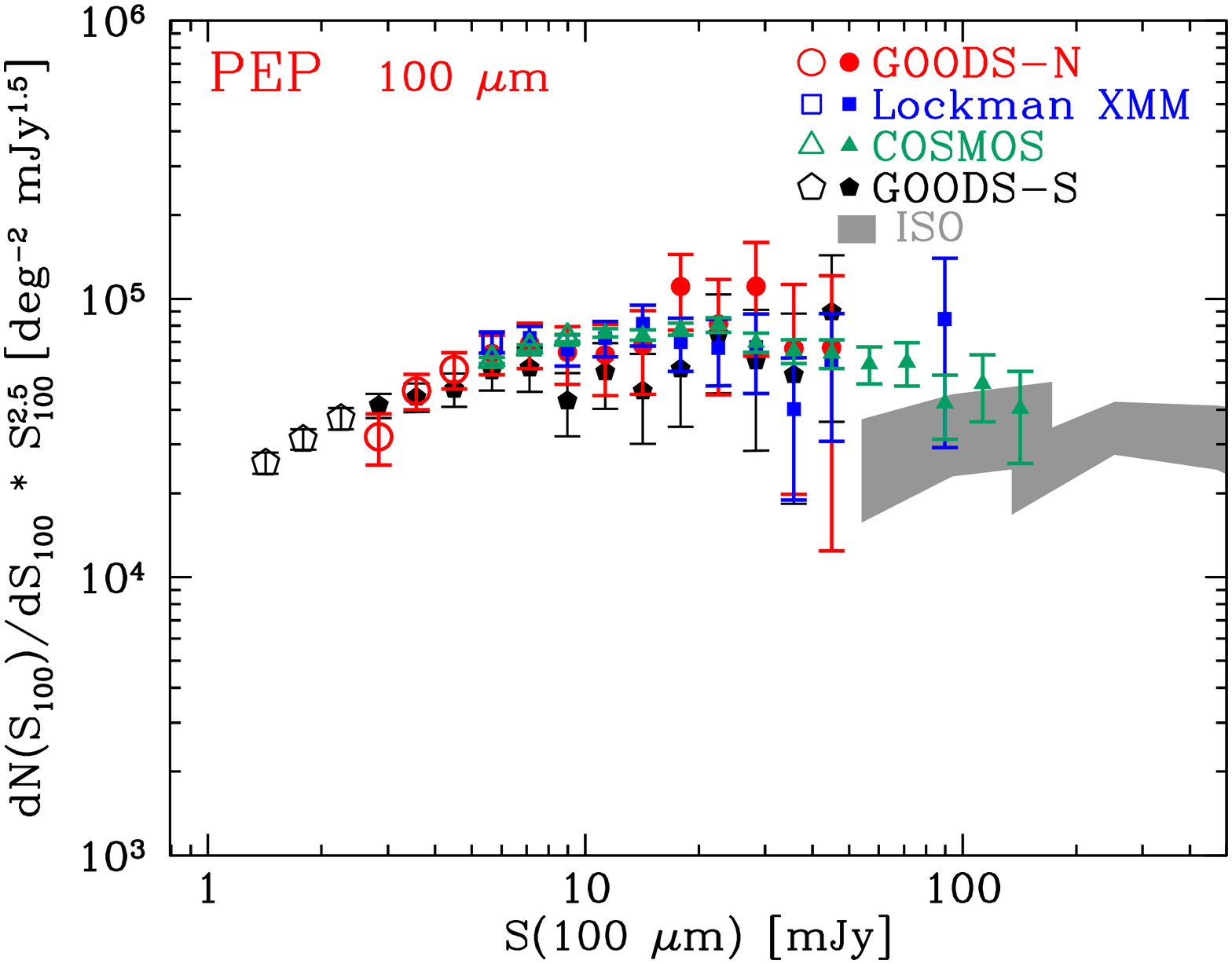}
}
\rotatebox{-90}{
\includegraphics[height=0.45\textwidth]{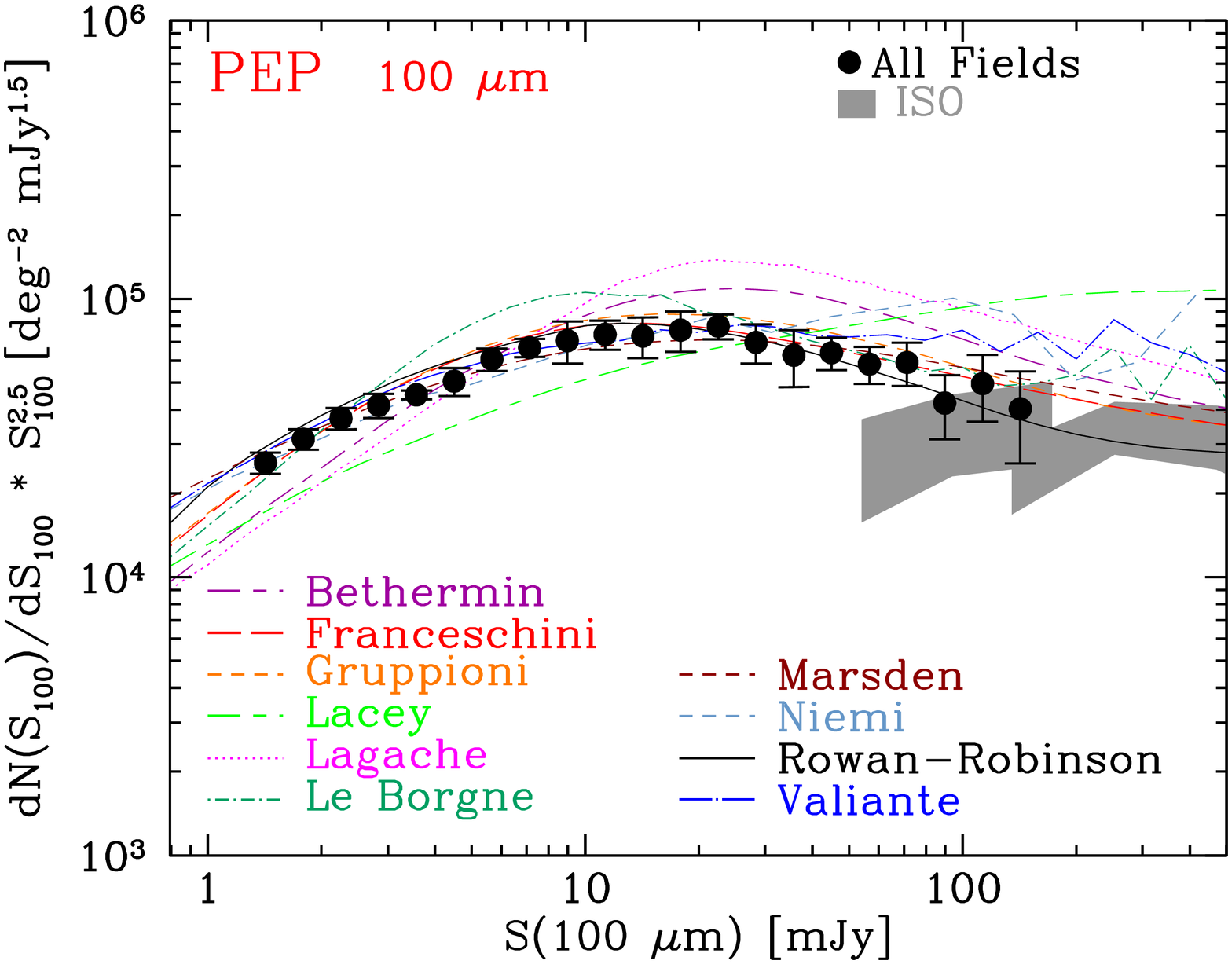}
}
\rotatebox{-90}{
\includegraphics[height=0.45\textwidth]{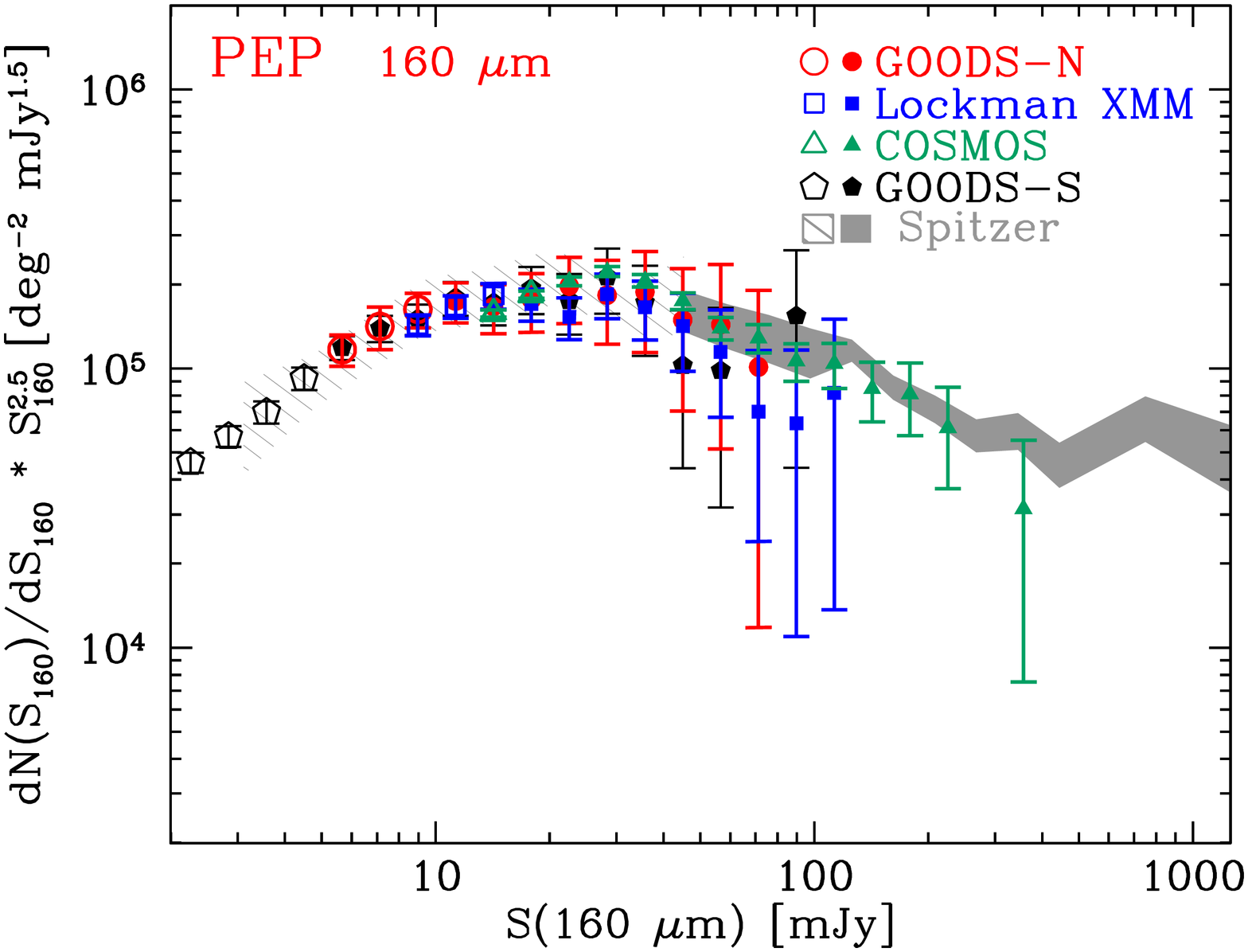}
}
\rotatebox{-90}{
\includegraphics[height=0.45\textwidth]{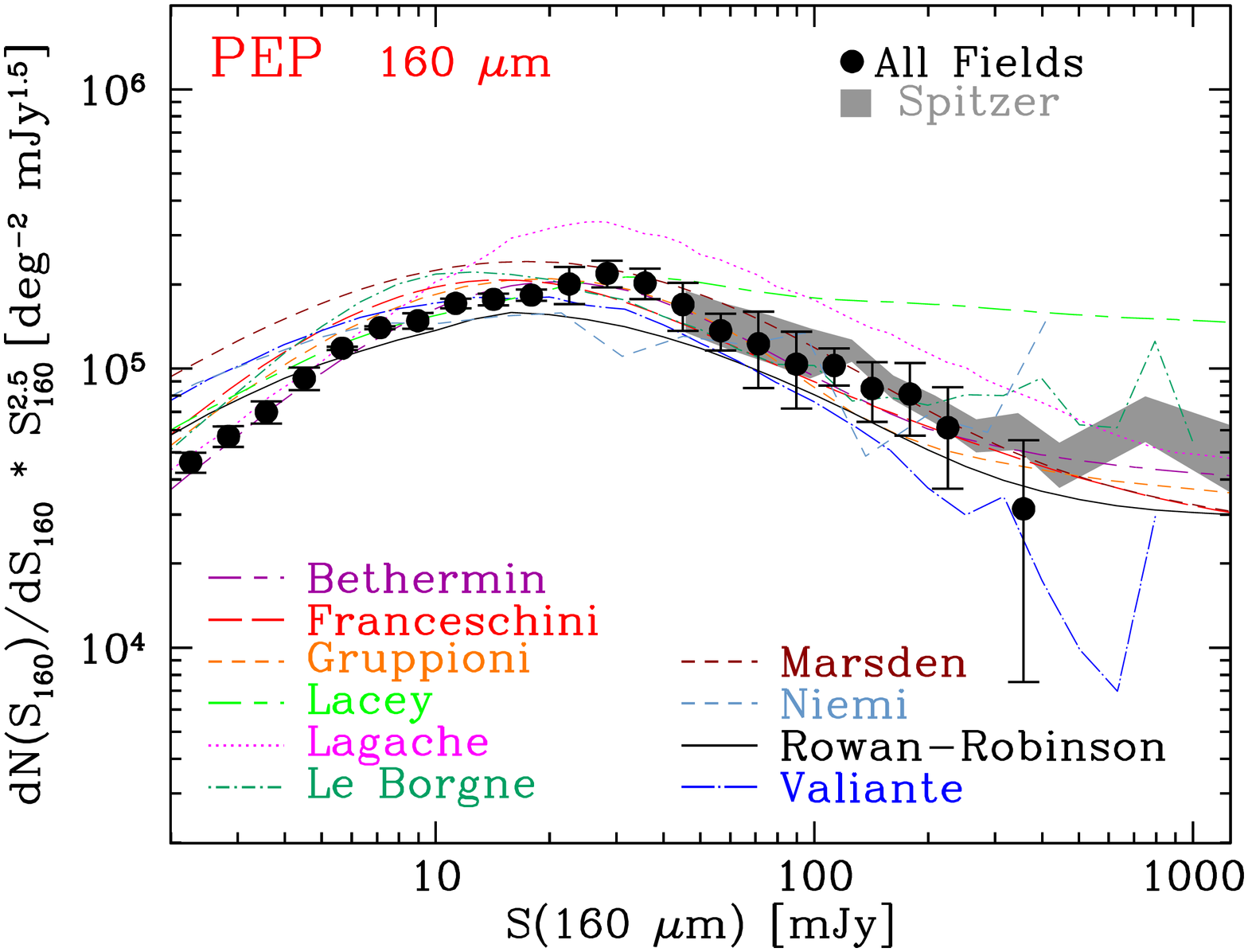}
}
\caption{Differential number counts in the three PACS bands, 
normalized to the Euclidean slope ($dN/dS\propto
S^{-2.5}$). Filled/open symbols belong to flux bins
above/below the 80\% completeness limit. Models belong to
\citet{lagache2004}, \citet{franceschini2010}, \citet{rowanrobinson2009}, 
\citet{leborgne2009}, \citet{valiante2009}, \citet{lacey2009},
\citet{bethermin2010c}, \citet{marsden2010}, Gruppioni et al. (\citeyear{gruppioni2011}, in prep.), 
Niemi et al. (\citeyear{niemi2011}, in prep.). 
Shaded areas represent ISO and Spitzer data
\citep{rodighiero2004,heraudeau2004,bethermin2010a}; hatched areas belong to Spitzer 24 $\mu$m
stacking \citep{bethermin2010a}. {\em Left} and {\em right} panels present 
individual fields and averaged counts, respectively.}
\label{fig:pep_counts}
\end{figure*}

We applied the method described by \citet{chary2004} and \citet{smail1995} to
correct number counts for incompleteness, on the basis of simulations. 
The distribution of input and output fluxes in simulations
is organized in a matrix $P_{ij}$, so that $i$ represents the input flux and 
$j$ the output flux. In other words, the $ij$-th
element of the matrix gives the number of sources with $i$-th input flux and
$j$-th output flux. 
The way $P_{ij}$ is built implies that $\sum_j P_{ij} \le 1$ represents the 
completeness correction factor at the $i$-th input flux in simulations.
In order to correct the observed counts, $P_{ij}$ is re-normalized 
such that the $\sum_i P_{ij}$ equals the number of real sources detected in the
$j$-th flux bin. Finally, the completeness-corrected counts in the given $i$-th
bin are given by $\sum_j P_{ij}^{\ renorm}$. 

Figure \ref{fig:pep_counts} presents PACS number counts at 70, 100 and 160
$\mu$m, normalized to the Euclidean slope (i.e. the slope expected for a uniform
distribution of galaxies in Euclidean space, $dN/dS\propto S^{-2.5}$). 
Error bars include Poisson statistics, flux calibration uncertainties, and
photometric errors. The latter have been 
propagated into number counts via $10^4$ 
realizations of random Gaussian flux errors applied to each PACS source, 
using a dispersion equal to the local measured noise. It is worth to note 
that in most cases the faint end of counts derived in shallow fields is consistent with 
results from deeper fields, thus confirming the validity of completeness 
corrections. Number counts, their corresponding uncertainties and
completeness values are reported in Tabs. \ref{tab:counts_070}, 
\ref{tab:counts_100}, and \ref{tab:counts_160}.

\begin{figure*}[!ht]
\centering
\rotatebox{-90}{
\includegraphics[height=0.45\textwidth]{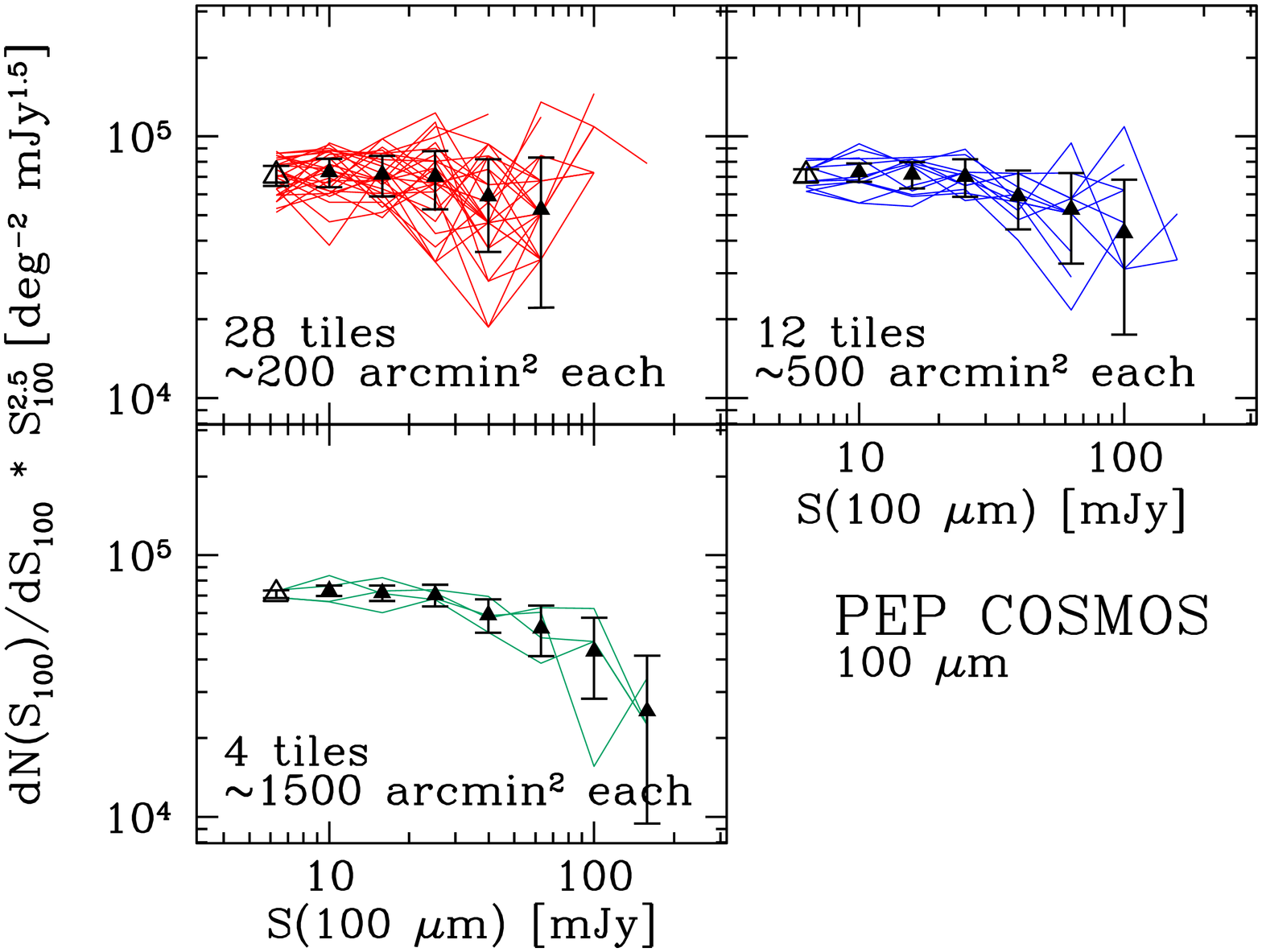}
}
\rotatebox{-90}{
\includegraphics[height=0.45\textwidth]{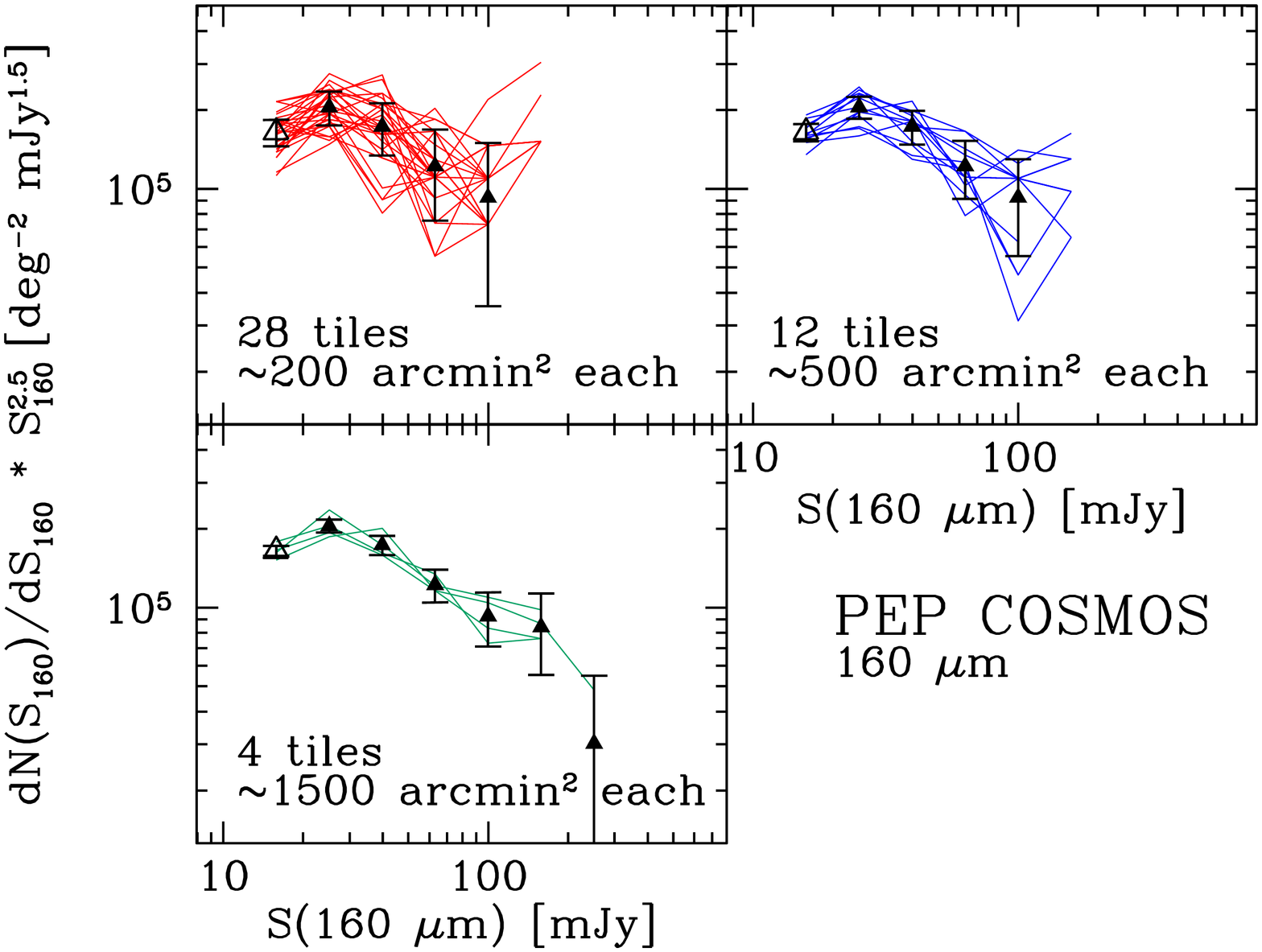}
}\\
\rotatebox{-90}{
\includegraphics[height=0.45\textwidth]{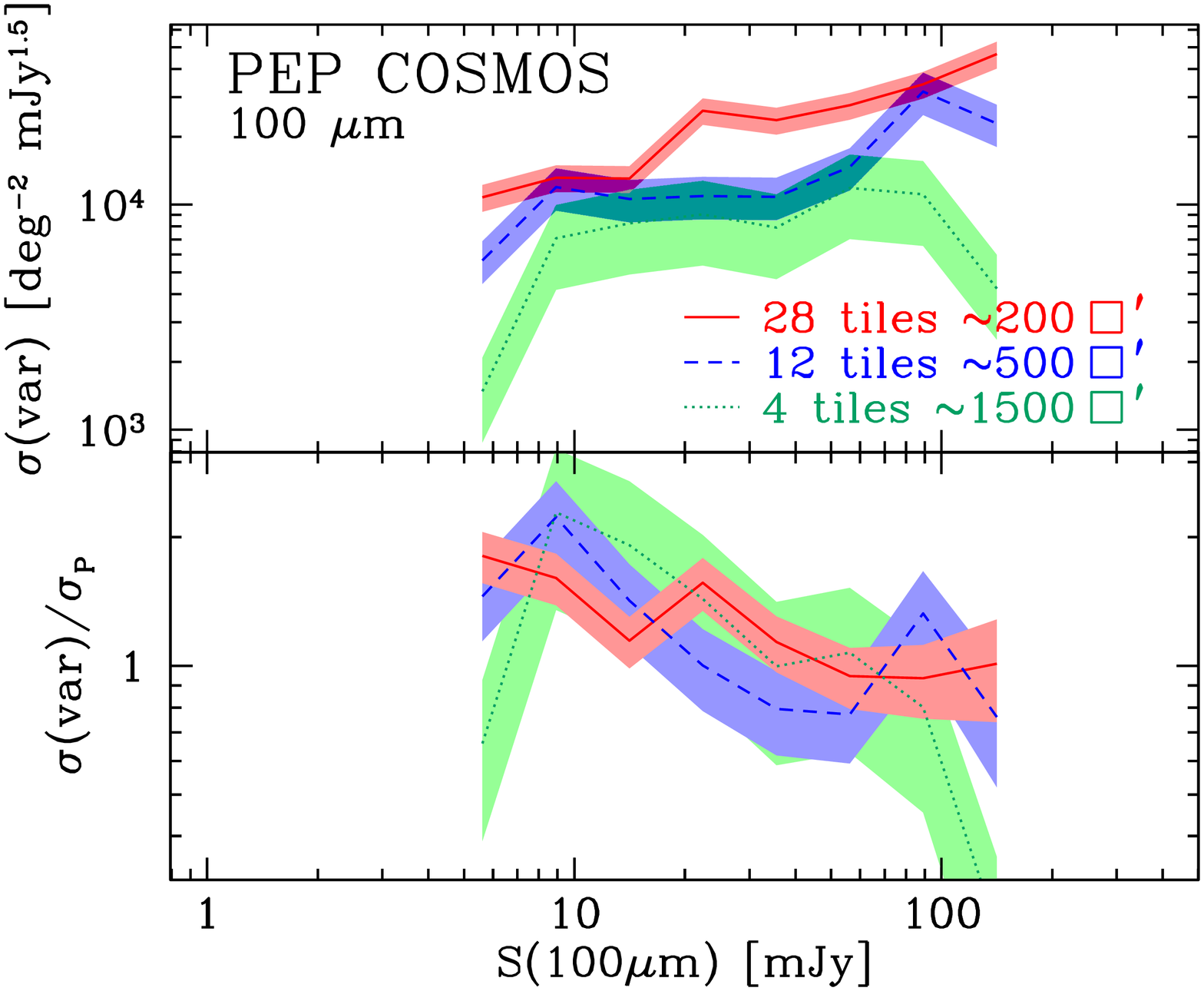}
}
\rotatebox{-90}{
\includegraphics[height=0.45\textwidth]{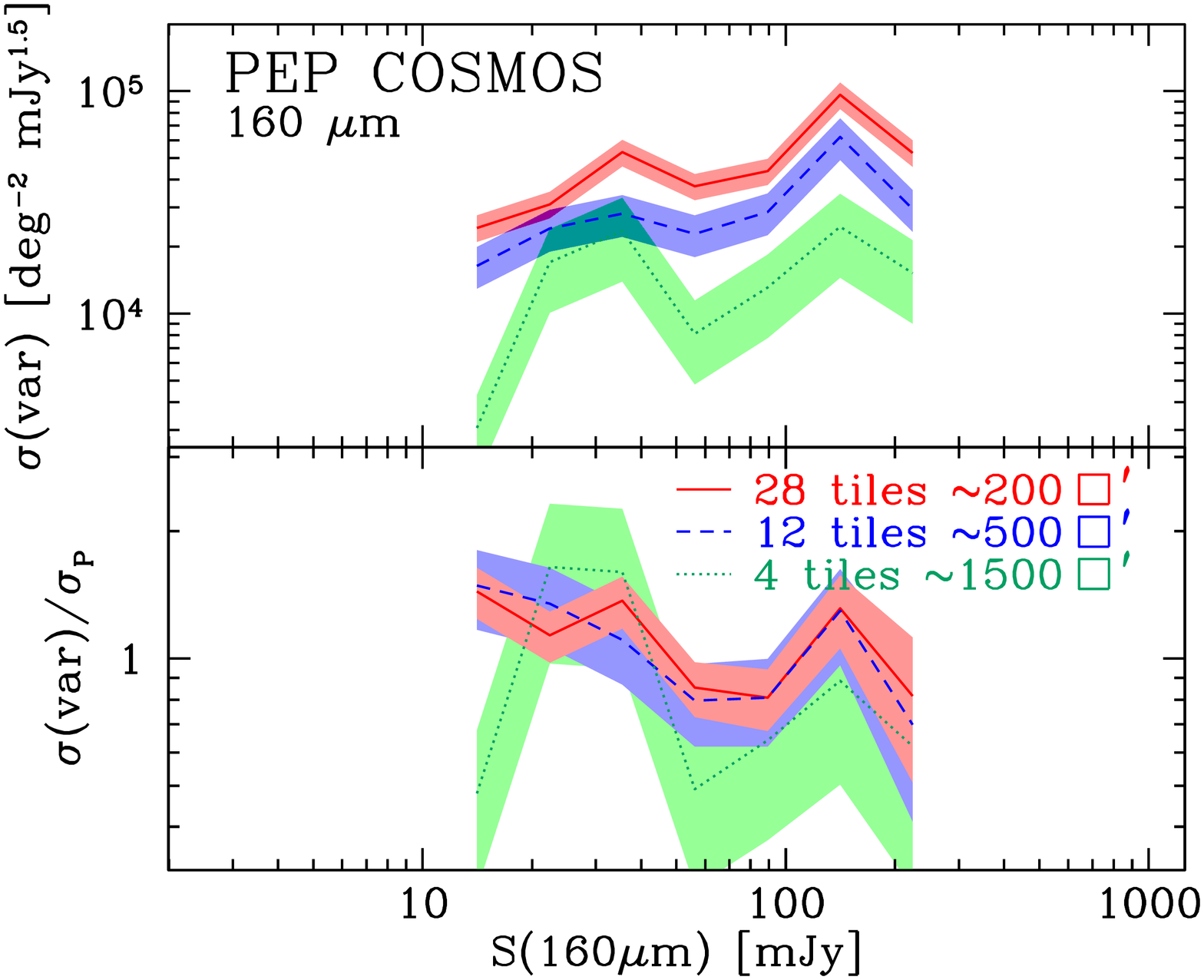}
}
\caption{Field-to-field variations of number counts in COSMOS. {\em Top panels}:
number counts split into sub-areas (lines) compared to full-field counts (black
dots). {\em Bottom diagram}: standard deviation among the sub-fields (top), as a function
of flux. In the very bottom panels, we show the comparison between the field-to-field $\sigma(var)$
and the uncertainty $\sigma_P$ obtained combining Poisson statistics, photometric errors
and systematics. Shaded areas represent 1$\sigma$ uncertainties on $\sigma(var)$ 
and $\sigma(var)/\sigma_P$ determinations.}
\label{fig:counts_field_to_field}
\end{figure*}

The left panels in Fig. \ref{fig:pep_counts} show the counts for each 
field separately.
Filled symbols represent data 
above 80\% completeness, while open symbols extend down to 
the 3$\sigma$ detection threshold.
The grey shaded areas in Fig. \ref{fig:pep_counts} belong to pre-Herschel number
counts. We include the \citet{rodighiero2004} and \citet{heraudeau2004} 95
$\mu$m ISO data, and the 70 $\mu$m and 160 $\mu$m Spitzer number counts 
by \citet{bethermin2010a}, based on GOODS/FIDEL, COSMOS and SWIRE fields.
Both individual detection and stacked Spitzer counts (hatched areas) are shown.
PACS counts and previous results are in good agreement, over the flux
range in common. The PEP deepest
field, GOODS-S, extends the knowledge on far-IR number 
counts one order of magnitude deeper in flux than Spitzer individual detections
at 160 $\mu$m and roughly 5 times deeper at 70 $\mu$m. The 3$\sigma$ limit 
in GOODS-S (1.2 mJy and 2.4 mJy at 100 and 160 $\mu$m, respectively) is
very close to the effective depth reached by \citet{altieri2010} in the Abell
2218 cluster when studying lensed background galaxies, but our improved source
statistics provide much tighter uncertainties.

In order to provide a single reference counts description, the number counts 
belonging to the four PEP fields studied have been combined via a 
simple average, weighted by their respective 
uncertainties in each flux bin. Results are included in Tabs. \ref{tab:counts_070}, 
\ref{tab:counts_100}, \ref{tab:counts_160}, and are shown 
in the right panels of Fig. \ref{fig:pep_counts}, as compared to a collection 
of model predictions (see Sect. \ref{sect:models}).

The depth reached by PACS/PEP allows us to accurately probe the faint-end of
counts. The peak in the normalized counts is well sampled and turns out
to lie at $\sim$4 mJy at 70 $\mu$m, $\sim$10 mJy at 100 $\mu$m, and $\sim$20-30 mJy
at 160 $\mu$m. 
The differential counts are reproduced by a broken power law ($dN/dS\propto
S^{\alpha}$), characterized by a break at flux $S_{break}$ and two 
distinct slopes at the at faint/bright sides of the break. A weighted least
squares fit was performed on the data, and the results are presented in Tab.  
\ref{tab:slopes1}, for different fields and flux ranges. Breaks happen at
$\sim$3.5 mJy at 70 $\mu$m, $\sim$5.0 mJy at 100 $\mu$m, and $\sim$9.0 at 160
$\mu$m. 
Uncertainties at the bright end are dominated by Poisson
statistics, and nearly-Euclidean slopes are allowed.


\subsection{Field to field variations}\label{sect:field_to_field}

The large area probed by COSMOS ($\sim$2 deg$^2$) can be used to test the effect
of field-to-field density variations on the bright-end of number counts. We 
adopt a fully empirical method, describing the variance in number counts 
coming from the inferred variations, while a full clustering quantification 
goes beyond the scope of this paper. 
To this aim, we split the COSMOS field into a number of fully independent tiles, probing
different angular scales: 28 tiles with size $\sim$200 arcmin$^2$, similar to
the size of GOODS fields, 12 Lockman Hole like tiles ($\sim$500 arcmin$^2$),
and 4 tiles of $\sim$1500 arcmin$^2$ each.

\begin{figure}[!ht]
\centering
\includegraphics[width=0.45\textwidth]{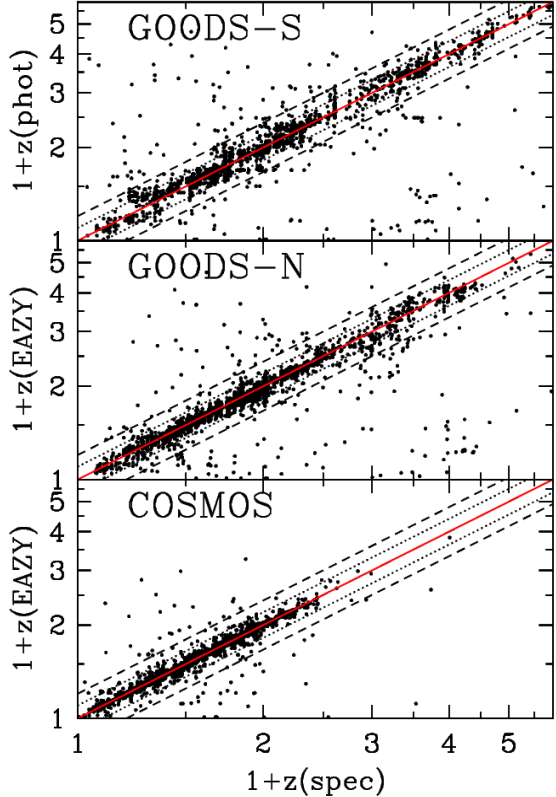}  
\caption{Comparison between photometric and available spectroscopic redshifts in
GOODS-S, GOODS-N and COSMOS, regardless of PACS detections. Dotted and dashed lines mark
10\% and 20\% uncertainty levels.}
\label{fig:photoz_specz}
\end{figure}

Number counts in each tile were computed as described before and are plotted in
Fig. \ref{fig:counts_field_to_field} as thin solid lines. Black symbols
represent the average counts in each flux bin, and the associated error bars are
computed in the usual manner accounting for Poisson statistics, systematics and
photometric uncertainties.

The properties of field-to-field standard deviation are studied in the bottom diagrams 
of Fig. \ref{fig:counts_field_to_field}.
As expected, the $\sigma(var)$ in number counts increases as a function of flux 
(upper panels), both because more luminous sources are
rarer in the sky and because lower-redshift objects are dominating at these
fluxes, hence probing a smaller volume.
Over the whole flux range covered by this analysis, 
the field-to-field deviation
is comparable to the uncertainty $\sigma_P$ obtained combining Poisson statistics, photometric errors
and systematics (bottom panels). Overall, $\sigma(var)/\sigma_P$ is slightly larger than unity within the 
errors. This effect can be explained considering that neighboring sub-tiles are not fully independent: because 
of clustering, a non-negligible correlation term contributes to $\sigma(var)$.
At larger scales (dotted green lines), $\sigma(var)/\sigma_P$ is more noisy because 
of the limited number of tiles available.


\subsection{Ancillary data and multi-wavelength catalogs}\label{sect:ancillary}

The fields observed with PACS as part of the PEP survey benefit from a plethora
of ancillary data, spanning from the x-rays to radio frequencies. We took
advantage of these data to build reliable multi-wavelength catalogs and
associate a full spectral energy distribution (SED) and a redshift estimate to
each PACS-detected object.

As described in Paper~I, a PSF-matched catalog was created in GOODS-N, including
photometry from GALEX far-UV to Spitzer IRAC and MIPS 24 $\mu$m.
The Southern GOODS field is rich in coverage as well. Here we adopt the
PSF-matched catalog built by \citet{grazian2006}, to which we add the 24 $\mu$m
photometry by \citet{magnelli2009} and a collection of spectroscopic redshifts
for more than 3000 sources
\citep{balestra2010,popesso2009,santini2009,vanzella2008,lefevre2005,mignoli2005,
doherty2005,szokoly2004,dickinson2004,vanderwel2004,stanway2004a,stanway2004b,
strolger2004,bunker2003,croom2001,cristiani2000}.
Finally, we browsed the COSMOS public
database\footnote{http://irsa.ipac.caltech.edu/data/COSMOS/} and
combined the U-to-K broad- and intermediate-band photometry \citep[][containing 2,017,800
sources]{capak2007,ilbert2009}, the public IRAC catalogs, the 24 $\mu$m data
\citep{lefloch2009} and the available photometric 
\citep{ilbert2009} and spectroscopic \citep{lilly2009,trump2009} redshifts. 
As far as the Lockman Hole is concerned, an extensive ancillary catalog is
currently on the make by Fotopoulou et al. (\citeyear{fotopoulou2011}, in prep.), 
but is not available at the time of this analysis, hence this field will not be used in this piece of analysis.

PACS catalogs were matched to the ancillary source lists by means of a 
multi-band maximum likelihood procedure \citep{sutherland1992}, starting
from the longest wavelength available (160 $\mu$m, PACS) and progressively
matching 100 $\mu$m (PACS), 70 $\mu$m (PACS, GOODS-S only) and 24 $\mu$m
(Spitzer/MIPS) data.

When no spectroscopic redshift is available, a photometric estimate is
necessary. In GOODS-S we make use of the available photometric
redshifts by \citet{grazian2006}, while new
photo-$z$'s were produced in GOODS-N, exploiting the wealth of multi-wavelength
data collected, and adopting the EAZY \citep{brammer2008} code. 
Up to 14 photometric bands were used, depending on the data available. 
The top panel in Fig.
\ref{fig:photoz_specz} presents the comparison between photometric and
the available spectroscopic redshifts. The fraction of outliers, defined as
objects having $\Delta(z)/(1+z_{spec})\ge0.2$, is $\sim$6\% over the whole sample
of spectroscopic redshifts, and decreases to $\sim$2\% for sources with a PACS
detection. Most of these outliers are sources with few
photometric points available, or SEDs hardly reproduced by the available
templates. The median absolute deviation
(MAD\footnote{$MAD(x)=\textrm{median}\left(\left|x-\textrm{median}\left(x\right)\right|\right)$})
of the $\Delta (z)$ distribution\footnote{where here $\Delta$ denotes the
difference between photometric and spectroscopic redshift.} is 0.040 for the
whole catalog, and 0.038 for PACS-detected objects with spec-$z$ available.

\begin{figure*}[!ht]
\centering
\includegraphics[width=0.32\textwidth]{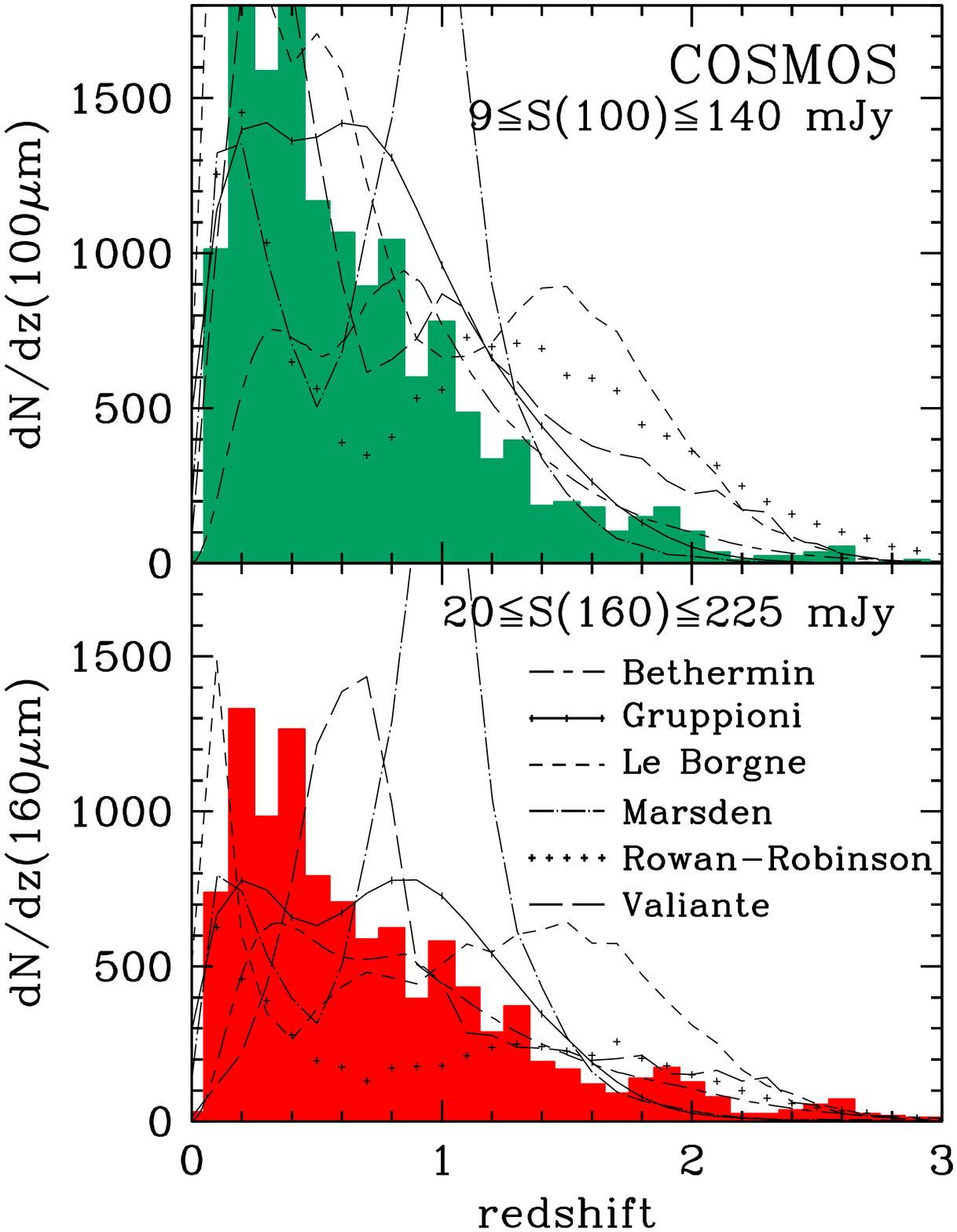}
\includegraphics[width=0.32\textwidth]{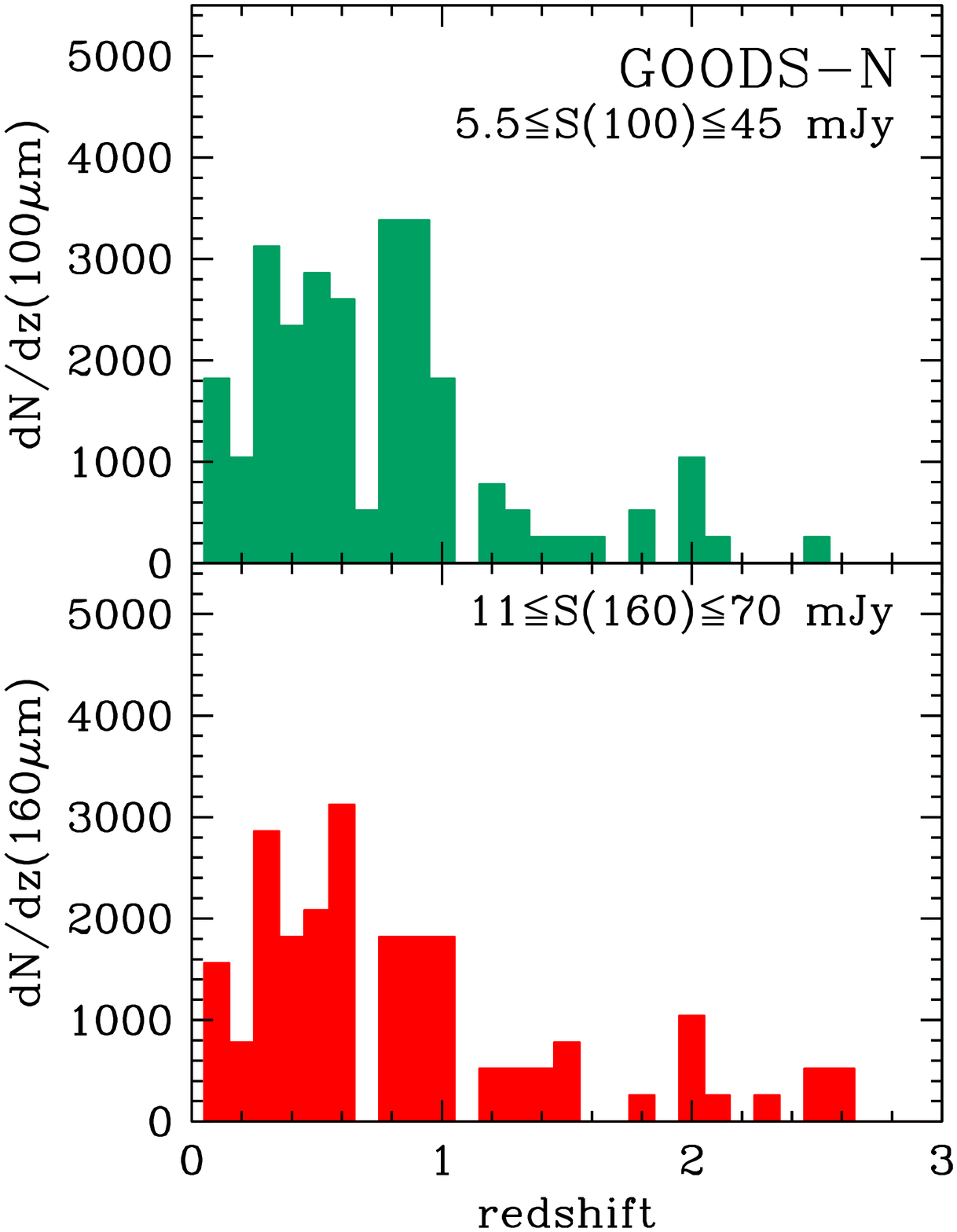}
\includegraphics[width=0.32\textwidth]{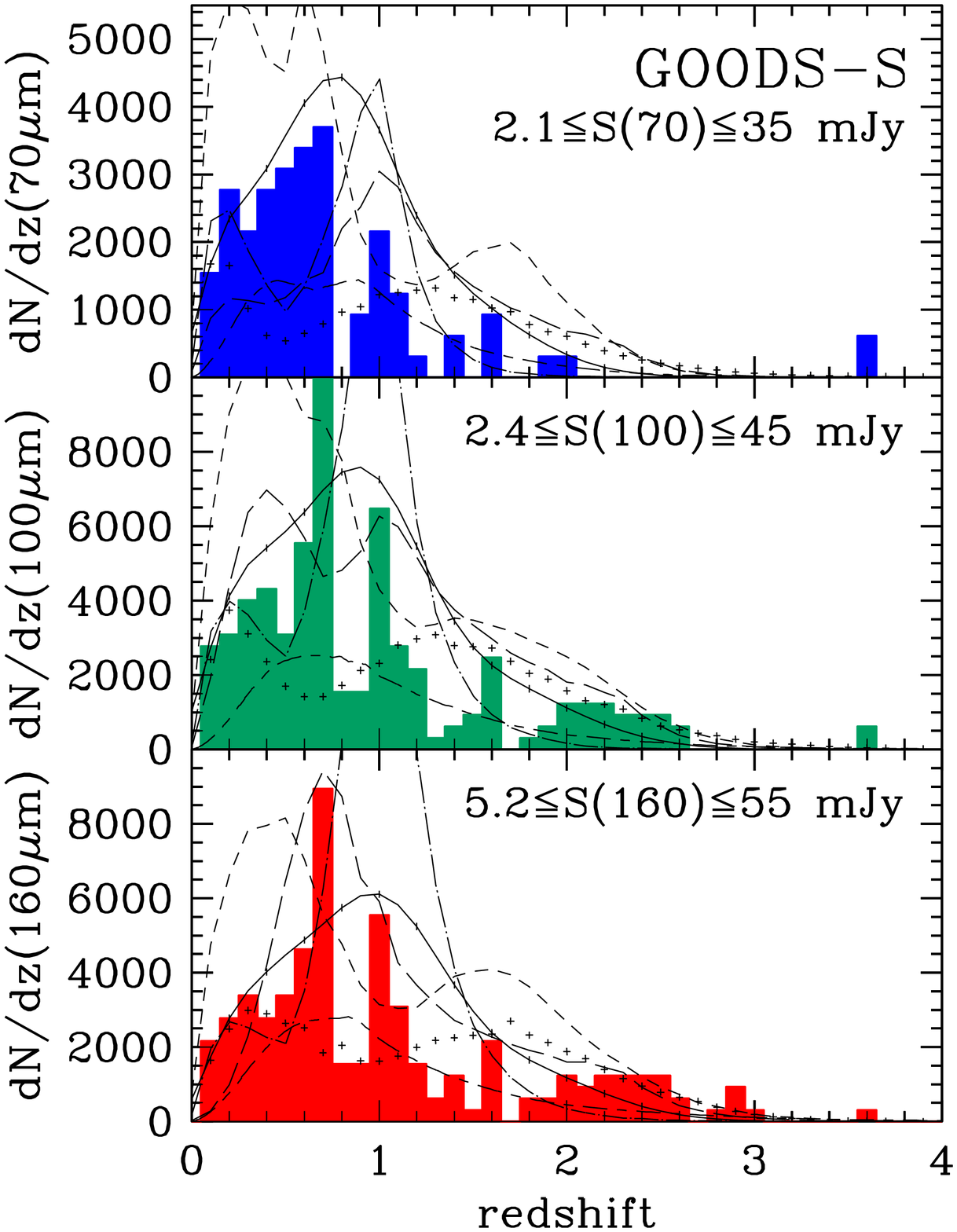}
\caption{Redshift derivative $dN/dz$ for PACS-detected sources in COSMOS ({\em left}),
GOODS-N ({\em center}) and GOODS-S ({\em right}), normalized to a 1 deg$^2$
area, and above the 80\% photometric completeness limit. Black lines refer to models 
and are reported only at the GOODS-S and COSMOS depths for clarity sake.}
\label{fig:zdistr}
\end{figure*}

In the area covered by ancillary data, roughly 60-65\% of GOODS-N sources
detected by PACS in either band have a spectroscopic redshift estimate. In the
GOODS-S MUSIC area \citep{grazian2006} this fraction is as high as $\sim$80\%. 
Roughly 95\% of these spectroscopic redshifts in GOODS-S lie at $z<2.0$, with 
an almost complete coverage. For the remaining PACS sources
we adopt photometric redshifts, obtaining a 100\% redshift completeness above
the 80\% photometric completeness threshold (see Tab. \ref{tab:fields}).
As far as COSMOS is concerned, the public photometric redshift catalog
\citep{ilbert2009} is
limited to a magnitude $I\le25$, thus producing a redshift incompleteness
in the PACS-selected sample. Only $\sim$75\% of PACS 
sources have a photometric redshift estimate in the \citet{ilbert2009} catalog. 
This incompleteness is independent of PACS fluxes.
We derived new photometric redshifts using the EAZY code and exploiting
the public COSMOS datasets. The bottom panel of Fig. \ref{fig:photoz_specz} shows
the results. The fraction of outliers is $\sim$1\% in this case, and MAD$(\Delta(z))
\simeq0.01$. This result is similar to that by \citet{ilbert2009} for the objects 
in common.

Given the non-null fractions of outliers in Fig. \ref{fig:photoz_specz}, 
it is possible that the number of sources at high redshift (e.g. $z\ge2$)
in PEP catalogs is in part contaminated by ``catastrophic photometric-redshift
failures'', mainly represented by low-$z$ objects with wrong, high photo-$z$.
Because of the paucity of $z\ge2$ PACS sources benefiting from a spectroscopic 
follow-up, the only viable approach to test this effect is to 
assume that the distribution of $\Delta(z)/(1+z_{spec})$ of PACS-detected objects 
is similar to that of the general galaxy population. For this reason, the 
resulting contamination fraction has to be considered as an upper limit only.

The fraction of potential contaminants at $z\ge2$ is computed 
in two steps. First the fraction of $z<2$ objects having $z_{spec}<2$, but 
$\Delta(z)/(1+z_{spec})\ge0.2$ and $z_{phot}>2$ is derived; then this is 
re-scaled to the ratio of $z\gtrless2$ PACS sources. It turns out that 
up to $\sim$25\% of GOODS-S $z\ge2$ PACS objects might have an improperly attributed 
high redshift. 
Similar results were obtained in GOODS-N, while this effect cannot be 
properly tested in COSMOS, because public $z\ge2$ spectroscopic redshifts are 
currently still lacking. 

We also note that the opposite phenomenon --- namely high-redshift sources with wrong 
photo-$z$ potentially being shifted at low-$z$ --- does not play a significant role here, 
because the vast majority of $z<2$ PEP/PACS objects benefits from a spectroscopic redshift 
measurement.

\subsection{Contribution to the counts from different epochs}\label{sect:counts_redshift}

We exploit the rich ancillary information described in the previous Section to
perform a detailed study of counts across cosmic time.
The redshift derivative $dN/dz$ of PACS sources above the 80\% photometric
completeness limits, normalized to 1 deg$^2$, is shown in Fig. \ref{fig:zdistr} for the three fields and
bands considered here. The covered flux ranges are quoted on each panel. 
In the COSMOS area, we sample the bright end of
PACS counts. The distribution in this field peaks at $z\le0.5$ in both 100 and
160 $\mu$m bands, with 70-80\% of all sources lying below $z=1$ and $\sim$20\% between 
$z=1$ and $z=2$. 
For the deeper GOODS-N data, the peak of the redshift distribution shifts to $z\sim0.7$.
Although the small sampled area limits source statistics, objects
at high redshift (up to $z\simeq5$) start to pop up. Our deepest field, GOODS-S,
covered in all three PACS bands, displays some remarkable features. Overall
$dN/dz$ is now peaked around $z\sim1$. It is possible to recognize
two well known structures at $z\simeq0.7$ and $z\simeq1.1$
\citep[e.g.][]{gilli2003,vanzella2005}, which produce narrow
and intense spikes in the distribution at 100 and 160 $\mu$m. On the other hand,
at 70 $\mu$m these structures are barely seen. At higher redshift, a broad
``bump'' is detected, between $z=2-3$. This peak cannot be identified in the
shallower 70 $\mu$m data, but is outstanding at the other two PACS wavelengths.
Cutting the GOODS-S catalog at the depth
reached by GOODS-N (5.5 mJy and 11.0 mJy at 100 and 160 $\mu$m, respectively),
the high-$z$ feature  
disappears and the redshift distribution resembles that of GOODS-N, with
only a few sources left above $z\ge2$. Similarly, when cutting GOODS-S at the
COSMOS 80\% depth, we retrieve a distribution peaked at $z=0-0.5$, obviously
with much poorer statistics than in the COSMOS field itself. 
An extensive analysis of PACS GOODS-S large scale structure at $z=2-3$ and of a
$z=2.2$ filamentary over density is being presented by Magliocchetti
et al. (\citeyear{magliocchetti2011}, sub.).
Table \ref{tab:zdistr} reports $dN/dz$ $[$deg$^{-2}]$, as derived in these three PEP fields. 
The median redshifts of the sources detected in GOODS-S are $z\sim0.6$, $\sim0.7$, $\sim0.8$ 
in the three PACS bands, but would shift if the known spikes in $dN/dz$ were not present.

Finally, it is possible to split
number counts in GOODS-S, GOODS-N and COSMOS into redshift bins, similarly to what was done in Paper~I for
GOODS-N. Counts are then constructed in four broad bins, in order to allow for
a sufficient number of sources in the small fields. In every redshift bin we apply the same completeness correction derived
for the total number counts at the given flux. 
Similarly to what was done for the total counts, 
we build combined number counts through a weighted mean in each flux bin. The 
average number counts, sliced in redshift intervals are reported in Tabs. 
\ref{tab:counts_redshift_070}, \ref{tab:counts_redshift_100}, and \ref{tab:counts_redshift_160}, 
and are shown in Fig. \ref{fig:counts_zbins}.
PACS observations in the three fields nicely complement each other: 
the southern GOODS field reaches deep flux densities, not probed previously, and also 
traces the high redshift population that was missing in GOODS-N. 
The bright end of counts is sampled by observations in the COSMOS area, which 
not only contribute to the low-redshift counts, but also nicely trace the bright 
component up to $z=2$ and beyond.


\subsection{Comparison to model predictions for counts across time}\label{sect:models}

Number counts and redshift distributions encode the evolution of galaxies: 
the upturn at intermediate fluxes and the over-Euclidean slope at the bright side of the 
peak are usually interpreted as signatures of strong evolution in
the properties of the underlying galaxy populations, and have stimulated a
plethora of model interpretations.

We are comparing here our results to examples of two basic classes of models 
attempting to reproduce these observables.
{\em Backward evolutionary models} transform the statistical properties of far-IR
galaxies observed at known redshift (e.g. local luminosity functions) into
observables at any redshift (e.g. galaxy counts, CIB brightness, etc.) 
assuming a library of template SEDs and parametric laws of luminosity and/or density
evolution. These models do not implement fundamental physical information but simply
attempt to describe the evolution of galaxy populations 
\citep[e.g.][]{lagache2004,franceschini2010,rowanrobinson2009,
leborgne2009,valiante2009,bethermin2010c,marsden2010,gruppioni2011}.

\begin{figure}[!ht]
\centering
\rotatebox{-90}{
\includegraphics[height=0.49\textwidth]{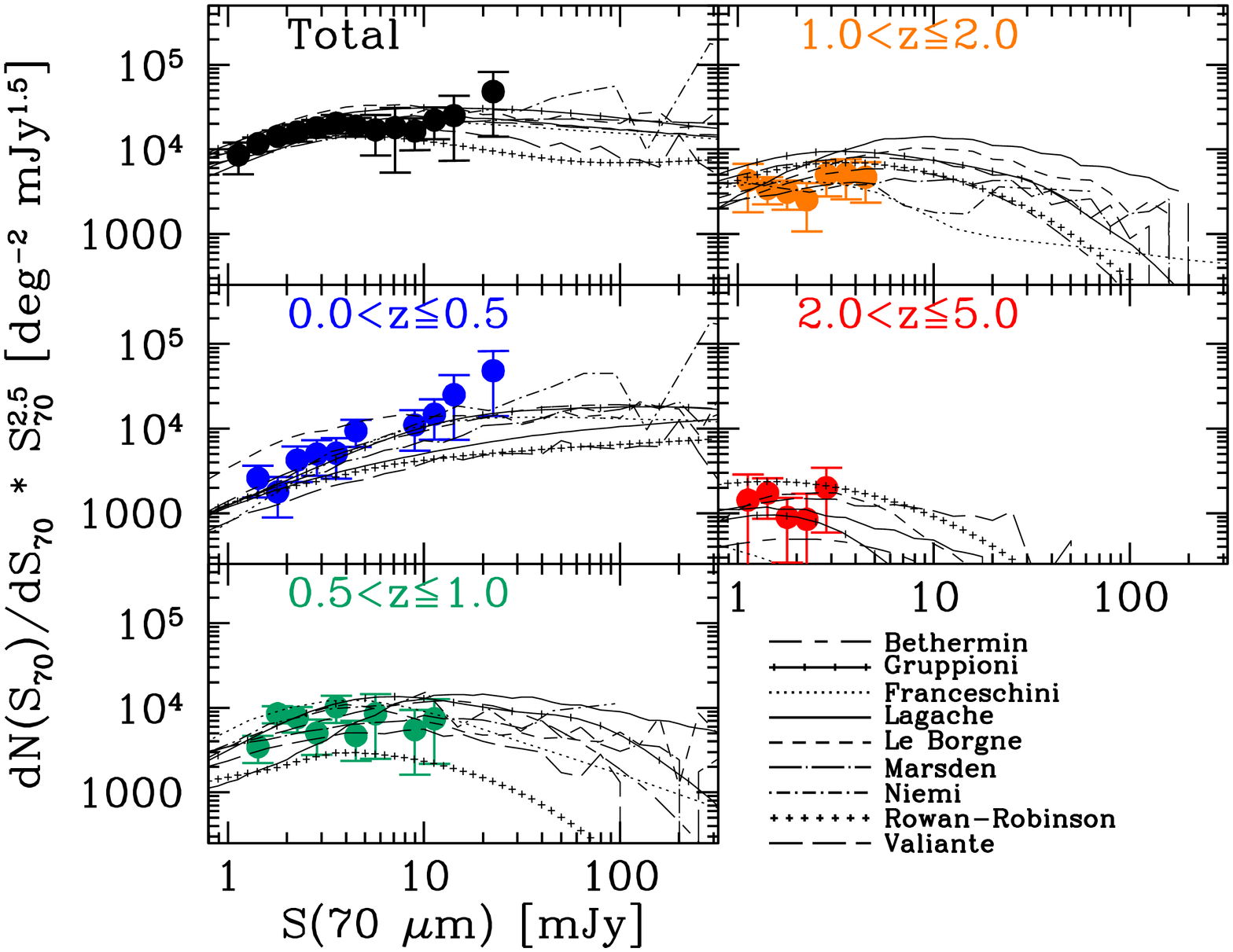}
}
\rotatebox{-90}{
\includegraphics[height=0.49\textwidth]{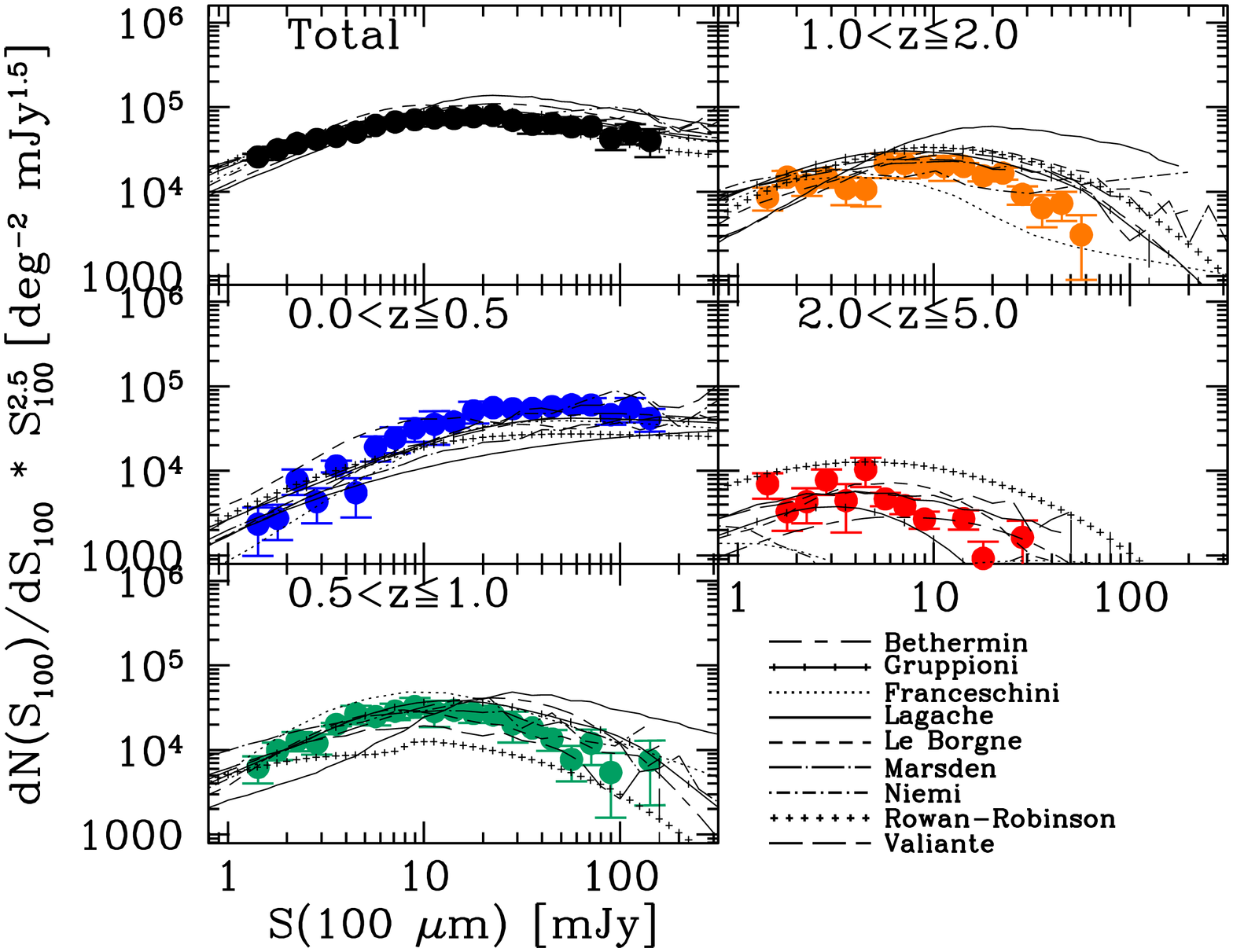}
}
\rotatebox{-90}{
\includegraphics[height=0.49\textwidth]{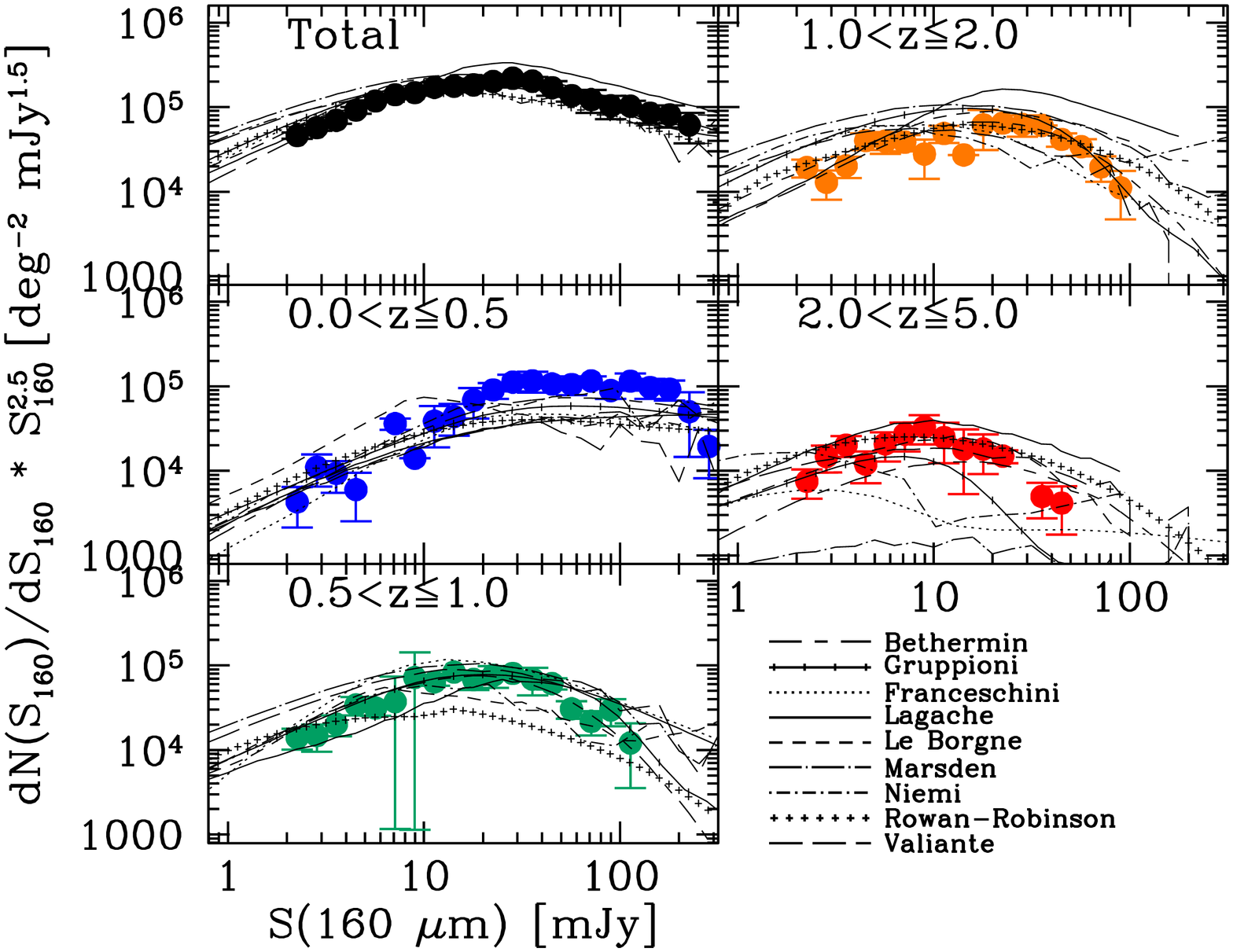}
}
\caption{Differential number counts in the three PACS bands, normalized to the
Euclidean slope and split in redshift bins. The 70 $\mu$m counts belong to the 
GOODS-S field, while those at 100 and 160 $\mu$m were obtained via a weighted average between 
GOODS-S, GOODS-N and COSMOS. See Fig. \ref{fig:pep_counts} for models 
references.}
\label{fig:counts_zbins}
\end{figure}

Very simple in their principles, backward models 
are thus parameterizations embedding the knowledge obtained from previous observations. 
The different adopted recipes were optimized to reproduce available observables such as 
mid-IR ISO and Spitzer number counts and far-IR Spitzer counts based on detections 
and  --- in some cases --- stacking, sub-mm counts, as well as redshift distributions.
A couple of models took advantage of early Herschel results, namely SDP PACS and SPIRE 
number counts \citep{bethermin2010c} and PACS luminosity functions up to $z\sim3$ 
\citep{gruppioni2011}. 
Generally speaking, confronting backward models to the new, detailed Herschel data is a test to their 
flexibility in producing reliable predictions at far-IR wavelengths.

At the opposite side, {\em forward evolution} models simulate the physics of galaxy 
formation and evolution forward in time, from the Big Bang to present days. 
Current implementations are based on semi-analytic
recipes (SAM) to describe the dissipative and non-dissipative processes influencing 
galaxy evolution, framed into $\Lambda$-CDM dark matter numerical simulations 
\citep[e.g.][Niemi et al. in prep.]{lacey2009}. Galaxy radiation, including 
far-IR emission, is computed with spectrophotometric synthesis and radiation transfer 
dust reprocessing, given the fundamental properties of the galaxies in the model.

Figures \ref{fig:pep_counts} and 
\ref{fig:counts_zbins} show a collection of models overlaid to the observed PACS
counts.
Most of the models reproduce fairly well the observed 100 and 160 $\mu$m
normalized counts, on average the most successful being \citet{franceschini2010}, 
\citet{marsden2010}, \citet{rowanrobinson2009}, \citet{valiante2009} and 
\citet{gruppioni2011}.
Very different assumptions produce relatively similar number counts predictions. 
The \citet{franceschini2010} model adopts four different galaxy populations, including 
normal galaxies, luminous infrared galaxies, AGNs and a
class of strongly-evolving ULIRGs, which dominate above $z\ge1.5$, but is
negligible at later epochs, resembling high-$z$ sub-mm galaxies. 
Also \citet{rowanrobinson2009} uses four galaxy populations (cirrus-dominated
quiescent galaxies, M82-like starbursts, Arp220-like extreme starbursts, AGN dust
torii), but employs analytic evolutionary functions without discontinuities. 
\citet{valiante2009} describe galaxy population taking into account the observed
local dispersion in dust temperature, the local observed distribution of AGN
contribution to $L_{TIR}$ as a function of luminosity, and an evolution of 
the local luminosity-temperature relation for IR galaxies.
\citet{marsden2010} base their SEDs on \citet{draine2007} prescriptions and tune them to 
reproduce the local color-dependent far-IR luminosity functions; they 
include luminosity, density and color evolutions to fit sub-mm counts, 
redshift distributions and EBL.
Finally, \citet{gruppioni2011} include a significant Seyfert-2 
population, based on a fit to Herschel LFs \citep[see also][]{gruppioni2010}. 

Semi-analytical approaches surely represent a more complete view of galaxy 
evolution, including a wide variety of physics in a single coherent model. 
They cover a wide range of observational data: UV, optical, near-IR luminosity 
functions, galaxy sizes, metallicity, etc. Moreover, at far-IR wavelengths, 
even in the case that global properties such as star formation and AGN activity 
are correctly modeled, a further complication arises from the assumptions 
about dust content and structure, which need to be invoked. As a result, the large 
number of parameters involved goes at the expense of inference precision:
the performance of SAM models with respect to PACS observables 
needs still substantial tuning.

As described above, it seems that pre-Herschel models, with only a few exceptions, 
are quite successful at reproducing total number counts 
despite the range and diversity of the employed solutions, thus showing that 
the discriminatory power of this observable is rather limited. 

The right answer comes from the redshift information available 
in the selected PEP fields. Figure \ref{fig:zdistr} presents 
the redshift distribution $dN/dz$ of PACS galaxies, and Fig. \ref{fig:counts_zbins}
shows number counts split in redshift slices ($d^{\,2}N/dS/dz$).
The combination of the two is a real discriminant: model predictions are 
now dramatically different. 

None of the available models 
provides a convincing coherent prediction of the whole new set of observables 
(both $dN/dz$ and $d^{\,2}N/dS/dz$), over the flux range covered.
One common source of discrepancy seems to be the well-know degeneracy between 
dust temperature (and hence luminosity and total dust mass) and 
redshift \citep[e.g.][]{blain2003}: red objects can be either cool local galaxies 
or warmer distant galaxies. 
What is observed is a mis-prediction 
of the redshift distribution, reflected for example in an overestimation of high redshift counts 
and an underestimation at later epochs --- or vice versa. 
Several models seem to systematically over predict the number of galaxies above $z\simeq1$
in the deep regime.
A variety of SED libraries are adopted in these model recipes, based on different 
assumptions and templates. These differences in SED shapes and their evolution --- as well as the implementation 
of luminosity and density evolution for the adopted galaxy populations --- 
indeed produce significantly different results.

Among all, the \citet{bethermin2010c} distribution is probably 
the closest to observations, although it presents a significant 
underestimation of $dN/dz$ at 70 $\mu$m. This model was optimized 
taking into account differential number counts between 15 $\mu$m and 1.1 mm, 
including PACS \citep{berta2010} and SPIRE \citep{oliver2010} early results, 
mid-IR luminosity functions (LF) up to $z=2$, the far-IR local LF, and CIB 
measurements.
This success demonstrates the need to include in model tuning (ideally by means 
of a proper automated fit) not only Herschel data, but also the detailed redshift
information (e.g. evolving LF, number counts split in redshift bins, redshift distributions, etc.).

In summary, while several pre-Herschel backward evolutionary models already provide reasonable
descriptions of the total counts in at least some of the PACS bands, they tend to fail
in the synopsis of all counts and in particular redshift distributions. 
The new Herschel dataset is a treasure box, allowing to explore 
the evolution of counts and CIB with unprecedented detail and much deeper than previous 
data, on which models were calibrated.
Significant modifications on model
assumptions for SEDs and/or evolution will be needed for a satisfactory fit to this new
quality of data.


\section{Level 2: stacking of 24 $\mu$m sources}\label{sect:stacking_counts}

Limiting the number counts analysis to individually-detected sources only, we
miss a significant fraction of the information stored in PACS maps.
It is possible to recover part of this information by performing 
stacking of sources from deeper data 
\citep[typically at shorter wavelengths, e.g.][]{dole2006,marsden2009,bethermin2010a}.
The 24 $\mu$m catalog in the GOODS-S field, extending down to $\sim$20 $\mu$Jy
\citep{magnelli2009}, provides the ideal priors to perform stacking of
faint sources on PACS maps.

\begin{figure}[!ht]
\centering
\includegraphics[width=0.45\textwidth]{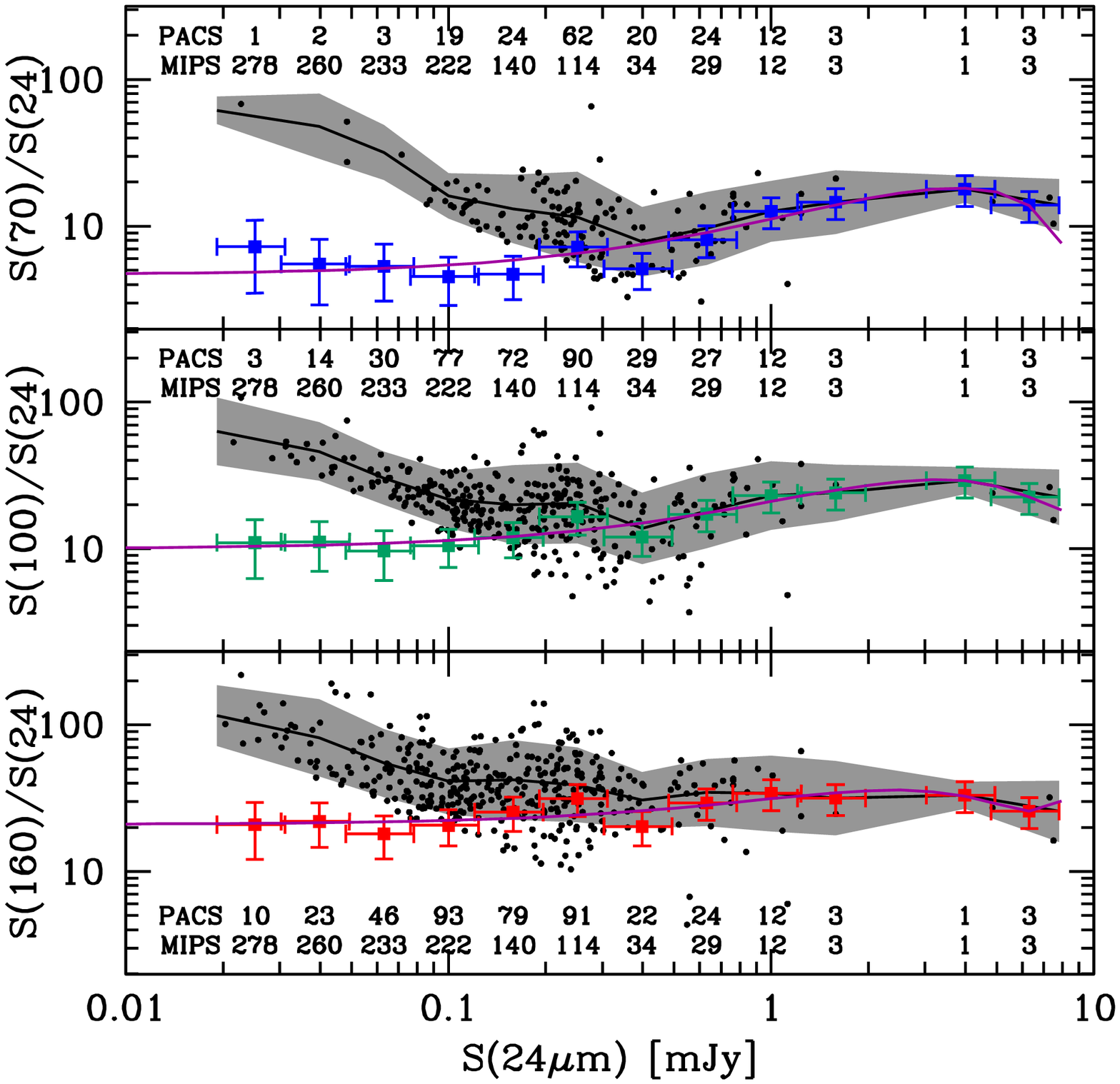}
\rotatebox{-90}{
\includegraphics[height=0.45\textwidth]{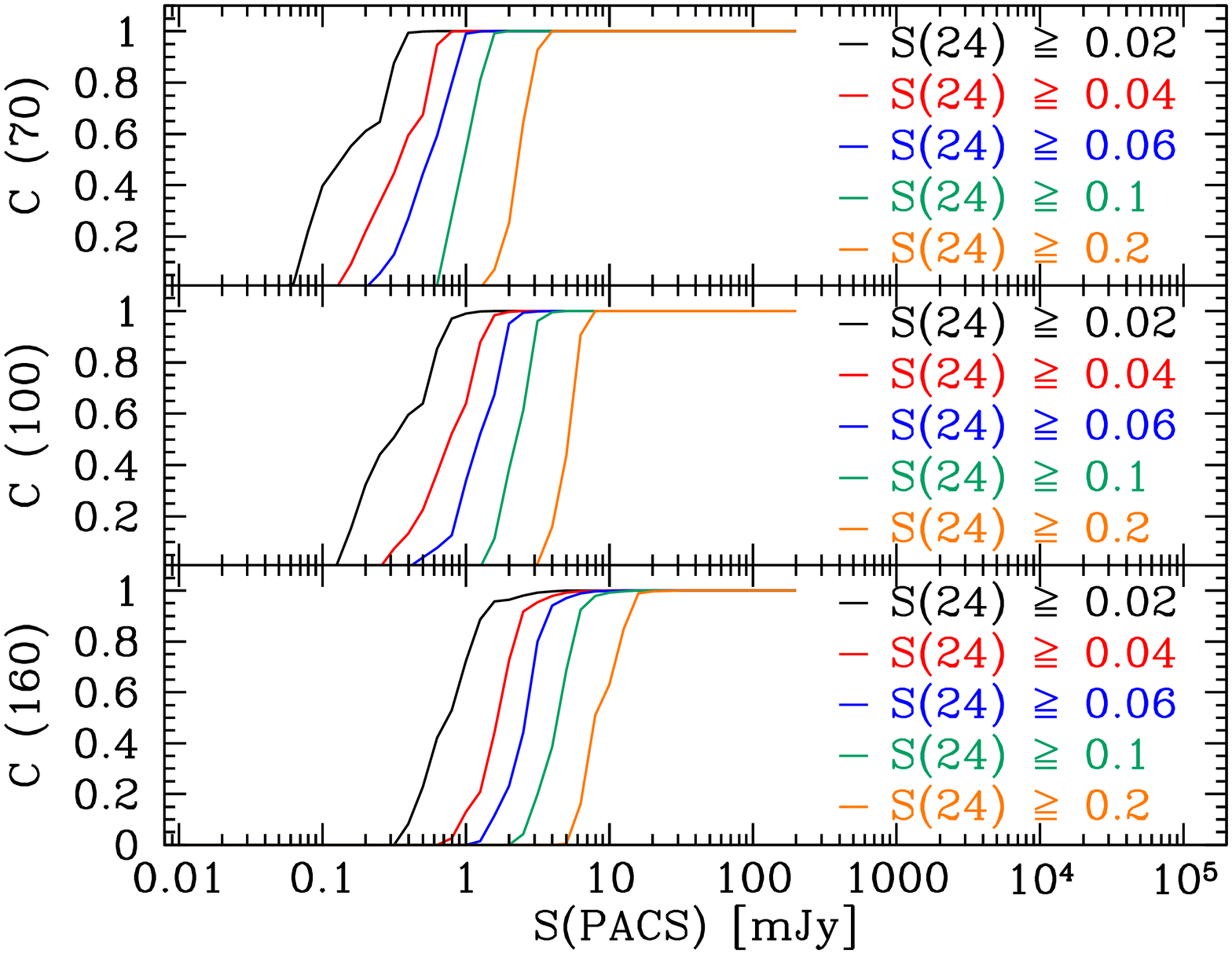}
}
\caption{Stacking of 24 $\mu$m sources on PACS maps in GOODS-S. 
{\em Top panel}: PACS/24$\mu$m colors of sources, as measured from individual detections (black
dots) and detections+stacking (colored symbols). Black solid lines and grey shaded areas
represent average colors and dispersion of PACS-detected sources. 
Solid purple lines are polynomial fits to the stacked points. For each 24 $\mu$m
flux bin the number of sources detected by PACS and the total number of MIPS 24
$\mu$m objects are quoted.
{\em Bottom panel}: Completeness analysis, based on the \citet{leborgne2009}
backward-evolution model; the quoted 24 $\mu$m fluxes are in $[$mJy$]$ units.}
\label{fig:stacked_colors}
\end{figure}

\begin{figure}[!ht]
\centering
\includegraphics[width=0.45\textwidth]{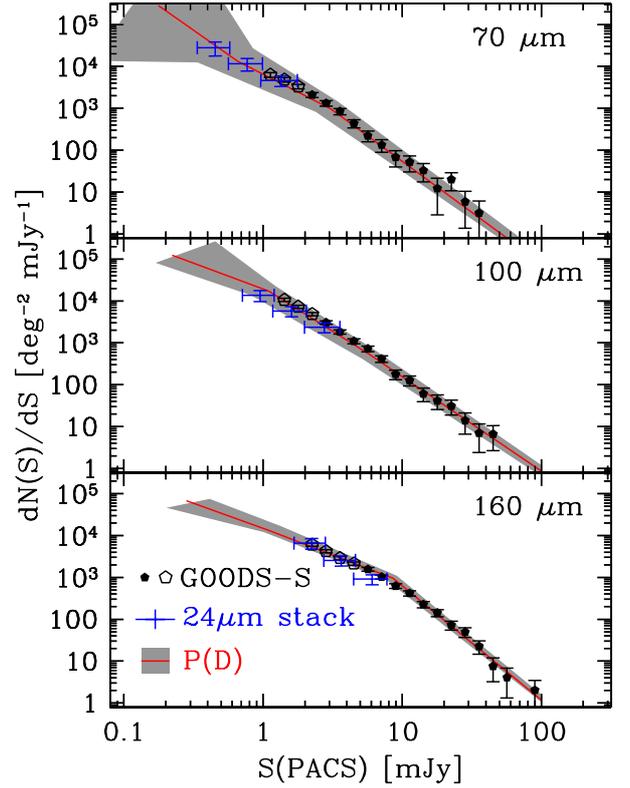}
\caption{Differential number counts $dN/dS$ in GOODS-S. Resolved counts (black
symbols), results of stacking (blue crosses) and $P(D)$ results (red solid line
and 3$\sigma$ grey shaded area) are shown.}
\label{fig:counts_not_norm}
\end{figure}

The aim here is to transform 24 $\mu$m faint number counts into 
PACS counts, given the average PACS/24$\mu$m colors of galaxies.
The overall procedure consists in building such colors as a function of 
24 $\mu$m flux, including both sources detected by PACS 
and un-detected ones. Then the derived colors are used to transform 
24 $\mu$m counts into PACS counts, probing the faint end of the distribution, below 
the PACS detection limit.

To reach this goal, we select 24 $\mu$m sources not detected in PACS maps, and
bin them by their 24 $\mu$m flux. 
Following the standard technique described by \citet{dole2006} and
\citet{bethermin2010a}, in each $S(24\mu$m$)$ flux bin, we pile-up
postage stamps at the position of these sources and produce stacked
frames at 70, 100 and 160 $\mu$m. 
We measure the flux density of the stacked sources by performing PSF-fitting in
the same way as for the individually-detected objects \citep[see][]{berta2010}.
Uncertainties on the stacked fluxes are computed through a simple bootstrap
procedure. Stacking was performed on PACS residual images, i.e. after removal of
individually-detected sources, and the fluxes of the latter were then added back
to stacking results. 
Finally, we checked that no significant signal was detected when stacking at
random positions, and that stacks of PACS-detected sources retrieved their actual
total summed flux. Flux corrections due to losses during the high-pass filtering 
process were tested via simulations and turned out to have a negligible 
effect on our results (see Lutz et al. \citeyear{lutz2011} for a description).

We derived average flux densities simply dividing by the total
number of sources in each 24 $\mu$m flux bin, and then we built PACS/24$\mu$m average
colors. Figure \ref{fig:stacked_colors} shows the results.
Stacking of mid-IR sources actually allows us to retrieve missed PACS fluxes
down to faint regimes, thus recovering the actual average colors, that
would not be possible to derive otherwise. On the bright end, average stacked+detected fluxes and
individual detections progressively converge, and become consistent within errors
when $\ge$55\% of 24 $\mu$m sources in the given flux bin are detected by PACS.

Stacked points were fit with a polynomial function and then used to transform the
observed 24 $\mu$m differential counts into PACS counts, following the recipe
suggested by \citet{bethermin2010a}:
\begin{equation}
\frac{dN}{dS_{PACS}} = \frac{dN}{dS_{24}}\times
\frac{dS_{24}}{dS_{FIR}}\textrm{,}
\end{equation}
where the right terms are computed in the given 24 $\mu$m flux bin and the left
term is computed at the corresponding $S_{FIR}=f(S_{24})$, as defined by the
average color functions. The derivative $dS_{24}/dS_{FIR}$ is estimated
numerically.

Although this method has been widely and successfully used by several authors, it
is worth to note that stacking results are affected by completeness
limitations. In this specific case, a flux cut in the 24 $\mu$m priors (e.g. the
Magnelli et al. \citeyear{magnelli2009} catalog reaches 20 $\mu$Jy at the
3$\sigma$ detection threshold) induces incompleteness in the PACS stacked fluxes,
because red objects (faint at 24 $\mu$m) are not included in the stack.
Consequently, stacking of 24 $\mu$m sources on the PACS maps provides only a lower
limit to the PACS number counts below a given PACS flux.
The bottom panel of Fig. \ref{fig:stacked_colors} is based on
\citet{leborgne2009} mock catalogs, and shows that, for a 24 $\mu$m cut of 20
$\mu$Jy, the PACS stacked counts are 90\% complete only 
above $\sim$0.35, $\sim$0.7 and $\sim$1.3 mJy at 70, 100 and 160 
$\mu$m, respectively. To this effect, one should add the intrinsic properties of
the 24 $\mu$m parent catalog, which reaches $\sim$90\% completeness at $S(24)\sim35$
$\mu$Jy. 

Figure \ref{fig:counts_not_norm} shows the differential number counts $dN/dS$ 
and includes the results of stacking 
above the 80\% stacking completeness (big blue crosses), in good agreement with
the faint-end of the resolved counts presented before (open symbols in Figs. 
\ref{fig:counts_not_norm} and \ref{fig:pep_counts}). The result of a power-law
fit is reported in Tab. \ref{tab:slopes1}: at 70 and 100 $\mu$m these slopes are
slightly flatter, while at 160 $\mu$m slightly steeper, than those of resolved
counts.


\section{Level 3: $P(D)$ analysis}\label{sect:pdd_counts}

\begin{figure*}[!ht]
\centering
\rotatebox{-90}{
\includegraphics[height=0.32\textwidth]{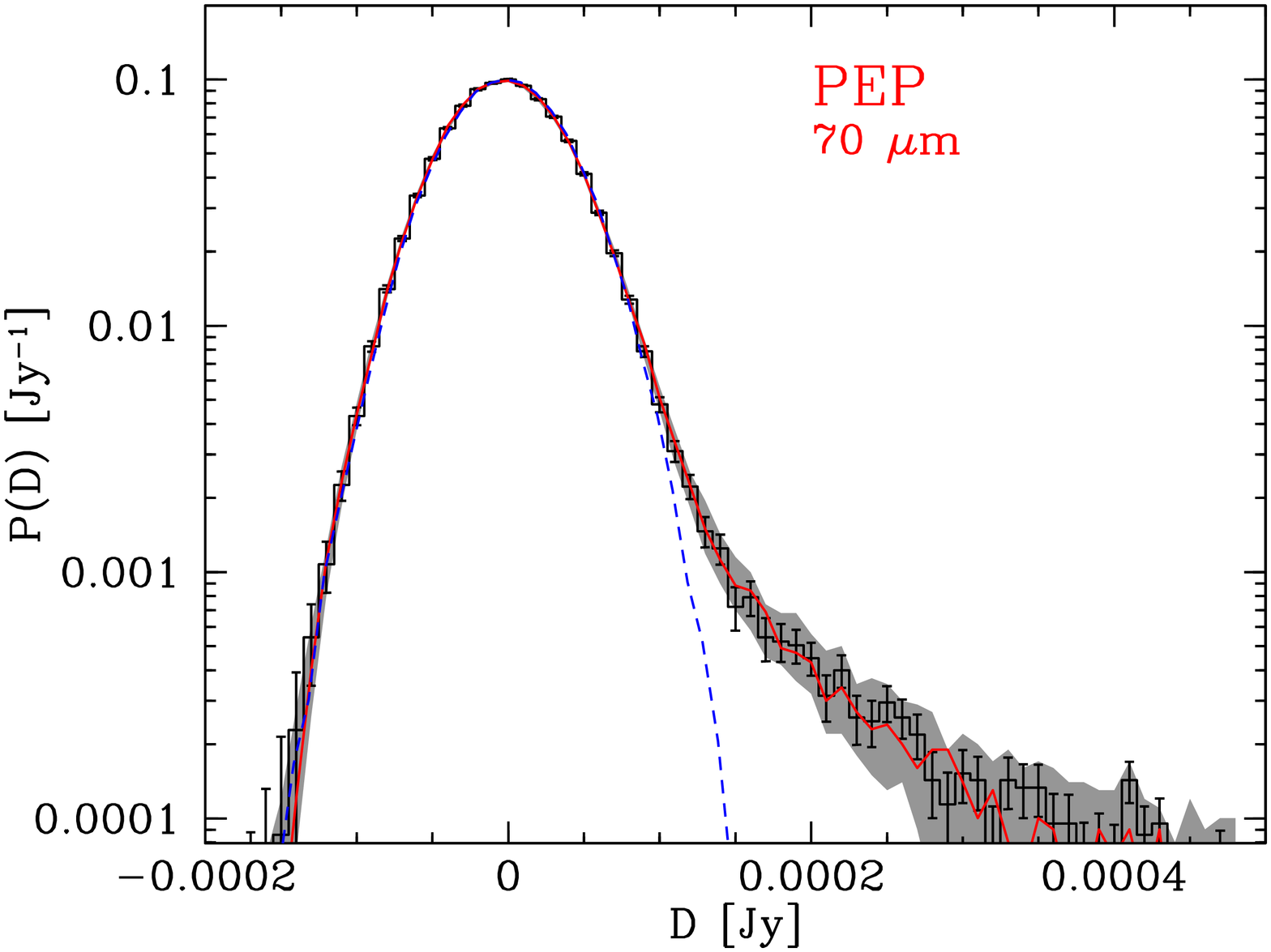}
}
\rotatebox{-90}{
\includegraphics[height=0.32\textwidth]{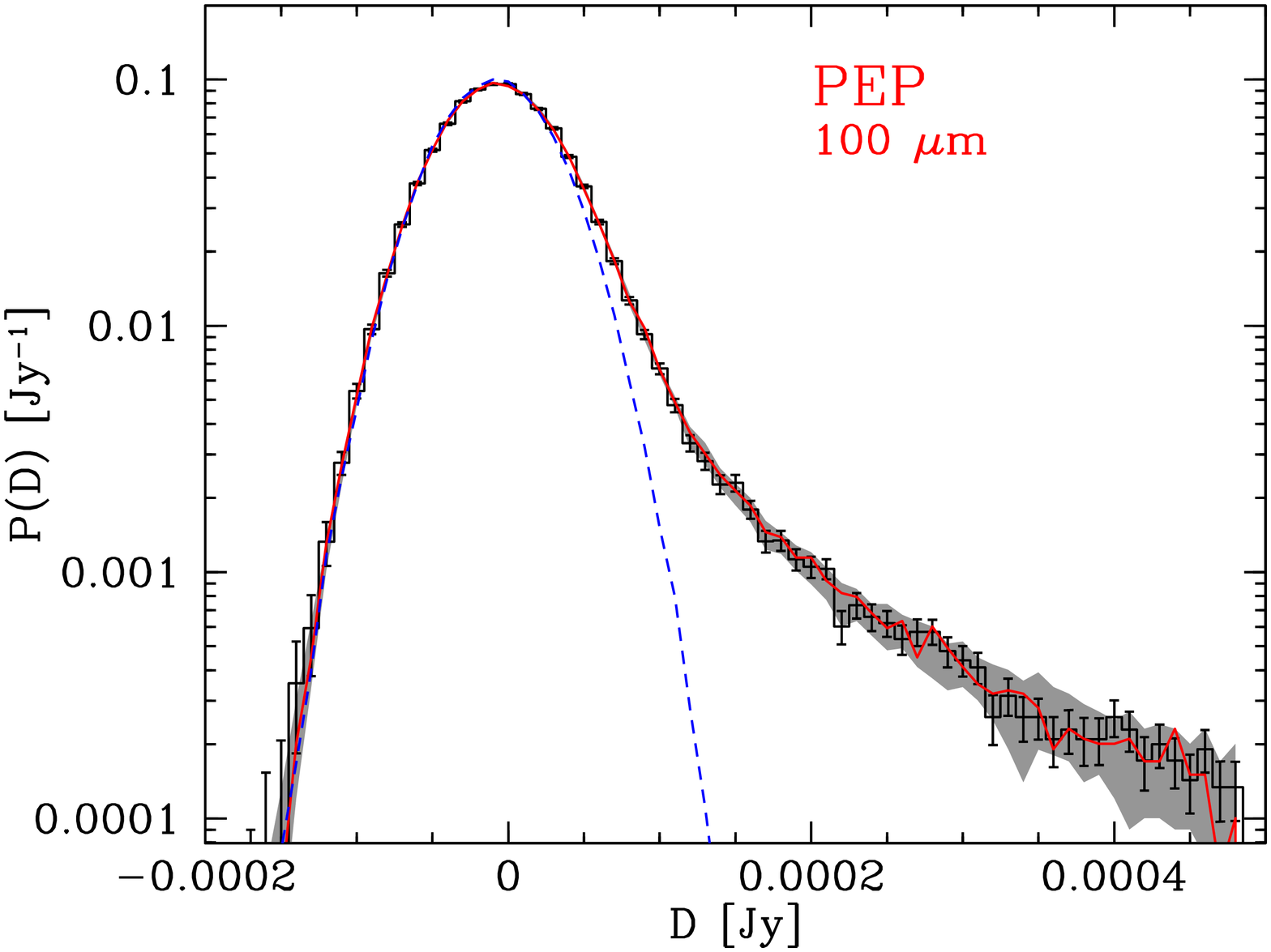}
}
\rotatebox{-90}{
\includegraphics[height=0.32\textwidth]{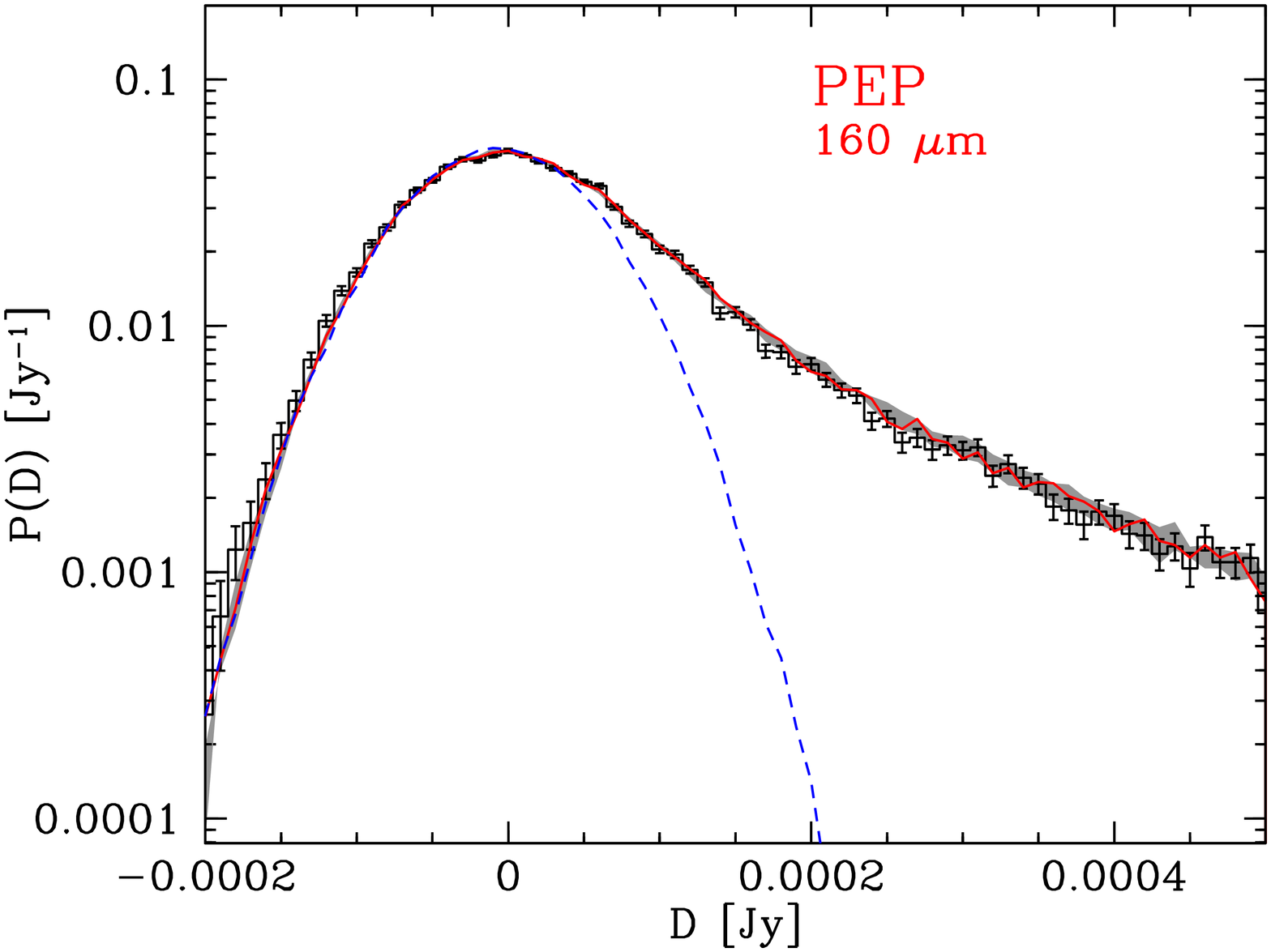}
}
\caption{$P(D)$ distributions in GOODS-S. The black histogram and error
bars belong to the observed $P(D)$, while the red solid line and grey shaded area
represent the best fit and its 3$\sigma$ confidence interval. The Gaussian noise 
contribution is depicted with blue dashed lines.}
\label{fig:p_of_d}
\end{figure*}

It is possible to extend the study of number counts and CIB to even
deeper flux regimes, beyond the confusion limit, exploiting the so-called 
``probability of deflection'' statistics, or $P(D)$ distribution 
\citep[e.g.][]{scheuer1957,condon1974,franceschini1989,franceschini2010,oliver1997,patanchon2009,glenn2010}. 
In a simplified notation, this is a representation of the
distribution of pixel values in a map: its shape and width are mainly driven by 
three components: number counts, the instrumental spatial
response function, and the instrumental noise. Once the three are 
known, it is possible to reproduce the observed pixel flux probability
distribution, or --- vice versa --- the observed $P(D)$ can be used to
constrain the underlying $dN/dS$ number counts.

The first $P(D)$ applications were carried out at radio frequencies
\citep[e.g.][]{scheuer1957,condon1974,franceschini1989}, but have subsequently been applied to
different wavelengths, spanning across the whole electromagnetic spectrum (e.g.
X-ray, Scheuer et al. \citeyear{scheuer1974}, Barcons et al.
\citeyear{barcons1994}; sub-mm, Hughes et al.
\citeyear{hughes1998}, Wei{\ss} et al. \citeyear{weiss2009}; Scott et al.
\citeyear{scott2010} Patanchon et al. \citeyear{patanchon2009}; infrared, Oliver et al.
\citeyear{oliver1997}).
Since the Herschel satellite was launched, a number of analyses have made use of
$P(D)$ techniques to describe the properties of real maps
\citep{glenn2010,oliver2010}, as well as
produce model predictions to be compared to actual far-IR observations
\citep{bethermin2010c,franceschini2010}.

We performed the $P(D)$ analysis on GOODS-S PACS maps, modeling 
differential counts $dN/dS$ as a broken power-law with three sections and two nodes.
To this aim, we developed a new, numeric method to
estimate the $P(D)$, given the counts model, the observed PSF and the observed
noise. The position of nodes and amplitude of counts at these positions are free
parameters. The best combination of parameters reproducing the observed $P(D)$
is sought by minimizing the difference between model prediction and data, via a
Monte Carlo Markov Chain (MCMC) engine.  Appendix \ref{sect:pdd_method} describes
the details of this method.

Figure \ref{fig:p_of_d} shows the fit to the observed $P(D)$ in the three PACS
bands; Fig. \ref{fig:counts_not_norm} and Tab. \ref{tab:pow_law_pdd} report on
the number counts model producing the best fit to the observed $P(D)$.
Note that no constraint was set to number counts, but only 
the observed $P(D)$ distribution was fit. It is remarkable to note 
the very good agreement between $P(D)$ results and the resolved number counts
at the bright end, and over the whole flux range covered by the latter.
At the very faint end, the number count model derived from $P(D)$ tends to
diverge, because of degeneracies. This is dramatically evident at 70 $\mu$m,
where the faintest section in $dN/dS$ is totally undefined. For this reason, we
limit the 70 $\mu$m result to $\sim0.35$ mJy, when computing CIB surface
brightnesses in the next Sections. At 100 and 160 $\mu$m, the $P(D)$ approach
allows the knowledge of PACS number counts to be extended down to $\sim$0.2 mJy,
one order of magnitude deeper than individual detections.
The resulting slopes are consistent with resolved and stacked counts, over the
flux range in common.


\section{Confusion noise}\label{sect:confusion}

The extreme depth reached with the $P(D)$ analysis, and the detailed view of
number counts coming from individually-detected sources, make worthwhile to take a
digression and study the properties of confusion noise in PACS maps.

Following the \citet{dole2003} formalism, two different approaches can be adopted
to describe the confusion noise due to extragalactic sources, i.e. the effect of
fluctuations due to the presence of point sources in the beam \citep[see
also][among others]{frayer2006,condon1974,franceschini1989}: the {\em photometric} 
and the {\em density} criteria. 

Given a detection threshold $S_{lim}$, the former is derived from the
signal fluctuations due to sources below $S_{lim}$, while the latter consists in
deriving the density of sources detected above $S_{lim}$ and limiting the
probability that objects are missed because blended with neighbors to a small
fraction (e.g. 10\%).

Here we apply both criteria to PACS/PEP data and directly measure the confusion
noise and limits of Herschel deep extragalactic imaging.

\begin{figure}[!ht]
\centering
\includegraphics[width=0.45\textwidth]{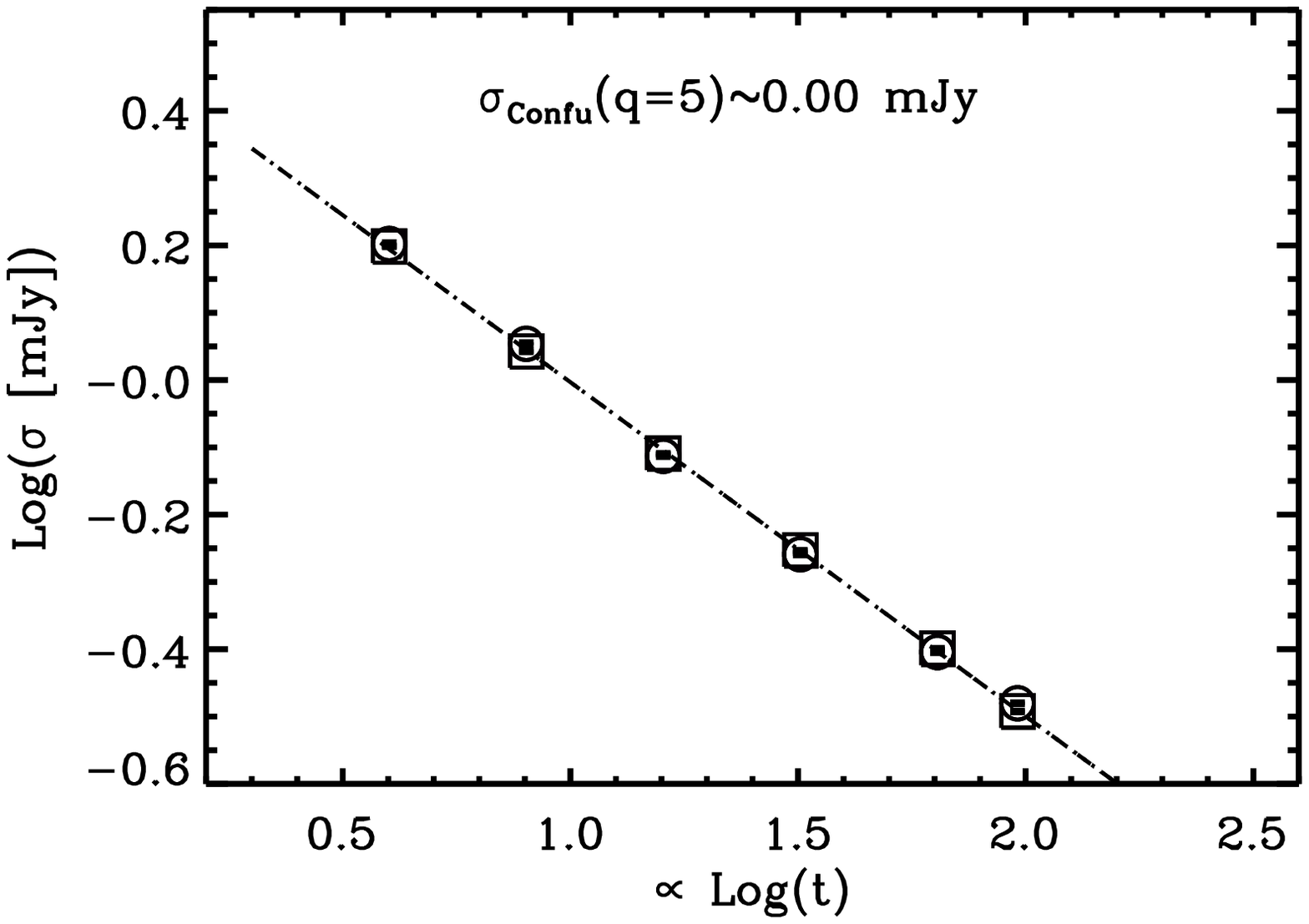}
\includegraphics[width=0.45\textwidth]{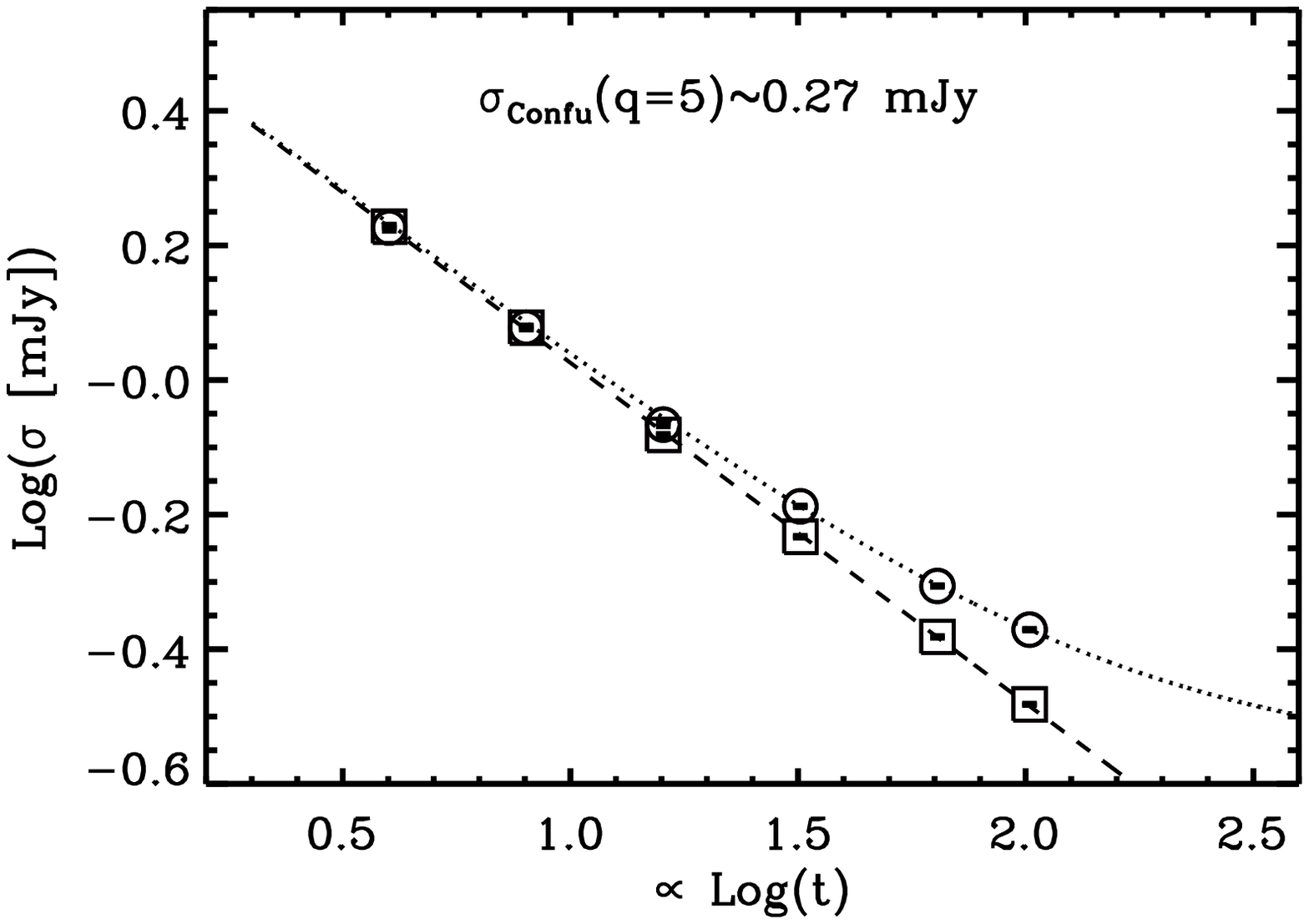}
\includegraphics[width=0.45\textwidth]{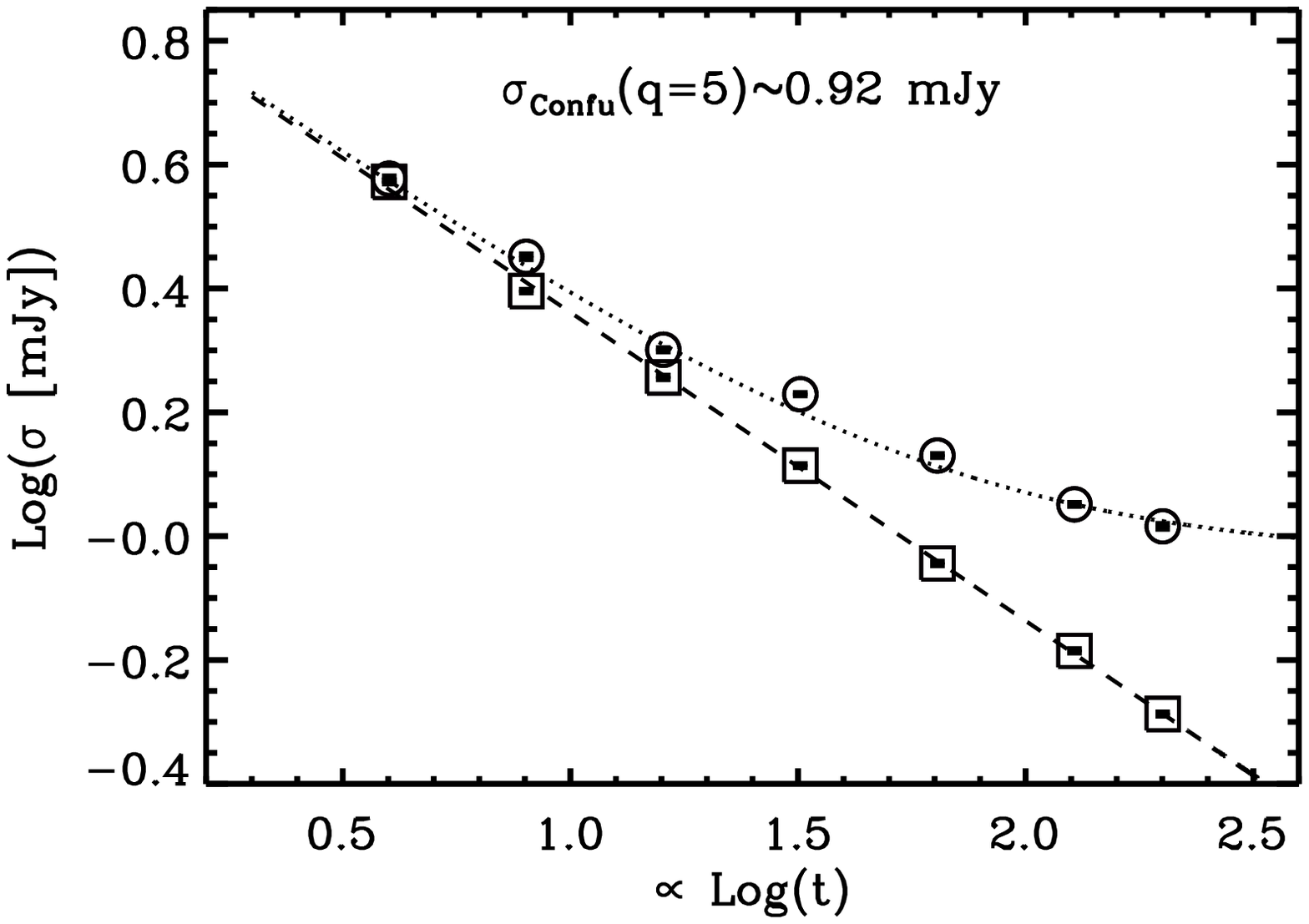}
\caption{Noise in PACS GOODS-S maps at 70, 100, 160 $\mu$m (top, middle, bottom
panels), as a function of exposure time. The dashed
line represent the trend $\sigma_I\propto t^{-0.5}$ followed by instrumental
noise; while the dotted line is $\sigma_T$ obtained by summing $\sigma_I$ and the
confusion noise $\sigma_c$ in quadrature. Squares mark the expected
pure-instrumental noise at the
exposure times considered, while circles denote the measured $\sigma_T$.}
\label{fig:phot_conf}
\end{figure}

\subsection{Photometric confusion}\label{sect:confusion_phot}

Photometric confusion is defined by setting the desired 
signal-to-noise ratio $q$ between the limiting flux 
$S_{lim}$ and the noise $\sigma_c$ describing beam-to-beam fluctuations 
by sources fainter than $S_{lim}$, i.e. $q=S_{lim}/\sigma_c$.
We adopt a value $q=5$ for our analysis. 

Generally speaking, the noise measurable on PACS images is given by 
the combination of the instrumental noise $\sigma_I$ (including photon noise,
detector noise and data-processing noise) and $\sigma_c$.
The instrumental noise was estimated empirically, similarly to
\citet{frayer2006}. 
We build partial-depth maps in the GOODS-S field, exploiting only a fraction of
the available PACS observations. By progressively increasing total exposure time,
we drew the diagrams in Fig. \ref{fig:phot_conf}. 
Sources were detected above $S_{lim}$ in the usual way
\citep[see][and Lutz et al. in prep.]{berta2010} and subtracted prior to measuring
the noise on the maps.
For short exposure times, noise is dominated by instrumental
effects, and is proportional to $t^{-0.5}$. This allows to estimate $\sigma_I$ and
extrapolate it to deeper regimes. When a long effective exposure time is used, 
the measured total noise $\sigma_T$ deviates from the $t^{-0.5}$ trend, testifying that
confusion noise plays an increasingly significant role.
Since both $\sigma_T$ (after source extraction) and $\sigma_I$ have nearly Gaussian distributions,
the confusion noise can also be approximated by
$\sigma_c=\sqrt{\sigma_T^2-\sigma_I^2}$.
We iterated between source extraction at different depths
and noise measurements, until convergence at $q=5$ was reached.  

No deviation from the instrumental $\sigma_I\propto t^{-0.5}$ is detected at 70
$\mu$m, at the depth of PEP GOODS-S observations, while we derive $\sigma_c=0.27$
and 0.92 mJy at 100 $\mu$m and 160 $\mu$m, respectively (with $q=5$).
Table \ref{tab:confusion} summarizes the derived values.

\subsection{Density criterion for confusion}\label{sect:confusion_dens}

The second criterion to determine the influence of confusion on extragalactic
observations is defined by requiring a minimum completeness level in the
detection of sources brighter than $S_{lim}$, driven by the fraction of objects
missed because blended to their (detected) neighbors.
Source number counts provide a neat one-to-one relationship between source
density and flux, so that the flux at which the source density confusion (SDC)
limit is reached is univocally defined, given $dN/dS$.

Adopting the \citet{lagache2003} definition of beam (i.e., $\Omega = 1.14 \times
\theta_{FWHM}^2$), and allowing a probability $P=10$\% that sources are tightly
blended and thus cannot be extracted, \citet{dole2003} derived a SDC limit of
16.7 beams/source. In what follows we use this value in order to facilitate 
comparisons with previous works and future exploitations of this PACS confusion limit 
estimate. Nevertheless, it is important to recall 
that such threshold depends on the steepness of source counts. Following the general 
treatment of counts and confusion by \citet{franceschini1982}, the density of sources 
with flux $S>S_{lim}$ can be expressed as $n(>S_{lim})\simeq \frac{3+\alpha}{(-1-\alpha)q^2}$, where 
$\alpha$ is the slope of number counts, and $q$ was defined in Sect. \ref{sect:confusion_phot}.
The canonical density of 16.7 beams/source is retrieved for $\alpha=-1.8$, which lies within the 
range of slopes estimated by power-law fitting of PEP data (see Tab. \ref{tab:slopes1}), but 
the SDC limit can significantly vary between 8.3 and 25 beams/source for $\alpha=-1.5$ 
to $-2.0$.

The depth reached in PACS maps, especially in the GOODS fields, is such that the
high density of detected sources hinders the extraction of fainter
objects. In Paper~I we have already treated the case of GOODS-N, showing that 
the observed catalog is already hitting the density confusion limit at 160
$\mu$m.

At 70 $\mu$m, despite deep observations, GOODS-S is far from being confused:
even at the 3$\sigma$ limit of 1.1 mJy, PACS catalogs have a density of 
32 beams/source including completeness correction. The SED shape in the far-IR is such that,
while a large number of sources is detected at 100 and 160 $\mu$m at the PEP
effective exposure time, positive $k$-correction strongly suppresses the number of
detections at 70 $\mu$m. The typical 70/160 color of galaxies is red enough to require a deeper
threshold in order to detect a comparable number of sources in the two bands, but PACS
sensitivity at 70 $\mu$m is not sufficient to compensate this effect.
Based on $P(D)$ results, the 16.7 beams/source SDC limit would be reached at
$\sim0.4$ mJy in this band.

At longer wavelengths, GOODS-N and GOODS-S completeness-corrected number counts
and $P(D)$ modeling agree in indicating 
that the SDC limit is reached between 1.5 and 2.0 mJy at 100 $\mu$m and around 8.0
mJy at 160 $\mu$m (Tab. \ref{tab:confusion}). At the GOODS-S 3$\sigma$ levels (1.1 mJy at 100 $\mu$m and 
2.4 mJy at 160 $\mu$m) the density of sources in our catalogs is 10 and 5 beams/source, 
after completeness correction.
It is worth to note that the \citet{dole2003} SDC limit
corresponds to a blending-completeness of 90\%, while sources can be reliably
extracted at lower completeness also below the 16.7 beams/source threshold.
Furthermore, photometric completeness further flattens the detected counts, thus
allowing for sporadic very faint objects to be extracted. All values quoted above
here are nevertheless based on completeness-corrected number counts.
For comparison, the depth reached by $P(D)$ at 100 and 160 $\mu$m digs deep into
density confusion, down to levels corresponding to 2-4 beams/source.


\section{Discussion: the Cosmic Infrared Background}\label{sect:CIB}

Since its discovery by COBE \citep{puget1996,hauser1998}, several attempts have been 
carried out to derive the surface brightness of the CIB, based on direct measurements, 
integration of number counts, statistical analyses, and 
$\gamma-\gamma$ opacity to TeV photons. The PACS data exploited so far can now be
used to derive new, stringent lower limits to the actual far-IR background.

\subsection{CIB measurements: state of the art}

The integral of number counts provides a lower limit to the cosmic IR 
background, to be compared to direct measurements from COBE maps. 
Here we adopt the DIRBE measurements by \citet{dole2006}:
$14.4\pm6.3$, $12.0\pm6.9$, and $12.3\pm2.5$ $[$nW m$^{-2}$ sr$^{-1}]$, at 100, 140, 
and 240 $\mu$m, respectively. Performing the geometrical average  between the upper 
and lower limits to the CIB given by the \citet{fixsen1998} fit to the FIRES spectrum, 
one obtains a 160 $\mu$m CIB brightness estimate of $12.8\pm6.4$ $[$nW m$^{-2}$ sr$^{-1}]$, 
consistent with \citet{dole2006} measurements at 140 and 240 $\mu$m.
\citet{dole2006} thoroughly describe the calibration and uncertainty 
details for CIB direct measurements. The most significant problems arise from the effective brightness of zodiacal light, 
currently known only within a factor of two accuracy. 

\citet{odegard2007} used the Wisconsin H$\alpha$ Mapper (WHAM) Northern Sky Survey as a tracer 
of the ionized medium, in order to study the effects of the foreground interplanetary 
and Galactic dust on DIRBE CIB measurements. The authors find CIB surface brightness values in 
agreement with those by \citet{dole2006}, once renormalized to the FIRAS photometric scale.
\citet{juvela2009} derive the CIB surface brightness at 170 $\mu$m from ISO maps, obtaining 
a value $\sim1.5$ times higher than COBE results, with 30\% systematics and 30\% statistical 
uncertainties. Based on recent observations of the AKARI Deep Field South (ADF-S), carried out 
with this satellite at 65, 90, 140 and 160 $\mu$m, \citet{matsuura2010} detected and measured 
the absolute brightness and spatial fluctuations of the CIB, deriving values consistent to our 
\citet{dole2006} 160 $\mu$m reference, but detecting a significant excess at $90-140$ $\mu$m.

Finally, the most recent estimate of the far-IR CIB comes from the fluctuation analysis
carried out by Penin et al. (\citeyear{penin2011}) on Spitzer 160 $\mu$m maps of the SWIRE ELAIS-N1
field and reprocessed IRAS 60 and 100 $\mu$m data \citep{miville2002}. The newly-derived 
CIB brightness is $6.6\pm2.7$ and $14.4\pm2.3$ $[$nW m$^{-2}$ sr$^{-1}]$, at 100 and 160 $\mu$m, 
respectively. The 160 $\mu$m measurement is consistent with the previous COBE interpolation, 
but the 100 $\mu$m value is almost halved.

No direct measurement is available at 70
$\mu$m, except for the \citet{miville2002} fluctuation analysis on the IRAS 60 $\mu$m map 
($\nu I_\nu=9.0$ $[$nW m$^{-2}$ sr$^{-1}]$) and the AKARI estimate at 65 $\mu$m 
($\nu I_\nu=12.4\pm1.4\pm9.2$ $[$nW m$^{-2}$ sr$^{-1}]$, including statistical and 
zodiacal-light uncertainties, Matsuura et al. \citeyear{matsuura2010}). 

Further constraints on the actual value of the CIB come from the cosmic 
photon-photo opacity: very high-energy photons suffer opacity effects 
by $\gamma-\gamma$ interactions with local radiation backgrounds, 
producing particle pairs \citep[e.g.][]{franceschini2008,nikishov1962}.
Because of the large photon density of the cosmic microwave background (CMB), 
any photon with energy $\epsilon>100$ TeV has a very short free path, and extragalactic sources 
are undetectable above this energy. Less energetic photons suffer from the opacity 
induced by extragalactic backgrounds, other than the CMB, including the CIB. 
Observations of absorption features and cut-offs in the high-energy spectra of 
BLAZARs have been successfully used to pose upper limits to 
the intensity of the EBL. \citet{mazin2007}, among many others, assembled a
compilation of 11 TeV BLAZARs (all those known at the time of their analysis)
and combined them obtaining constraints to the EBL over the whole 0.44-80 $\mu$m
wavelength range.

\subsection{The total CIB in PEP}\label{sect:cib_tot_pep}

Integration of number counts was performed over as wide a flux range as possible, i.e.
using the combined counts presented in Sect. \ref{sect:detected_counts}, including all 
the four fields analyzed so far: GOODS-S, GOODS-N, Lockman Hole and COSMOS. 
To this aim, GOODS-S data were extended down to the 3$\sigma$ detection threshold, 
thus reaching 1.2 mJy at 70 and 100 $\mu$m and 2.0 mJy at 160 $\mu$m. At the bright side, 
COSMOS allows the integration to be carried out to 140 mJy at 100 $\mu$m and 360 mJy 
at 160 $\mu$m. Including completeness corrections, the derived CIB locked in 
resolved number counts is $\nu I_\nu=7.82\pm0.94$ and 
$9.17\pm0.59$ $[$nW m$^{-2}$ sr$^{-1}]$, corresponding to $54\pm7$\% and
$72\pm5$\% of the COBE direct measurements at 100 and 160 $\mu$m, respectively. 
If the Penin et al. (\citeyear{penin2011}) 100 $\mu$m value were taken as reference, then 
PACS sources would produce $\sim$100\% of the CIB.
The resolved CIB at 70 $\mu$m is $\nu I_\nu=3.61\pm1.12$ $[$nW m$^{-2}$ sr$^{-1}]$.

The PEP fields are missing the very bright end of number counts, that can be probed only 
over areas even larger than COSMOS. This lack of information is particularly relevant at 
70 $\mu$m, because this band was employed only for GOODS-S observations. We therefore add previous 
bright counts measurements to our data, in order to derive the CIB brightness emitted {\em
above} PEP flux limits, integrated all the way to $+\infty$. The available data taken into 
account here are the 70 and 160 $\mu$m Spitzer counts by \citet{bethermin2010a}, built over a
more than of $\sim$50 deg$^2$ and extending to $\sim$1 Jy, and 100 $\mu$m ISO and IRAS number counts
\citep{rodighiero2004,heraudeau2004,rowanrobinson2004,efstathiou2000,oliver1992,bertin1997}, 
extending to $\sim$60 Jy.
Beyond the flux range covered by these past surveys, we extrapolate the number counts 
with an Euclidean law, $dN/dS\propto S^{-2.5}$, normalized so to match the very bright end of
observed counts, and extended to infinity.

\begin{figure}[!ht]
\centering
\rotatebox{-90}{
\includegraphics[height=0.4\textwidth]{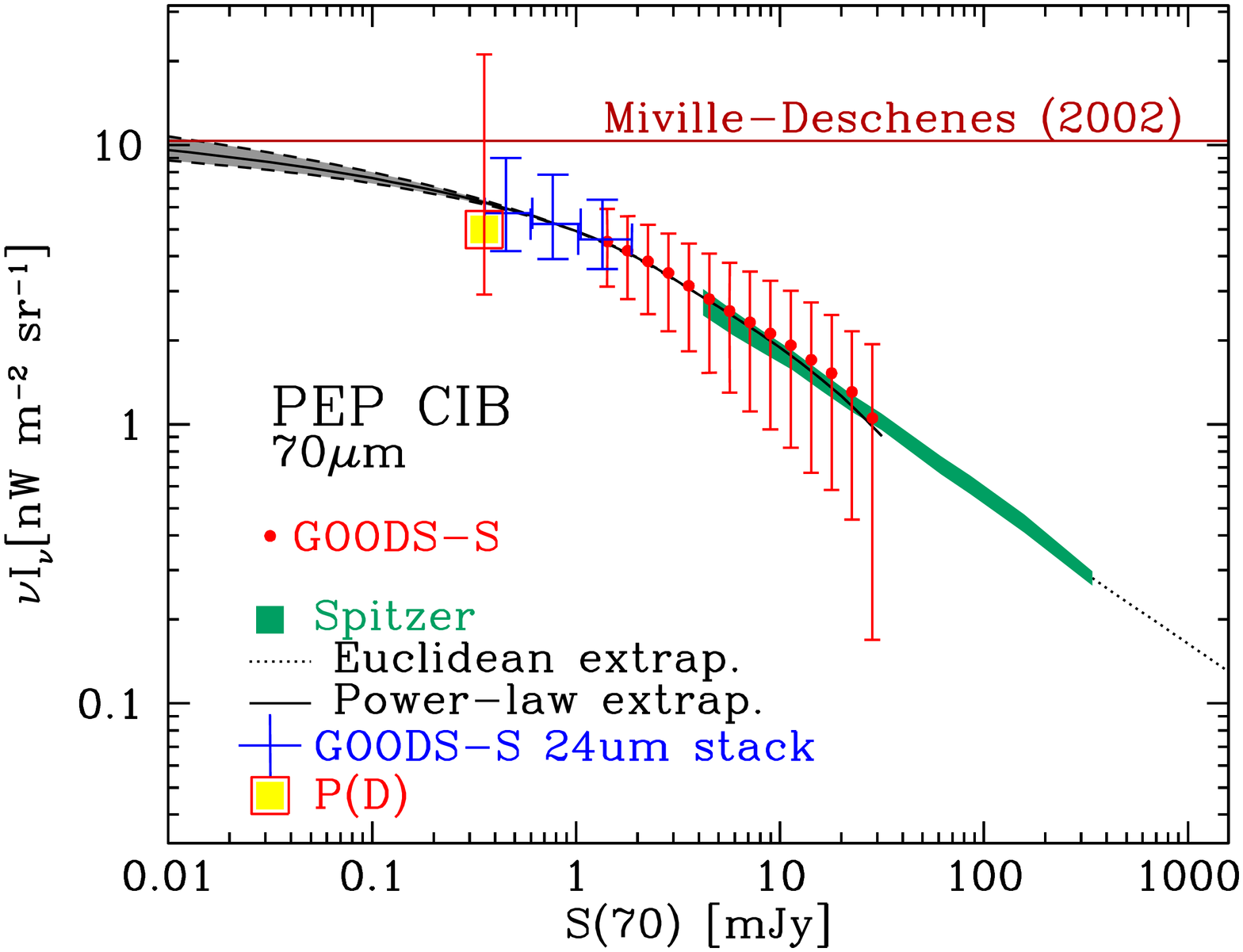}
}
\rotatebox{-90}{
\includegraphics[height=0.4\textwidth]{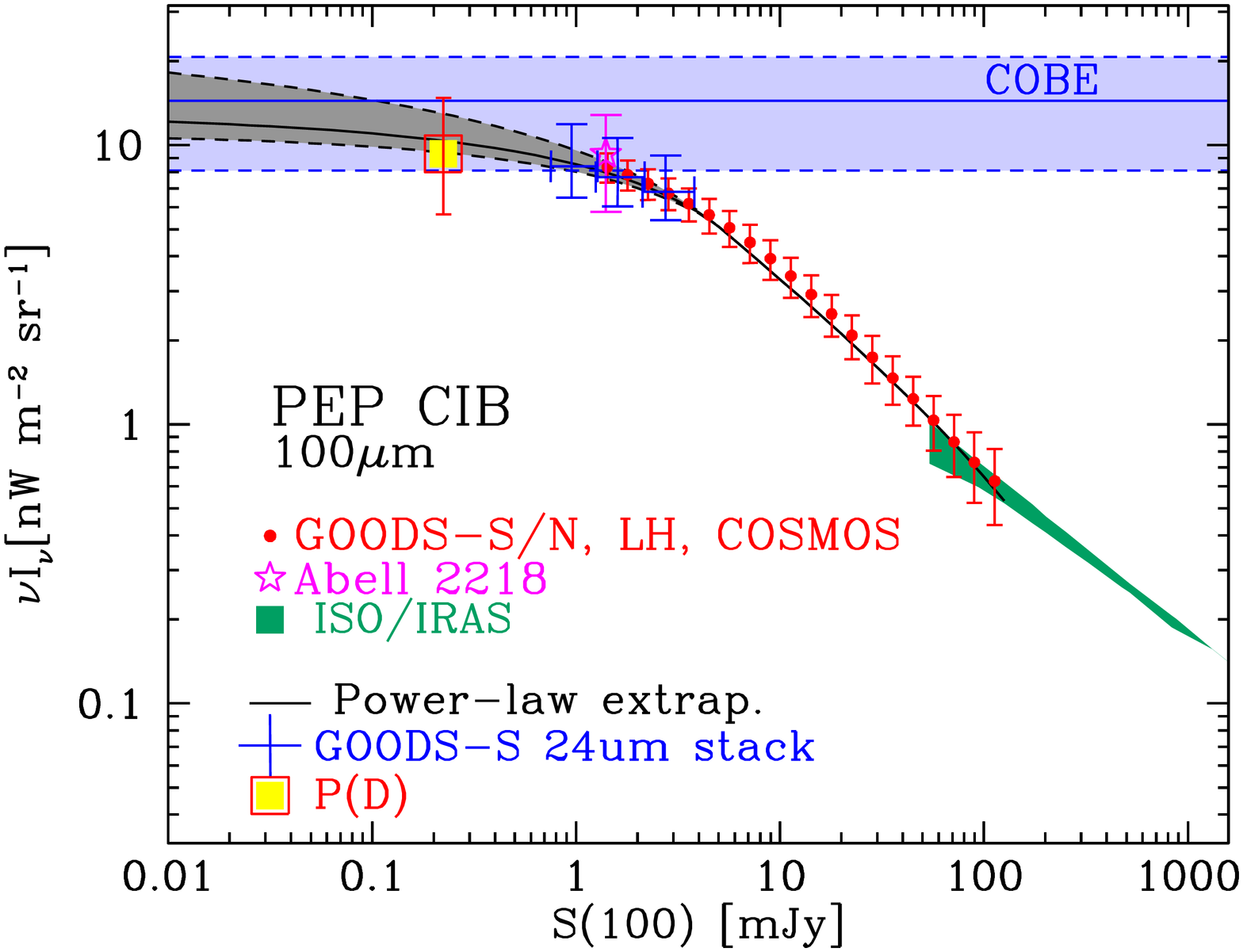}
}
\rotatebox{-90}{
\includegraphics[height=0.4\textwidth]{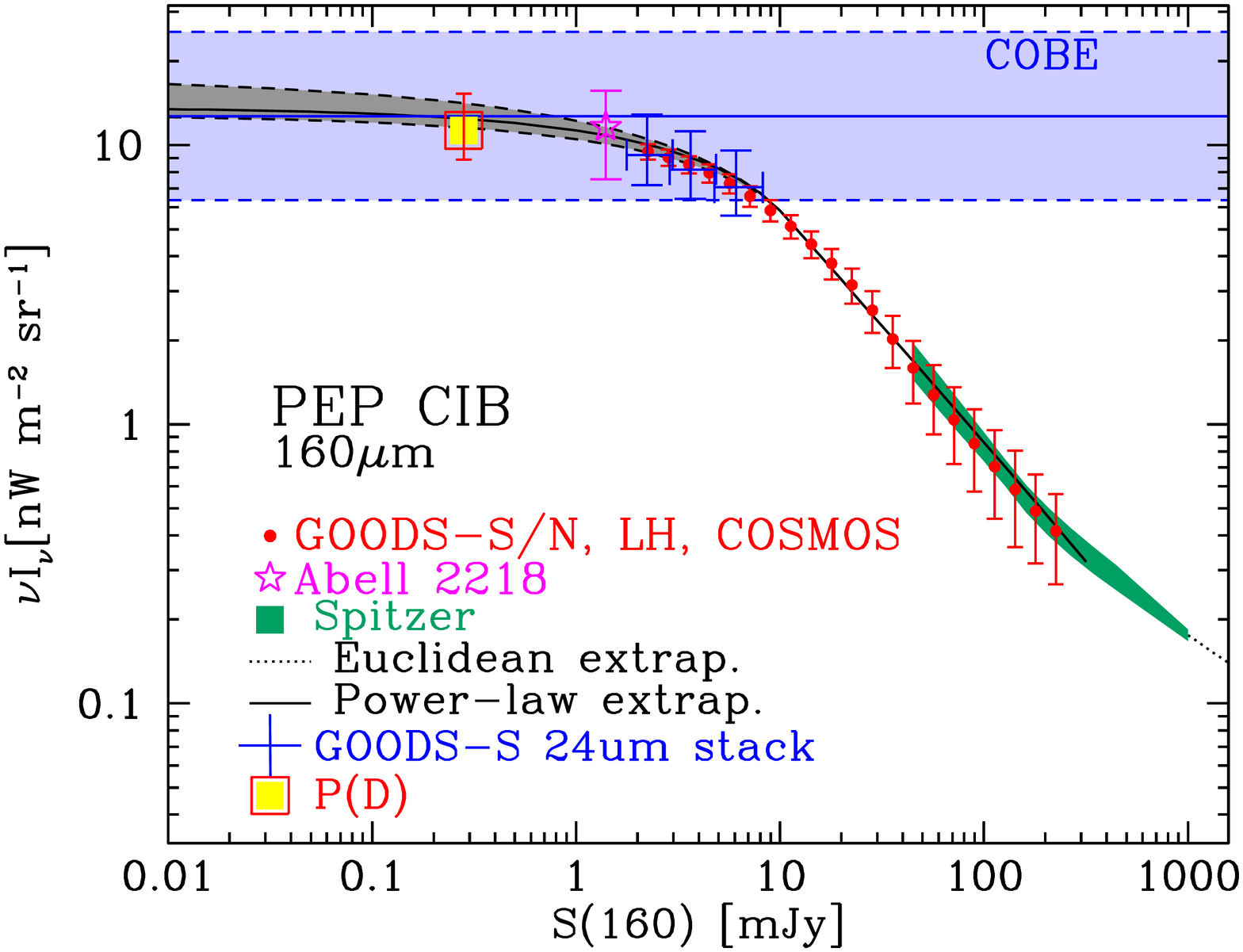}
}
\caption{Cumulative CIB as a function of flux. Red circles, blue crosses
and the yellow square belong to completeness-corrected counts, stacking
and $P(D)$ analysis, respectively.  The black line solid and grey
shaded area are based on power-law fit to the resolved number counts.
The star symbol indicates the CIB derived from gravitational lensing in
Abell 2218 \citep{altieri2010}. The horizontal lines and shades mark the
reference direct measurements of the CIB by \citet[][at 100 and 160
$\mu$m, including 1$\sigma$ uncertainty]{dole2006} and \citet[][at 70
$\mu$m]{miville2002}. The green shaded area belongs to previous surveys carried out with IRAS,
ISO and Spitzer
\citep{oliver1992,bertin1997,efstathiou2000,rowanrobinson2004,rodighiero2004,heraudeau2004}. 
At the very bright end, an Euclidean extrapolation is used (dotted lines).}
\label{fig:cumulative_CIB}
\end{figure}

Figure \ref{fig:cumulative_CIB} shows the cumulative CIB surface brightness 
as a function of flux, as derived from the combination of PACS, Spitzer, ISO and IRAS data. 
The curve describing the resolved
CIB  (red circles) rapidly converges to the COBE measurements at 100
and 160 $\mu$m, over the flux range 
covered by PACS/PEP. Our data extend the knowledge of counts and CIB by over an order of 
magnitude in flux, with respect to previous estimates based on individual detections. 
The depth reached in GOODS-S is similar to the one obtained in Abell 2218
through gravitational lensing, and the corresponding CIB surface brightnesses are
consistent within the uncertainties. The total CIB emitted above PEP 3$\sigma$ flux limits 
(1.1 mJy at 70 $\mu$m, 1.2 mJy at 100 $\mu$m, 2.0 mJy at 160 $\mu$m), is 
$\nu I_\nu=4.52\pm1.18$, $8.35\pm0.95$ and $9.49\pm0.59$ $[$nW m$^{-2}$ sr$^{-1}]$ at 70, 100 and 160 $\mu$m,
respectively. Table \ref{tab:cib} lists all the values obtained.

Roughly 10\% of the CIB down to the 70 $\mu$m adopted threshold is emitted by bright galaxies,
out of reach for our survey, due to the limited volume sampled. On the other hand, only 
a few percent is produced at fluxes brighter than PEP COSMOS upper limit at 100 and 160 $\mu$m. 
Including 100 $\mu$m ELAIS and IRAS data (up to 60 Jy), only a negligible fraction ($<0.05$\%) is missed at the 
bright end, while beyond Spitzer counts (reaching $\sim$1 Jy at 70 and 160 $\mu$m) the Euclidean extrapolation 
provides roughly 1-2\% of the total CIB.

The power-law fit to resolved counts (see Sect. \ref{sect:detected_counts} and Tab.
\ref{tab:slopes1}) provides a new estimate of the total expected CIB. We extrapolate 
power laws down to 1.0 $\mu$Jy and obtain 
$\nu I_\nu\ge11.09$, $\nu I_\nu=12.61^{+8.31}_{-1.74}$ and $\nu I_\nu=13.63^{+3.53}_{-0.85}$
$[$nW m$^{-2}$ sr$^{-1}]$ in the three PACS bands.
At 70 $\mu$m our data provide only a lower limit, because the curve obtained 
from our power-law fit is not fully converging at 1 $\mu$Jy (see Fig. \ref{fig:cumulative_CIB}).
The uncertainty on the 100 $\mu$m CIB is still large because discordant slopes 
were found at the faint end in the different PEP fields. The results are fully consistent 
with COBE data, but PEP and PACS pinpoint the total CIB values with unprecedented precision, 
thanks to the high quality of observations, survey strategy, maps, and --- last but not least
--- thanks to Herschel capabilities (grey shaded areas in Fig. \ref{fig:cumulative_CIB} denote the 3$\sigma$ 
uncertainty on power-law fits).

Stacking results are fully consistent with resolved number counts, and the $P(D)$
analysis extends down by another decade in flux with the exception of 70 $\mu$m,
for which we truncated the computation of CIB because of a strong divergence in 
$P(D)$ uncertainties (see Sect. \ref{sect:pdd_counts}). 
Consequently, the lower limit set to the CIB by PACS data is further improved:
above the flux level reached by $P(D)$ statistics, the contributions to the CIB are 4.98, 9.32
and 11.31 $[$nW m$^{-2}$ sr$^{-1}]$ in the three bands, 
thus recovering $\sim$65\% and 89\% of the total \citet{dole2006} CIB at 100 and 160 $\mu$m, 
respectively. 
When referred to the total values obtained through power-law extrapolations, the $P(D)$ analysis
recovers 45\%, 74\% and 83\% of the total background at 70, 100 and 160 $\mu$m.

\subsection{CIB contributions from different cosmic epochs}


\begin{figure*}[!ht]
\centering
\begin{minipage}{0.45\textwidth}
\centering
\rotatebox{-90}{\resizebox{0.825\textwidth}{!}{
\includegraphics{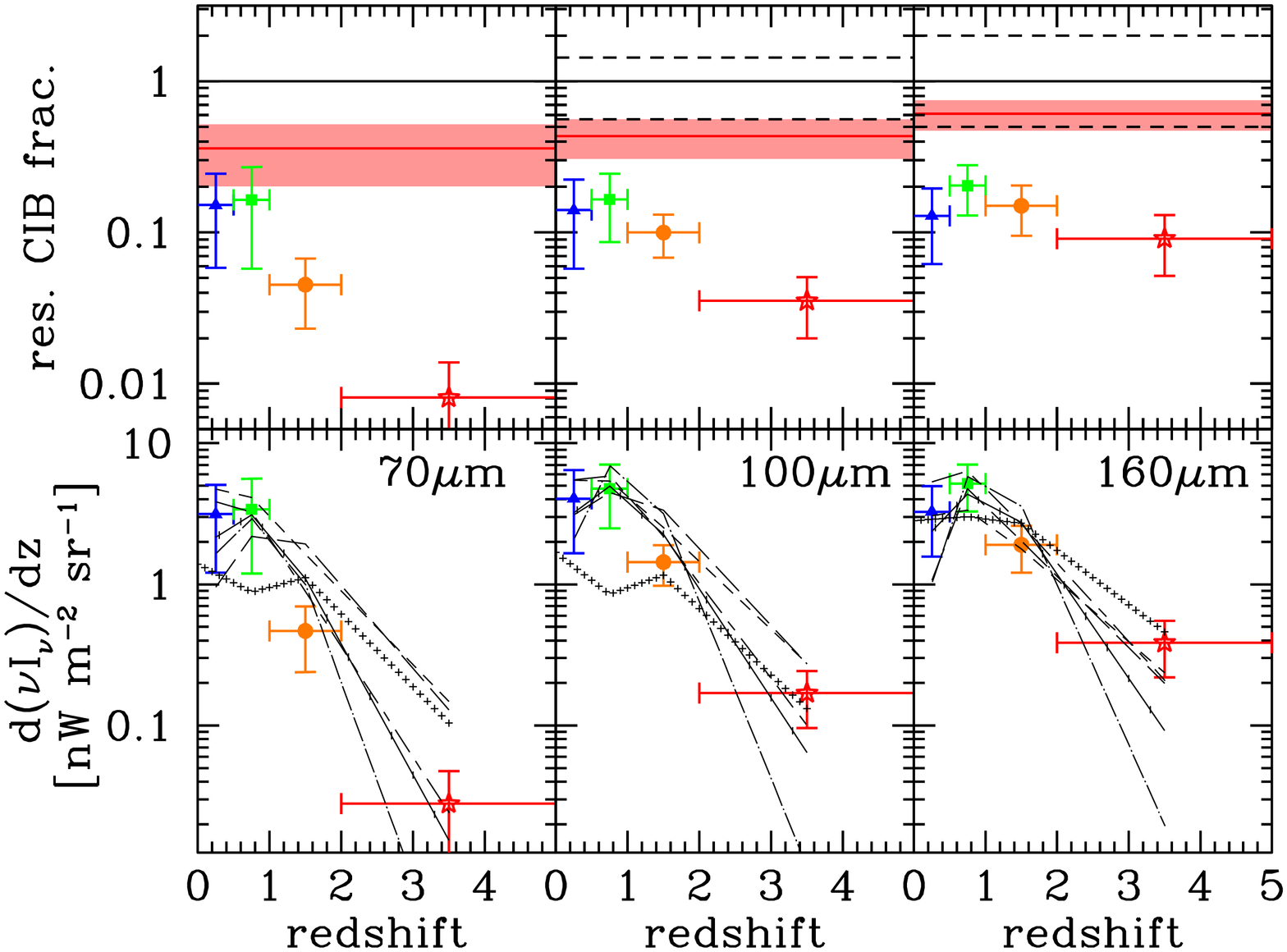}}
}
\end{minipage}
\begin{minipage}{0.34\textheight}
\centering
\resizebox{0.825\textwidth}{!}{
\includegraphics{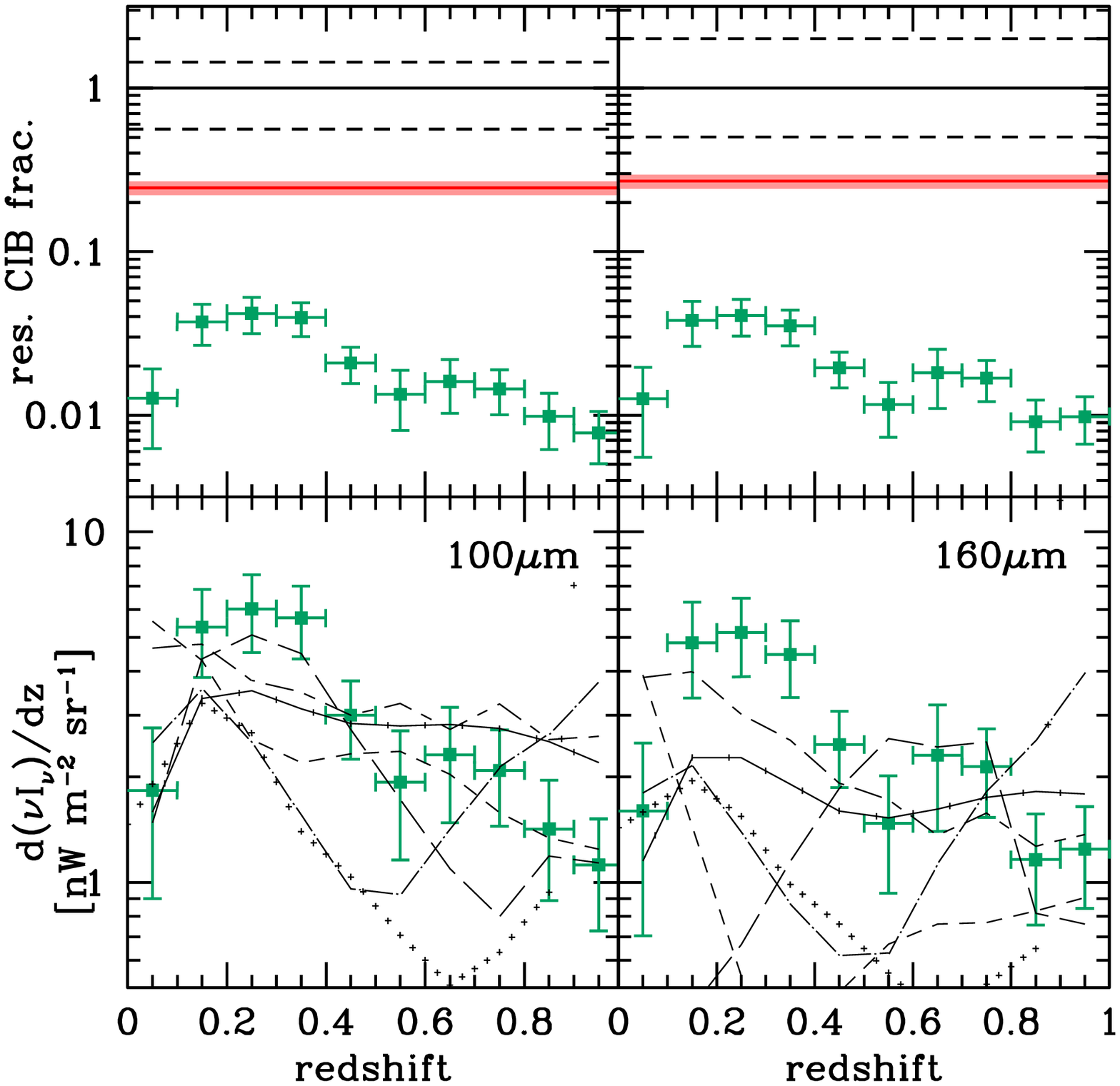}
}
\end{minipage}
\caption{Redshift distribution of CIB surface brightness, in 
GOODS-S ({\em left}) and COSMOS ({\em right}). {\em Top panels} depict 
the resolved fraction of CIB emitted at different epochs. The horizontal black
lines (solid and dashed) represent the total value and its uncertainty
\citep{dole2006,miville2002}. The red horizontal lines and shaded areas belong to
the fraction resolved into individual sources (plus completeness correction) by
PEP. {\em Bottom panels}: Redshift derivative, compared to available models
\citep{marsden2010,bethermin2010c,valiante2009,leborgne2009}. See Fig
\ref{fig:counts_zbins} for model lines notation.}
\label{fig:CIB_redshift}
\end{figure*}

Thanks to the rich ancillary datasets available in PEP fields, it is possible to
estimate the amount of CIB emitted at different epochs. In order to reach
this goal, we integrate the number counts split into redshift bins. The same
completeness correction was applied in each redshift bin, as derived from total
counts. Table \ref{tab:cib_redshift} reports the results for GOODS-S and for 
the combination of GOODS-S, GOODS-N and COSMOS.

Figure \ref{fig:CIB_redshift} shows the fraction of resolved CIB emitted at
different cosmic epochs in the GOODS-S and COSMOS fields. In this case, we 
keep the two fields separate, in order to study the details of different flux regimes
and aid the comparison to model predictions.  As stated above, deep
observations resolve most of the CIB at 100 and 160 $\mu$m, and thus give a fairly 
complete census of the redshift dependence of the background. 
At the PEP 3$\sigma$ detection threshold, more than half of the 
resolved CIB is emitted by objects lying at $z\le1$. 
The new results are in line with our Paper~I analysis, based on shallower GOODS-N
detections and stacking of 24 $\mu$m sources. As expected,
at the depth of GOODS-S, galaxies in the highest redshift bin come into play and provide a 
non negligible contribution to the CIB, as high as $\sim$15\% of the resolved amount in GOODS-S.

For comparison, based on SED template extrapolations from 15 $\mu$m, \citet{elbaz2002} estimated that 
the contribution of ISO 15 $\mu$m sources to the 140 $\mu$m CIB was of the order of 65\%, 
with a median redshift of $z=0.6$, and mostly emitted below $z=1$. Exploiting Spitzer 24 $\mu$m 
observations in the COSMOS field, \citet{lefloch2009} showed that $\sim$50\% of the 24 $\mu$m 
background intensity originates at $z\simeq1$. 

The comoving (IR) luminosity (or SFR) density dependence on redshift (Madau, Pozzetti \& Dickinson \citeyear{madau1998}, see
Gruppioni et al. \citeyear{gruppioni2010} for a recent determination based on Herschel data) is known to peak between redshift
$z=1.5$ and 3.0. \citet{harwit1999} showed that when transforming the resulting energy generation rate into unit redshift
interval, the high redshift component is strongly suppressed, by a factor $(1+z)^{5/2}$ in a $q_0=0.5$ cosmology. 
Moreover, the effective energy reception rate, thus the background observed locally, is further suppressed by a factor $(1+z)$,
because bandwidths are reduced. Based on these and other simple considerations, it is easy to 
demonstrate that the distribution of the integrated CIB should be dominated by $z<1$ galaxies \citep{harwit1999}.

\begin{figure}[!ht]
\centering
\includegraphics[width=0.45\textwidth]{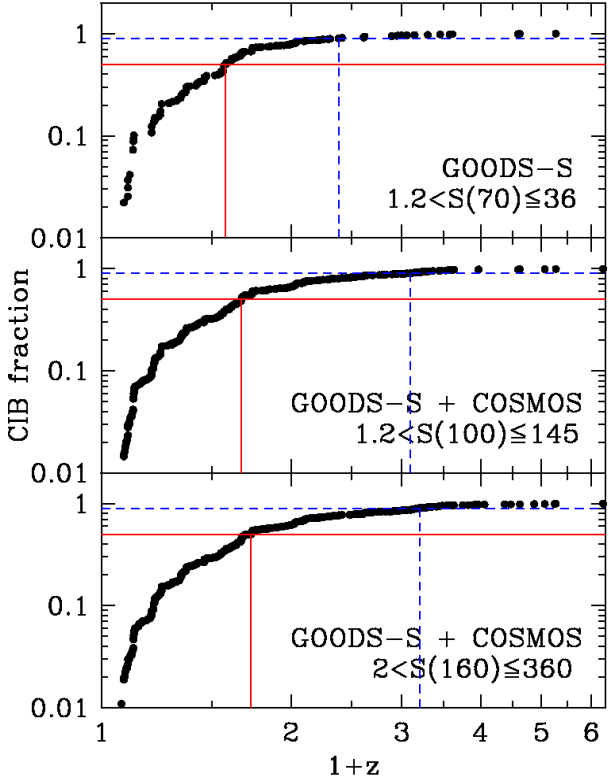}
\caption{Cumulative CIB fraction as a function of redshift, as obtained summing 
the contribution of individual sources in GOODS-S and COSMOS, and accounting 
for completeness correction. Top, middle and bottom panels belong to the 70, 
100, 160 $\mu$m bands, respectively. Red solid lines and dashed blue lines mark the 
50th and 90th percentiles, i.e. identify the redshifts below which 50\% and 90\% 
of the detected CIB was emitted. Flux ranges covered by observations are quoted in 
each panel.}
\label{fig:cib_cumul_z}
\end{figure}

The relative fraction among the four bins moves from low to high redshift, as 
wavelength increases: Fig. \ref{fig:cib_cumul_z} shows the cumulative CIB fraction as a 
function of redshift, computed on the combination of GOODS-S and COSMOS, and including 
completeness correction. 
Half of the CIB detected over the flux ranges quoted on each panel
was emitted below redshift $z=0.58$, 0.67 and 0.73 at 70, 100, 160 $\mu$m, respectively.

While at 70 $\mu$m $\sim$80\% of the resolved CIB was emitted at $z\le1.0$,  at 160
$\mu$m this fraction decreases to $\sim$55\%. This trend is mainly due to the
positive $k$-correction of far-IR SEDs short ward of their peak, and will 
become even more relevant in SPIRE and sub-mm bands, where negative $k$-correction
comes into play \citep{oliver2010,marsden2009}. Note also that including the contribution 
from 70 $\mu$m unresolved galaxies would shift the distribution to slightly higher redshifts
(see also discussion about 70 $\mu$m depth and confusion in Sect. \ref{sect:confusion_dens}). 
On the other hand, adding the bright end of number counts, sampled by the observations in the COSMOS field, 
does not significantly influence the high redshift bins in the CIB, because most of the 
contribution goes into filling the lower redshift slices (see Tab. \ref{tab:cib_redshift}).
The 90th percentiles of the redshift-cumulative CIB fall at $z=1.38$, 2.03 and 2.20. Excluding the bright 
end covered by COSMOS (only at 100 and 160 $\mu$m), we witness only a marginal shift of the 
90\%-light redshift by $\Delta(z)\le0.05$.

Performing SED fitting of each individual object detected in GOODS-S in the PACS bands, 
\citet{rodighiero2010} derived IR luminosities (8-1000 $\mu$m) of the sources contributing to 
far-IR counts and CIB. Combining the CIB distribution found above and the notion of Malmquist bias, 
it is no surprise that the resolved background is dominated by normal galaxies ($L<10^{11}$ L$_\odot$) at 
low redshift and more luminous sources dominate the high redshift bins. Quantitatively, 95\% of 
the resolved 160 $\mu$m CIB at $z\le0.5$ is produced by normal galaxies, $>$90\% of the contribution at $0.5<z\le1.0$ 
arises from luminous infrared galaxies (LIRGs, $10^{11}\le L<10^{12}$ L$_\odot$) and ultra-luminous infrared galaxies (ULIRGs, $10^{12}\le L<10^{13}$ L$_\odot$) 
provide 50\% of the background at $1.0<z\le2.0$ and 88\% above. 
Globally, roughly 50\% of the CIB resolved in the three PACS bands in GOODS-S was produced by LIRGs. At 160 $\mu$m, the remainder 
is equally distributed between normal IR galaxies and ULIRGs, while at shorter wavelengths the former prevail.

COSMOS shallow observations probe the bright end of PACS number counts, and
hence are limited to lower redshifts, but the large area grants a greater detail in
the distribution of CIB, allowing a much finer binning. At the COSMOS 80\%
completeness limits (9.0 and 20.0 mJy at 100 and 160 $\mu$m, respectively), the
resolved CIB peaks at $z\simeq0.3$ and exhibits a weak secondary ``bump'' at $z\simeq0.7$
in both bands. Recently, \citet{jauzac2011} obtained similar 
results based on stacking of 24 $\mu$m sources on Spitzer MIPS 70 and 160 $\mu$m.
These authors find a dip at $z\simeq0.5$ in the differential CIB brightness in
COSMOS, i.e. the same secondary peak a $z\simeq0.7$ found in PACS data. This 
features are certainly due to the $z=0.73$ structure known in the COSMOS field. 
Only a small fraction ($\le$10\%) of the total CIB is locked in
$z<0.5$ bright objects, which are dominated by the non-evolutionary component of
number counts \citep[see for example][]{franceschini2010, berta2010}.

The bottom panels in Fig. \ref{fig:CIB_redshift} present the redshift derivative of
the background surface brightness, and compare it to a set of model predictions.
The derivative $d(\nu I_\nu)/dz$ peaks in the $0.5<z\le1.0$ bin and then drops at
higher redshifts. \citet{lefloch2009} find a similar behavior at 24 $\mu$m, based on 
deep COSMOS data. The decrement is very steep at 70
$\mu$m, while it becomes milder at longer wavelengths. 
While over broad redshift binning the five models considered are overall 
consistent to the data, at the low redshift bright-end, where detailed information is 
available, they significantly differ, especially at longer wavelengths.
Again, this might be a sign of possible differences in 
SED color- and temperature-luminosity assumptions and evolution. 

The fine details of the comparison between models and data in the COSMOS field 
are affected by the known $z\sim0.7$ overdensity. In order to study the impact of 
these features, the COSMOS and GOODS-S catalogs have been cut to a common (overlapping)
flux range. In this case, the $d(\nu I_\nu)/dz$ derived in the two areas
are consistent to each other within the errors, although the fine structure of GOODS-S cannot be 
studied. If then the same broad redshift binning is adopted,
the features due to COSMOS large scale structure are washed out, and the trends seen in 
the two fields fully resemble each other. 

\begin{figure*}[!ht]
\centering
\includegraphics[width=0.7\textwidth]{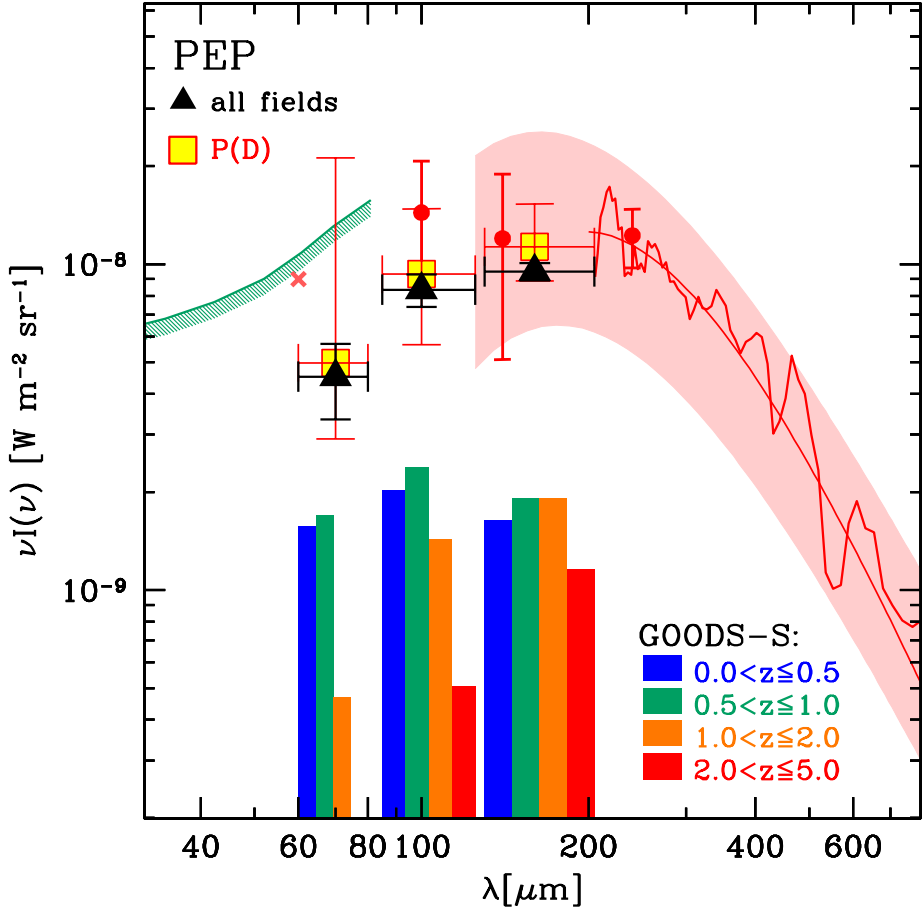}
\caption{The cosmic infrared background. 
Black filled triangles represent the total CIB emitted above the PEP flux limits, based on resolved number counts in GOODS-S, GOODS-N, Lockman Hole and
COSMOS, evaluated as described in Sect. \ref{sect:cib_tot_pep}. Yellow squares belong to the $P(D)$
analysis in GOODS-S. 
Histograms denote the contribution of different redshift bins to the CIB, over the flux range covered by GOODS-S. 
Literature data include: DIRBE measurements (filled circles, 1$\sigma$ errors, 
Dole et al. \citeyear{dole2006}), FIRAS spectrum (solid lines above 200 $\mu$m, Lagache et al.
\citeyear{lagache1999}, \citeyear{lagache2000}), \citet{fixsen1998} modified Black
Body (shaded area), 60 $\mu$m IRAS fluctuation analysis
\citep[cross,][]{miville2002}, and $\gamma$-ray upper limits \citep[green hatched
line below 80 $\mu$m][]{mazin2007}.}
\label{fig:cib}
\end{figure*}

Figure \ref{fig:cib} summarizes our findings and compares the PEP CIB estimates to
direct measurements and upper limits from $\gamma$-rays opacity. 
Our estimate of the background surface brightness emitted above the 
PEP detection and $P(D)$ limits is fully enclosed in the COBE 1$\sigma$ error bars. 
The uncertainties on the CIB values derived from PACS resolved counts, $P(D)$, and 
power-law extrapolation are overall significantly smaller than what achieved 
previously by COBE/DIRBE. 
Far-IR number counts are currently in the position to provide a more accurate 
estimate of the extragalactic background light than
direct photometric measurements.


\section{Summary and conclusions}\label{sect:summary}

Using data belonging to the {\em PACS Evolutionary Probe} Herschel survey
(Lutz et al. in prep.), we derived far-IR number
counts in the GOODS-S, GOODS-N, Lockman Hole and COSMOS fields. By employing stacking
and $P(D)$ analyses, it was possible to exploit the whole information in PACS
maps and extend the study to very deep fluxes. The wealth of ancillary data
in the GOODS and COSMOS fields allowed counts and CIB to be split into redshift
bins. Through integration of number counts, we computed an unprecedented estimate
of the contribution of IR galaxies to the extragalactic background light, across
cosmic epochs. Among the many analyses carried out here, the main
results that were presented are:

\begin{itemize}
\item resolved far-IR number counts at 70, 100, 160 $\mu$m extend from a few mJy to
$\sim$200 mJy. In Euclidean-normalized units, our combination of four fields
tightly defines the bright side of counts,  
the peak at intermediate fluxes, the downturn and the slope at the faint-end.
Stacking and $P(D)$ analyses push the knowledge of counts by a decade fainter in
flux, down to $\sim$0.2 mJy in GOODS-S.
\item based on the observed counts, we derive a photometric confusion noise estimate of $\sigma_c=0.27$ 
and 0.92 mJy (obtained with $q=5$) at 100 and 160 $\mu$m respectively, and 16.7 beams/source density confusion 
limits of 0.4, $\sim$2.0 and 8.0 mJy in the three PACS bands.
\item the fraction of CIB emitted at fluxes brighter that PEP 3$\sigma$ 
thresholds is 58\% (74\%) of the \citet{dole2006} reference value at 100 $\mu$m (160 $\mu$m).
Using $P(D)$ statistics,
this fraction increases to $\sim$65\% ($\sim$89\%); this estimate lies well 
within the 1$\sigma$ uncertainty of the direct CIB measurements, with a 3$\sigma$
confidence interval smaller than COBE's by a factor of 2 at 160 $\mu$m. 
\item exploiting power-law extrapolation of number counts at the faint side, and 
including past IRAS, ISO and Spitzer results and Euclidean extrapolations at the bright end, 
we derive new expectation values for the total CIB at 70, 100 and 160 $\mu$m: 
$\nu I_\nu\ge11.09$, $\nu I_\nu=12.61^{+8.31}_{-1.74}$ and $\nu I_\nu=13.63^{+3.53}_{-0.85}$
$[$nW m$^{-2}$ sr$^{-1}]$.
\item at the current 3$\sigma$ detection threshold, more than half of the 
resolved CIB was emitted at redshift $z\le1$. 
The half-light redshift lies at $z=0.58$, 0.67 and 0.73 at 70 (GOODS-S), 100 and 160 $\mu$m
(GOODS-S and COSMOS combined), respectively.
The balance moves towards higher
redshift at longer wavelengths: while at 70 $\mu$m roughly 80\% of the resolved CIB
is emitted at $z\le1.0$, at 160 $\mu$m the CIB budget is almost equi-partitioned
below/above $z=1.0$. Roughly 50\% of the CIB observed at $z=0$ and resolved in the three PACS
bands in GOODS-S is emitted by LIRGs.
\item most of the available evolutionary models fairly reproduce PACS total 
counts, despite the large variety of adopted assumptions. On the other hand, 
the detailed high-quality PACS data highlight dramatic differences between models and severe 
failures when compared to redshift distributions and CIB derivatives. Only models actually 
tuned taking into account redshift information and early Herschel data are successful in 
reproducing the new PEP dataset, pointing out the need to include these constraints in 
future modeling attempts.
\end{itemize}

Herschel and PACS set a new reference of the brightness of the cosmic far-IR
background. Expectantly, next analyses and missions, rather than 
trying to recover the missing fraction of CIB, envisage the reconstruction of the
bolometric background from the mid-IR to sub-mm frequencies. 
The contribution of galaxies at different redshift to the CIB is being investigated 
to an increasing details by current surveys. A further study of the redshift-dependent CIB, 
e.g. based on luminosity functions and properly including the effects of $k$-correction, goes 
beyond the purpose of this paper and is deferred to future works.
This information is not only important in the fine tuning of galaxy evolution models, but is also 
of fundamental importance to constrain the intrinsic spectra of distant TeV sources, such as BLAZARs,
whose $\gamma$-ray photons interact with the CIB. So far, such constraints have been based solely on 
model predictions, to be confirmed with direct estimates of the CIB as observed {\em from} 
different epochs.
Herschel source statistics have demonstrated to provide a precise estimate of the total CIB.
Future studies will necessitate to improve direct measurements and thus the 
study of foreground contaminants, in order to complete the understanding of the total background
and pinpoint its measurement.



\begin{table*}[!ht]
\centering
\begin{tabular}{l c c c c c}
\hline
\hline
Field     	& Eff. Area	& $t_{exp}$	& 3$\sigma$ 	& $N$ 		& 80\% compl	\\
Band		& arcmin	& h		& mJy		& $>3\sigma$	& mJy 		\\
\hline
GOODS-N 100 	& 300 arcmin$^2$	& 30	& 3.0		& 291		& 5.5 		\\
GOODS-N 160	& 300 arcmin$^2$ 	& 30	& 5.7		& 317		& 11.0 		\\
GOODS-S 70 	& 300 arcmin$^2$	& 113	& 1.1		& 375		& 2.1		\\
GOODS-S 100 	& 300 arcmin$^2$	& 113	& 1.2		& 717		& 2.4		\\
GOODS-S 160	& 300 arcmin$^2$	& 226	& 2.4		& 867		& 5.2		\\
LH 100		& 0.18 deg$^2$		& 35	& 3.6		& 909		& 5.9		\\
LH 160		& 0.18 deg$^2$ 		& 35	& 7.5		& 841		& 13.2		\\
COSMOS 100	& 2.04 deg$^2$		& 213	& 5.0		& 5360		& 9.0		\\
COSMOS 160	& 2.04 deg$^2$		& 213	& 10.2		& 5105		& 20.0		\\
\hline
\end{tabular}
\caption{Main properties of the PEP fields included in the number counts and CIB
analysis.}
\label{tab:fields}
\end{table*}

\begin{table*}[!ht]
\centering
\begin{tabular}{l | c c c | c c c }
\hline
\hline
\multicolumn{1}{c|}{Field} & \multicolumn{3}{c|}{Resolved counts} & \multicolumn{3}{c}{Stacking}\\
\hline
\hline
\& band & Flux range  & Slope & Error & Flux range  & Slope & Error\\
$[\mu$m$]$ & $[$mJy$]$ & $\alpha$ & $d\alpha$ & $[$mJy$]$ & $\alpha$ & $d\alpha$\\
\hline
GOODS-S 70  & 1.4$-$3.5 & -1.87 & $\pm$0.07 & 0.45$-$1.35 & -1.66 & $\pm$0.12\\
GOODS-S 70  & 3.5$-$36 	& -2.27 & $\pm$0.11 & -- & -- &-- \\
\hline
GOODS-S 100 & 1.4$-$5.6 & -1.96 & $\pm$0.06 & 0.95$-$2.75 & -1.66 & $\pm$0.05\\
GOODS-N 100 & 2.8$-$5.6 & -1.32 & $\pm$0.27 & -- & -- &-- \\
LH 100      & 6.0$-$89 	& -2.60 & $\pm$0.05 & -- & -- &-- \\
COSMOS 100  & 8.0$-$142 & -2.58 & $\pm$0.03 & -- & -- &-- \\
All fields 100 & 1.4-5.6& -1.92 & $\pm$0.05 & -- & -- &-- \\
All fields 100 & 6.0-142& -2.41 & $\pm$0.05 & -- & -- &-- \\
\hline
GOODS-S 160 & 2.8$-$9.0 & -1.49 & $\pm$0.08 & 2.20$-$6.00 & -1.98 & $\pm$0.03\\
GOODS-N 160 & 5.6$-$9.0 & -1.67 & $\pm$0.32 & -- & -- &-- \\
LH 160      & 9.0$-$113 & -2.67 & $\pm$0.05 & -- & -- &-- \\
COSMOS 160  & 17.0$-$350& -2.96 & $\pm$0.05 & -- & -- &-- \\
All fields 160 & 2.8-9.0& -1.64 & $\pm$0.08 & -- & -- &-- \\
All fields 160 & 9.0-350& -2.82 & $\pm$0.07 & -- & -- &-- \\
\hline
\end{tabular}
\caption{Power-law fit to PACS differential number counts in the form
$dN/dS\propto S^{\alpha}$. Results belonging to resolved number counts (i.e.
individually detected sources, plus completeness correction) and stacking of 24
$\mu$m sources are reported.}
\label{tab:slopes1}
\end{table*}

\begin{table}[!ht]
\centering
\begin{tabular}{r | c c c}
\hline
\hline
\multicolumn{1}{c|}{} & \multicolumn{3}{c}{GOODS-S} \\
\multicolumn{1}{c|}{$S_{center}$} & Counts & Err. & Compl. \\ 
\multicolumn{1}{c|}{$[$mJy$]$} & \multicolumn{2}{c}{$[10^4$ deg$^{-2}$ mJy$^{1.5}]$} & \\
\hline
  1.13  & 0.86 & 0.11 & 0.36\\
  1.42  & 1.17 & 0.14 & 0.54\\
  1.79  & 1.43 & 0.17 & 0.77\\
  2.25  & 1.61 & 0.22 & 0.80\\
  2.84  & 1.81 & 0.27 & 0.85\\
  3.57  & 2.05 & 0.36 & 0.92\\
  4.50  & 1.88 & 0.42 & 0.95\\
  5.66  & 1.70 & 0.48 & 0.97\\
  7.13  & 1.81 & 0.60 & 0.98\\
  8.97  & 1.65 & 0.69 & 1.00\\
 11.30  & 2.22 & 0.93 & 1.00\\
 14.22  & 2.51 & 1.18 & 1.00\\
 17.91  & 1.66 & 1.27 & 1.00\\
 22.54  & 4.83 & 2.22 & 1.00\\
 28.38  & 2.53 & 1.94 & 1.00\\
 35.73  & 2.41 & 2.28 & 1.00\\
\hline															
\end{tabular}
\caption{PEP 70 $\mu$m number counts, normalized to the Euclidean slope.}
\label{tab:counts_070}
\end{table}

\begin{landscape}

\begin{table}  
\caption{PEP 100 $\mu$m number counts, normalized to the Euclidean slope.}
\label{tab:counts_100}
\centering
\begin{tabular}{r | c c c c | c c c c | c c c c | c c c | c c }
\hline
\hline
\multicolumn{1}{c|}{} & \multicolumn{4}{c|}{GOODS-S} & \multicolumn{4}{c|}{GOODS-N} & \multicolumn{4}{c|}{Lockman Hole} & \multicolumn{3}{c|}{COSMOS}  & \multicolumn{2}{c}{All Fields} \\
\multicolumn{1}{c|}{$S_{center}$} & Counts & Err. & $\sigma(var)$ & Compl. & Counts & Err. & $\sigma(var)$ & Compl. & Counts & Err. & $\sigma(var)$ & Compl. & Counts & Err. & Compl. & Counts & Err. \\ 
\multicolumn{1}{c|}{$[$mJy$]$} & \multicolumn{3}{c}{$[10^4$ deg$^{-2}$ mJy$^{1.5}]$} & & \multicolumn{3}{c}{$[10^4$ deg$^{-2}$ mJy$^{1.5}]$} & & \multicolumn{3}{c}{$[10^4$ deg$^{-2}$ mJy$^{1.5}]$} & & \multicolumn{2}{c}{$[10^4$ deg$^{-2}$ mJy$^{1.5}]$} & & \multicolumn{2}{c}{$[10^4$ deg$^{-2}$ mJy$^{1.5}]$} \\
\hline
  1.42  & 2.58 & 0.22 &   -- & 0.24 &  --   &   -- &   -- &  --  &   --	     &	--  &   -- &  --  &  --  &  --  &  --  & 2.58 & 0.22 \\
  1.79  & 3.13 & 0.26 &   -- & 0.62 &  --   &   -- &   -- &  --  &   --	     &	--  &   -- &  --  &  --  &  --  &  --  & 3.13 & 0.26 \\
  2.25  & 3.72 & 0.33 &   -- & 0.64 &  --   &   -- &   -- &  --  &   --	     &	--  &   -- &  --  &  --  &  --  &  --  & 3.72 & 0.33 \\
  2.84  & 4.15 & 0.41 &   -- & 0.81 &  3.19 & 0.67 &   -- & 0.10 &   --	     &	--  &   -- &  --  &  --  &  --  &  --  & 4.15 & 0.41 \\
  3.57  & 4.44 & 0.52 &   -- & 0.92 &  4.67 & 0.68 &   -- & 0.49 &   --	     &	--  &   -- &  --  &  --  &  --  &  --  & 4.52 & 0.16 \\
  4.50  & 4.74 & 0.65 & 1.85 & 0.95 &  5.57 & 0.83 & 1.85 & 0.54 &   --	     &	--  & 0.96 &  --  &  --  &  --  &  --  & 5.06 & 0.58 \\
  5.66  & 5.52 & 0.84 & 1.41 & 0.97 &  6.37 & 1.03 & 1.41 & 0.77 &   7.00    & 0.60 & 0.74 & 0.65 & 6.02 & 0.15 & 0.27 & 6.07 & 0.56 \\
  7.13  & 5.65 & 1.03 & 1.36 & 0.99 &  6.90 & 1.28 & 1.36 & 0.84 &   7.22    & 0.72 & 1.06 & 0.78 & 6.67 & 0.18 & 0.48 & 6.67 & 0.50 \\
  8.97  & 4.31 & 1.10 & 1.80 & 1.00 &  6.43 & 1.51 & 1.80 & 0.90 &   6.58    & 0.84 & 1.65 & 0.86 & 7.23 & 0.22 & 0.70 & 7.08 & 1.22 \\
 11.30  & 5.47 & 1.45 & 2.14 & 1.00 &  6.29 & 1.81 & 2.14 & 0.97 &   7.24    & 1.06 & 1.42 & 0.93 & 7.55 & 0.27 & 0.83 & 7.45 & 0.89 \\
 14.22  & 4.67 & 1.66 & 1.86 & 1.00 &  6.81 & 2.27 & 1.86 & 0.99 &   8.13    & 1.36 & 1.54 & 0.96 & 7.42 & 0.32 & 0.95 & 7.35 & 1.22 \\
 17.91  & 5.61 & 2.13 & 1.98 & 1.00 &  11.10& 3.35 & 1.98 & 1.00 &   7.00    & 1.52 & 1.23 & 0.98 & 7.79 & 0.40 & 0.99 & 7.72 & 1.28 \\
 22.54  & 7.48 & 2.90 & 3.50 & 1.00 &  8.13 & 3.60 & 3.50 & 1.00 &   6.66    & 1.79 & 1.72 & 0.99 & 8.08 & 0.49 & 1.00 & 7.97 & 0.81 \\
 28.38  & 5.99 & 3.14 & 3.12 & 1.00 &  11.12& 4.85 & 3.12 & 1.00 &   6.70    & 2.13 & 1.52 & 1.00 & 6.97 & 0.55 & 1.00 & 6.97 & 1.11 \\
 35.73  & 5.34 & 3.50 & 3.83 & 1.00 &  6.63 & 4.65 & 3.83 & 1.00 &   4.01    & 2.12 & 1.44 & 1.00 & 6.50 & 0.64 & 1.00 & 6.27 & 1.45 \\
 44.98  & 8.97 & 5.35 & 3.21 & 1.00 &  6.68 & 5.43 & 3.21 & 1.00 &   5.97    & 2.89 & 1.88 & 1.00 & 6.37 & 0.76 & 1.00 & 6.40 & 0.86 \\
 56.62  &  --  &  --  & 4.13 &  --  &  --   &   -- & 4.13 &  --  & $\le$3.63 & --   & 2.25 & 1.00 & 5.83 & 0.88 & 1.00 & 5.83 & 0.88 \\
 71.29  &  --  &  --  & 4.22 &  --  &  --   &	-- & 4.22 &  --  & $\le$4.80 & --   & 3.10 & 1.00 & 5.91 & 1.05 & 1.00 & 5.91 & 1.05 \\
 89.74  &  --  &  --  & 5.30 &  --  &  --   &	-- & 5.30 &  --  &   8.46    & 5.55 & 5.14 & 1.00 & 4.22 & 1.10 & 1.00 & 4.38 & 1.08 \\
112.98  &  --  &  --  & 8.01 &  --  &  --   &	-- & 8.01 &  --  &    --     &  --  & 4.86 &  --  & 4.96 & 1.34 & 1.00 & 4.96 & 1.34 \\
142.23  &  --  &  --  & 6.91 &  --  &  --   &	-- & 6.91 &  --  &    --     &  --  & 5.20 &  --  & 4.03 & 1.47 & 1.00 & 4.03 & 1.47 \\
\hline															
\end{tabular}
\end{table}

\end{landscape}

\begin{landscape}

\begin{table}  
\caption{PEP 160 $\mu$m number counts, normalized to the Euclidean slope.}
\label{tab:counts_160}
\centering
\begin{tabular}{r | c c c c | c c c c | c c c c | c c c | c c }
\hline
\hline
\multicolumn{1}{c|}{} & \multicolumn{4}{c|}{GOODS-S} & \multicolumn{4}{c|}{GOODS-N} & \multicolumn{4}{c|}{Lockman Hole} & \multicolumn{3}{c|}{COSMOS} & \multicolumn{2}{c}{All Fields} \\
\multicolumn{1}{c|}{$S_{center}$} & Counts & Err. & $\sigma(var)$ & Compl. & Counts & Err. & $\sigma(var)$ & Compl. & Counts & Err. & $\sigma(var)$ & Compl. & Counts & Err. & Compl. & Counts & Err. \\ 
\multicolumn{1}{c|}{$[$mJy$]$} & \multicolumn{3}{c}{$[10^4$ deg$^{-2}$ mJy$^{1.5}]$} & & \multicolumn{3}{c}{$[10^4$ deg$^{-2}$ mJy$^{1.5}]$} & & \multicolumn{3}{c}{$[10^4$ deg$^{-2}$ mJy$^{1.5}]$} & & \multicolumn{2}{c}{$[10^4$ deg$^{-2}$ mJy$^{1.5}]$} & & \multicolumn{2}{c}{$[10^4$ deg$^{-2}$ mJy$^{1.5}]$} \\
\hline
  2.84  &   5.72    & 0.49 &  --   & 0.42 &  --   &  --  &  --   &  --  &  --   &  --  &  --  &  --  &  --	  &  --  &  --  & 5.72  & 0.49 \\
  3.57  &   6.99    & 0.63 &  --   & 0.66 &  --   &  --  &  --   &  --  &  --   &  --  &  --  &  --  &  --	  &  --  &  --  & 6.99  & 0.63 \\
  4.50  &   9.22    & 0.86 &  --   & 0.79 &  --   &  --  &  --   &  --  &  --   &  --  &  --  &  --  &  --	  &  --  &  --  & 9.22  & 0.86 \\
  5.66  &  11.90    & 1.16 &  --   & 0.85 & 11.68 & 1.51 &  --   & 0.45 &  --   &  --  &  --  &  --  &  --	  &  --  &  --  & 11.82 & 0.16 \\
  7.13  &  13.93    & 1.51 &  --   & 0.89 & 14.15 & 2.45 &  --   & 0.55 &  --   &  --  &  --  &  --  &  --	  &  --  &  --  & 13.99 & 0.16 \\
  8.97  &  15.05    & 1.89 &  6.19 & 0.90 & 16.30 & 2.31 &  6.19 & 0.61 & 14.31 & 1.23 & 4.19 & 0.56 &  --	  &  --  &  --  & 14.83 & 0.96 \\
 11.30  &  17.94    & 2.49 &  3.63 & 0.97 & 17.41 & 2.84 &  3.63 & 0.74 & 16.70 & 1.55 & 2.93 & 0.71 &  --	  &  --  &  --  & 17.11 & 0.70 \\
 14.22  &  17.22    & 2.94 &  2.93 & 0.99 & 16.65 & 3.33 &  2.93 & 0.82 & 18.25 & 1.93 & 1.92 & 0.85 & 15.74	  & 0.45 & 0.66 & 17.69 & 1.86 \\
 17.91  &  19.37    & 3.73 &  4.18 & 1.00 & 17.67 & 4.19 &  4.18 & 0.93 & 17.02 & 2.25 & 2.07 & 0.94 & 18.39	  & 0.57 & 0.73 & 18.32 & 0.86 \\
 22.54  &  17.53    & 4.30 &  4.79 & 1.00 & 19.75 & 5.26 &  4.79 & 0.98 & 15.29 & 2.62 & 3.53 & 0.97 & 20.52	  & 0.72 & 0.81 & 20.07 & 3.06 \\
 28.38  &  21.30    & 5.59 &  3.88 & 1.00 & 18.31 & 6.11 &  3.88 & 1.00 & 18.45 & 3.37 & 2.51 & 0.99 & 22.26	  & 0.92 & 0.95 & 21.91 & 2.40 \\
 35.73  &  17.24    & 6.14 &  6.27 & 1.00 & 18.82 & 7.42 &  6.27 & 1.00 & 16.59 & 3.95 & 4.00 & 1.00 & 20.63	  & 1.06 & 0.97 & 20.25 & 2.51 \\
 44.98  &  10.27    & 5.88 &  6.60 & 1.00 & 14.92 & 7.87 &  6.60 & 1.00 & 14.18 & 4.42 & 3.47 & 1.00 & 17.47	  & 1.20 & 0.99 & 16.95 & 3.32 \\
 56.62  &   9.84    & 6.66 &  6.12 & 1.00 & 14.36 & 9.22 &  6.12 & 1.00 & 11.44 & 4.75 & 3.61 & 1.00 & 13.95	  & 1.30 & 0.99 & 13.65 & 2.04 \\
 71.29  & $\le$9.98 & --   &  7.50 & 1.00 & 10.12 & 8.94 &  7.50 & 1.00 & 7.01  & 4.61 & 4.52 & 1.00 & 12.86	  & 1.49 & 1.00 & 12.26 & 3.75 \\
 89.74  &  15.45    & 11.04&  7.46 & 1.00 &  --   &  --  &  7.46 &  --  & 6.37  & 5.27 & 3.69 & 1.00 & 10.63	  & 1.64 & 1.00 & 10.36 & 3.18 \\
112.98  &  --	    &   -- &  8.23 &  --  &  --   &  --  &  8.23 &  --  & 8.19  & 6.83 & 7.32 & 1.00 & 10.41	  & 1.91 & 1.00 & 10.25 & 1.57 \\
142.23  &  --	    &   -- & 11.73 &  --  &  --   &  --  & 11.73 &  --  &  --   &  --  & 7.12 &  --  &  8.48	  & 2.05 & 1.00 & 8.48  & 2.05 \\
179.06  &  --	    &   -- & 11.50 &  --  &  --   &  --  & 11.50 &  --  &  --   &  --  & 7.24 &  --  &  8.10	  & 2.36 & 1.00 & 8.10  & 2.36 \\ 
225.42  &  --	    &   -- &  8.13 &  --  &  --   &  --  &  8.13 &  --  &  --   &  --  & 4.81 &  --  &  6.14	  & 2.43 & 1.00 & 6.14  & 2.43 \\ 
283.79  &  --	    &   -- &  9.86 &  --  &  --   &  --  &  9.86 &  --  &  --   &  --  & 6.45 &  --  & $\le$3.18  & --   & 1.00 & --	& --   \\ 
357.27  &  --	    &   -- &  --   &  --  &  --   &  --  &  --   &  --  &  --   &  --  &  --  &  --  &  3.14	  & 2.39 & 1.00 & 3.14  & 2.39 \\ 
\hline															
\end{tabular}
\end{table}

\end{landscape}

\begin{table*}[!ht]
\centering
\begin{tabular}{c | c c | c c | c c c}
\hline
\hline
$z_{cen}$ & \multicolumn{2}{c|}{COSMOS} & \multicolumn{2}{c|}{GOODS-N} & \multicolumn{3}{c}{GOODS-S} \\
& 100 $\mu$m & 160 $\mu$m & 100 $\mu$m & 160 $\mu$m & 70 $\mu$m & 100 $\mu$m & 160 $\mu$m \\
\hline
0.1 & 1.01e+03 & 7.37e+02 & 1.82e+03 & 1.56e+03 & 1.54e+03 & 2.77e+03 & 2.16e+03 \\ 
0.2 & 1.89e+03 & 1.33e+03 & 1.04e+03 & 7.81e+02 & 2.77e+03 & 3.08e+03 & 2.77e+03 \\  
0.3 & 1.59e+03 & 9.83e+02 & 3.12e+03 & 2.86e+03 & 2.16e+03 & 4.01e+03 & 3.39e+03 \\  
0.4 & 2.01e+03 & 1.26e+03 & 2.34e+03 & 1.82e+03 & 2.77e+03 & 4.31e+03 & 2.77e+03 \\  
0.5 & 1.17e+03 & 7.91e+02 & 2.86e+03 & 2.08e+03 & 3.08e+03 & 3.08e+03 & 3.39e+03 \\  
0.6 & 1.07e+03 & 7.07e+02 & 2.60e+03 & 3.12e+03 & 3.39e+03 & 5.55e+03 & 4.62e+03 \\  
0.7 & 8.93e+02 & 5.87e+02 & 5.20e+02 &    --    & 3.70e+03 & 1.11e+04 & 8.94e+03 \\  
0.8 & 1.04e+03 & 6.23e+02 & 3.38e+03 & 1.82e+03 &    --    & 1.54e+03 & 1.54e+03 \\  
0.9 & 5.99e+02 & 3.96e+02 & 3.38e+03 & 1.82e+03 & 9.24e+02 & 1.54e+03 & 1.54e+03 \\  
1.0 & 7.79e+02 & 5.81e+02 & 1.82e+03 & 1.82e+03 & 2.16e+03 & 6.47e+03 & 5.55e+03 \\  
1.1 & 4.85e+02 & 4.32e+02 &    --    &    --    & 1.23e+03 & 2.77e+03 & 3.08e+03 \\  
1.2 & 3.36e+02 & 2.88e+02 & 7.81e+02 & 5.20e+02 & 3.08e+02 & 2.16e+03 & 1.54e+03 \\  
1.3 & 3.96e+02 & 3.72e+02 & 5.20e+02 & 5.20e+02 &    --    & 3.08e+02 & 6.16e+02 \\  
1.4 & 1.86e+02 & 1.92e+02 & 2.60e+02 & 5.20e+02 & 6.16e+02 & 6.16e+02 & 1.23e+03 \\  
1.5 & 1.98e+02 & 1.68e+02 & 2.60e+02 & 7.81e+02 &    --    & 9.24e+02 & 3.08e+02 \\  
1.6 & 1.80e+02 & 1.20e+02 & 2.60e+02 &    --    & 9.24e+02 & 2.47e+03 & 2.16e+03 \\  
1.7 & 1.02e+02 & 8.99e+01 &    --    &    --    &    --    &    --    &    --    \\  
1.8 & 1.50e+02 & 1.38e+02 & 5.20e+02 & 2.60e+02 &    --    & 3.08e+02 & 6.16e+02 \\  
1.9 & 1.80e+02 & 1.74e+02 &    --    &    --    & 3.08e+02 & 6.16e+02 & 6.16e+02 \\  
2.0 & 1.02e+02 & 1.26e+02 & 1.04e+03 & 1.04e+03 & 3.08e+02 & 1.23e+03 & 1.23e+03 \\  
2.1 & 3.60e+01 & 7.79e+01 & 2.60e+02 & 2.60e+02 &    --    & 1.23e+03 & 9.24e+02 \\  
2.2 & 1.20e+01 & 2.40e+01 &    --    &    --    &    --    & 1.23e+03 & 1.23e+03 \\  
2.3 & 2.40e+01 & 2.40e+01 &    --    & 2.60e+02 &    --    & 9.24e+02 & 1.23e+03 \\  
2.4 & 2.40e+01 & 3.60e+01 &    --    &    --    &    --    & 9.24e+02 & 1.23e+03 \\  
2.5 & 3.60e+01 & 5.39e+01 & 2.60e+02 & 5.20e+02 &    --    & 9.24e+02 & 1.23e+03 \\  
2.6 & 5.39e+01 & 7.19e+01 &    --    & 5.20e+02 &    --    & 6.16e+02 & 6.16e+02 \\  
2.7 &    --    & 2.40e+01 &    --    &    --    &    --    &    --    &    --    \\  
2.8 & 5.99e+00 & 1.80e+01 &    --    &    --    &    --    &    --    & 3.08e+02 \\  
2.9 & 1.20e+01 & 1.20e+01 &    --    &    --    &    --    &    --    & 9.24e+02 \\  
3.0 & 5.99e+00 & 1.20e+01 &    --    &    --    &    --    &    --    & 3.08e+02 \\  
3.1 & 1.20e+01 & 1.20e+01 &    --    &    --    &    --    &    --    &    --    \\  
3.2 &    --    & 1.20e+01 &    --    &    --    &    --    &    --    &    --    \\  
3.3 & 1.20e+01 & 1.80e+01 & 2.60e+02 &    --    &    --    &    --    &    --    \\  
3.4 &    --    &    --    &    --    &    --    &    --    &    --    &    --    \\  
3.5 &    --    & 5.99e+00 &    --    &    --    &    --    &    --    &    --    \\  
3.6 &    --    &    --    &    --    &    --    & 6.16e+02 & 6.16e+02 & 3.08e+02 \\  
3.7 &    --    &    --    &    --    &    --    &    --    &    --    &    --    \\  
3.8 & 5.99e+00 & 5.99e+00 &    --    &    --    &    --    &    --    &    --    \\  
3.9 &    --    &    --    &    --    &    --    &    --    &    --    &    --    \\  
4.0 &    --    &    --    &    --    &    --    &    --    &    --    &    --    \\  
4.1 &    --    &    --    &    --    &    --    &    --    &    --    &    --    \\  
4.2 & 5.99e+00 &    --    &    --    &    --    &    --    &    --    & 3.08e+02 \\  
4.3 &    --    &    --    &    --    &    --    & 3.08e+02 & 3.08e+02 & 3.08e+02 \\  
4.4 &    --    &    --    & 2.60e+02 &    --    &    --    &    --    &    --    \\  
\hline															
\end{tabular}
\caption{Redshift derivative $dN/dz$ $[$deg$^{-2}]$ for PACS detected source in COSMOS, GOODS-N and
GOODS-S. Catalogs are cut at
80\% photometric completeness, see Fig. \ref{fig:zdistr} for actual flux ranges covered.}
\label{tab:zdistr}
\end{table*}

\begin{table*}[!ht]
\centering
\begin{tabular}{r | c c | c c | c c | c c }
\hline
\hline
\multicolumn{1}{c|}{} & \multicolumn{2}{c|}{$0.0<z\le0.5$} & \multicolumn{2}{c|}{$0.5<z\le1.0$} & \multicolumn{2}{c|}{$1.0<z\le2.0$} & \multicolumn{2}{c}{$2.0<z\le5.0$} \\
\multicolumn{1}{c|}{$S_{center}$} & Counts & Err. & Counts & Err. & Counts & Err. & Counts & Err. \\ 
\multicolumn{1}{c|}{$[$mJy$]$} & \multicolumn{2}{c|}{$[10^4$ deg$^{-2}$ mJy$^{1.5}]$} & \multicolumn{2}{c|}{$[10^4$ deg$^{-2}$ mJy$^{1.5}]$} & \multicolumn{2}{c|}{$[10^4$ deg$^{-2}$ mJy$^{1.5}]$} & \multicolumn{2}{c}{$[10^4$ deg$^{-2}$ mJy$^{1.5}]$}\\
\hline
   1.13 &  --  &  --  &  --  &  --  & 0.43 & 0.25 & $<$0.28 & -- 	\\
   1.42 & 0.26 & 0.11 & 0.35 & 0.12 & 0.35 & 0.12 & 0.17    & 0.09 	\\
   1.79 & 0.18 & 0.09 & 0.85 & 0.19 & 0.31 & 0.12 & 0.09    & 0.06 	\\
   2.25 & 0.42 & 0.19 & 0.76 & 0.25 & 0.25 & 0.15 & $<$0.18 & -- 	\\
   2.84 & 0.50 & 0.22 & 0.50 & 0.22 & 0.50 & 0.22 & 0.20    & 0.14 	\\
   3.57 & 0.51 & 0.26 & 1.02 & 0.36 & 0.51 & 0.26 &  --     & --  	\\
   4.50 & 0.94 & 0.33 & 0.47 & 0.24 & 0.47 & 0.24 &  --     & --  	\\
   5.66 &  --  &  --  & 0.85 & 0.60 &  --  &  --  &  --     & --  	\\
   7.13 &  --  &  --  &  --  &  --  &  --  & --   & --      & --  	\\
   8.97 & 1.10 & 0.55 & 0.55 & 0.39 &  --  & --   & --      & --  	\\
  11.30 & 1.48 & 0.74 & 0.74 & 0.52 &  --  & --   & --      & --  	\\
  14.22 & 2.51 & 1.78 &  --  &  --  &  --  & --   & --      & --  	\\
  17.91 &  --  &  --  &  --  &  --  &  --  & --   & --      & --  	\\
  22.54 & 4.83 & 3.41 &  --  &  --  &  --  & --   & --      & --  	\\
\hline															
\end{tabular}
\caption{GOODS-S 70 $\mu$m number counts, normalized to the Euclidean slope, and split in
redshift slices.}
\label{tab:counts_redshift_070}
\end{table*}

\begin{table*}[!ht]
\centering
\begin{tabular}{r | c c | c c | c c | c c }
\hline
\hline
\multicolumn{1}{c|}{} & \multicolumn{2}{c|}{$0.0<z\le0.5$} & \multicolumn{2}{c|}{$0.5<z\le1.0$} & \multicolumn{2}{c|}{$1.0<z\le2.0$} & \multicolumn{2}{c}{$2.0<z\le5.0$} \\
\multicolumn{1}{c|}{$S_{center}$} & Counts & Err. & Counts & Err. & Counts & Err. & Counts & Err. \\ 
\multicolumn{1}{c|}{$[$mJy$]$} & \multicolumn{2}{c|}{$[10^4$ deg$^{-2}$ mJy$^{1.5}]$} & \multicolumn{2}{c|}{$[10^4$ deg$^{-2}$ mJy$^{1.5}]$} & \multicolumn{2}{c|}{$[10^4$ deg$^{-2}$ mJy$^{1.5}]$} & \multicolumn{2}{c}{$[10^4$ deg$^{-2}$ mJy$^{1.5}]$}\\
\hline
      1.42  &  0.23 & 0.14 & 0.62 & 0.22 & 0.86 & 0.26 & 0.70 & 0.23 \\ 
      1.79  &  0.27 & 0.12 & 0.99 & 0.23 & 1.48 & 0.29 & 0.33 & 0.13 \\
      2.25  &  0.78 & 0.26 & 1.30 & 0.34 & 1.21 & 0.32 & 0.43 & 0.19 \\
      2.84  &  0.43 & 0.19 & 1.21 & 0.32 & 1.47 & 0.36 & 0.78 & 0.26 \\
      3.57  &  1.14 & 0.18 & 2.00 & 0.10 & 1.09 & 0.40 & 0.44 & 0.26 \\
      4.50  &  0.55 & 0.27 & 2.69 & 0.66 & 1.06 & 0.39 & 1.04 & 0.39 \\
      5.66  &  1.92 & 0.63 & 2.49 & 0.54 & 2.14 & 0.70 & 0.47 & 0.08 \\
      7.13  &  2.44 & 0.82 & 2.88 & 0.63 & 2.15 & 0.70 & 0.39 & 0.08 \\
      8.97  &  3.18 & 0.98 & 3.26 & 0.85 & 1.95 & 0.17 & 0.27 & 0.06 \\
     11.30  &  3.55 & 1.52 & 2.85 & 0.98 & 2.00 & 0.66 & 0.27 & 0.07 \\
     14.22  &  3.79 & 0.54 & 2.95 & 0.50 & 1.99 & 0.19 & 0.09 & 0.05 \\
     17.91  &  5.10 & 1.48 & 2.71 & 0.61 & 1.54 & 0.22 &  --  &  --  \\
     22.54  &  5.58 & 0.33 & 2.60 & 0.00 & 1.66 & 0.27 & 0.16 & 0.09 \\
     28.38  &  5.43 & 0.66 & 2.04 & 0.81 & 0.93 & 0.23 &  --  &  --  \\
     35.73  &  5.47 & 0.00 & 1.82 & 0.00 & 0.64 & 0.26 &  --  &  --  \\
     44.98  &  5.73 & 0.68 & 1.35 & 0.37 & 0.73 & 0.27 &  --  &  --  \\
     56.62  &  6.04 & 0.97 & 0.77 & 0.35 & 0.31 & 0.22 &  --  &  --  \\
     71.29  &  6.01 & 1.20 & 1.20 & 0.54 &  --  &  --  &  --  &  --  \\
     89.74  &  4.61 & 1.12 & 0.54 & 0.38 &  --  &  --  &  --  &  --  \\
    112.98  &  5.51 & 1.74 &  --  &  --  &  --  &  --  &  --  &  --  \\
    142.23  &  4.16 & 1.25 & 0.76 & 0.53 &  --  &  --  &  --  &  --  \\
\hline	        													 
\end{tabular} 
\caption{PEP 100 $\mu$m number counts, normalized to the Euclidean slope, and split in
redshift slices. These counts have been obtained with a weighted average between GOODS-S, 
GOODS-N and COSMOS areas.}
\label{tab:counts_redshift_100}
\end{table*}

\begin{table*}[!ht]
\centering
\begin{tabular}{r | c c | c c | c c | c c }
\hline
\hline
\multicolumn{1}{c|}{} & \multicolumn{2}{c|}{$0.0<z\le0.5$} & \multicolumn{2}{c|}{$0.5<z\le1.0$} & \multicolumn{2}{c|}{$1.0<z\le2.0$} & \multicolumn{2}{c}{$2.0<z\le5.0$} \\
\multicolumn{1}{c|}{$S_{center}$} & Counts & Err. & Counts & Err. & Counts & Err. & Counts & Err. \\ 
\multicolumn{1}{c|}{$[$mJy$]$} & \multicolumn{2}{c|}{$[10^4$ deg$^{-2}$ mJy$^{1.5}]$} & \multicolumn{2}{c|}{$[10^4$ deg$^{-2}$ mJy$^{1.5}]$} & \multicolumn{2}{c|}{$[10^4$ deg$^{-2}$ mJy$^{1.5}]$} & \multicolumn{2}{c}{$[10^4$ deg$^{-2}$ mJy$^{1.5}]$}\\
\hline
      2.84 &  1.11 & 0.21 & 1.48 & 0.52 & 1.29 & 0.49 & 1.48 & 0.52 \\ 
      3.57 &  0.93 & 0.45 & 2.02 & 0.56 & 2.02 & 0.56 & 2.02 & 0.56 \\
      4.50 &  0.60 & 0.38 & 3.41 & 0.83 & 4.01 & 0.90 & 1.20 & 0.49 \\
      5.66 &  2.13 & 0.35 & 3.14 & 0.60 & 3.94 & 1.13 & 2.08 & 0.79 \\
      7.13 &  3.62 & 0.56 & 3.73 & 3.62 & 3.85 & 0.83 & 2.71 & 1.02 \\
      8.97 &  1.41 & 0.81 & 7.20 & 7.09 & 2.78 & 1.36 & 3.29 & 1.24 \\
     11.30 &  3.87 & 1.98 & 6.38 & 0.65 & 4.88 & 1.10 & 2.47 & 1.24 \\
     14.22 &  4.39 & 1.80 & 8.35 & 0.34 & 2.72 & 1.57 & 1.81 & 1.28 \\
     17.91 &  6.83 & 2.69 & 6.82 & 1.70 & 6.18 & 3.07 & 1.81 & 0.90 \\
     22.54 &  9.04 & 1.93 & 7.64 & 2.23 & 6.47 & 0.50 & 1.49 & 0.26 \\
     28.38 & 11.20 & 2.48 & 8.05 & 0.22 & 6.21 & 1.74 &  --  &  --  \\
     35.73 & 11.57 & 3.21 & 6.90 & 2.49 & 6.08 & 0.78 & 0.50 & 0.22 \\
     44.98 & 10.65 & 0.23 & 6.10 & 0.92 & 4.16 & 0.76 & 0.42 & 0.24 \\
     56.62 & 10.49 & 2.62 & 3.07 & 0.72 & 3.41 & 0.76 &  --  &  --  \\
     71.29 & 11.56 & 1.59 & 2.18 & 0.69 & 1.96 & 0.65 &  --  &  --  \\
     89.74 &  8.91 & 1.82 & 2.97 & 1.05 & 1.11 & 0.64 &  --  &  --  \\
    112.98 & 11.50 & 2.64 & 1.21 & 0.86 &  --  &  --  &  --  &  --  \\
    142.23 &  9.67 & 2.58 &  --  &  --  &  --  &  --  &  --  &  --  \\
    179.06 &  9.24 & 2.47 &  --  &  --  &  --  &  --  &  --  &  --  \\
    225.42 &  5.00 & 3.53 &  --  &  --  &  --  &  --  &  --  &  --  \\
    283.79 &  1.94 & 1.12 &  --  &  --  &  --  &  --  &  --  &  --  \\
\hline	        													 
\end{tabular} 
\caption{PEP 160 $\mu$m number counts, normalized to the Euclidean slope, and split in
redshift slices. These counts have been obtained with a weighted average between GOODS-S, 
GOODS-N and COSMOS areas.}
\label{tab:counts_redshift_160}
\end{table*}

\begin{table}[!ht]
\centering
\begin{tabular}{l | c c c}
\hline
\multicolumn{4}{c}{Counts from $P(D)$} \\
\hline
Field & Flux range & Slope & Error \\
\& band & $[$mJy$]$ & $\alpha$ & $d\alpha$ \\
\hline
GOODS-S  70 & 0.2$-$0.7 & -2.33 & $^{+2.13}_{-3.63}$ \\
GOODS-S  70 & 0.7$-$3.1 & -1.70 & $^{+0.37}_{-0.62}$ \\
GOODS-S  70 & 3.1$-$100 & -2.44 & $^{+0.22}_{-0.09}$ \\
\hline
GOODS-S 100 & 0.2$-$1.1	& -1.23 & $^{+0.02}_{-1.59}$ \\
GOODS-S 100 & 1.1$-$5.8 & -2.09 & $^{+0.34}_{-0.02}$ \\
GOODS-S 100 & 5.8$-$100 & -2.25 & $^{+0.13}_{-0.17}$ \\
\hline
GOODS-S 160 & 0.3$-$5.8	& -1.20 & $^{+0.30}_{-0.46}$ \\
GOODS-S 160 & 5.8$-$8.5 & -1.28 & $^{+0.03}_{-0.18}$ \\
GOODS-S 160 & 8.5$-$100 & -2.73 & $^{+0.17}_{-0.08}$ \\
\hline
\end{tabular}
\caption{Results of $P(D)$ fit, including 3 $\sigma$ uncertainties.}
\label{tab:pow_law_pdd}
\end{table}

\begin{table}[!ht]
\centering
\begin{tabular}{l c c}
\hline
\hline
Band & $\sigma_c$ & $S_{lim}$ \\
     &            & 16.7 beams/src \\
\hline
70 $\mu$m  & --       & 0.4 mJy \\
100 $\mu$m & 0.27 mJy & 1.5-2.0 mJy\\
160 $\mu$m & 0.92 mJy & 8.0 mJy \\
\hline
\end{tabular}
\caption{Confusion photometric noise, obtained assuming $q=5$, and 16.7 beams/source density flux
limit, for PACS blank extragalactic surveys.}
\label{tab:confusion}
\end{table}

\begin{table}[!ht]
\centering
\begin{tabular}{l | c c c c}
\hline
\hline
Description & band   & Flux range & $\nu I_\nu$ &  Error\\
            & $\mu$m & $[$mJy$]$  & \multicolumn{2}{c}{$[$nW m$^{-2}$ sr$^{-1}]$} \\
\hline
GOODS-S		& 70	& 1.2-40 	& 3.61 	& $\pm$1.12	\\
GOODS-S		& 70	& $>$1.2 	& 4.52 	& $\pm$1.18	\\
GOODS-S $P(D)$	& 70	& $>$0.35	& 4.98	& $^{+16.2}_{-2.07}$ \\ 
Power-law	& 70	& Total		& $\ge$11.09	& -- \\ 
\hline
All fields	& 100	& 1.2-140	& 7.82	& $\pm$0.94	\\
All fields	& 100	& $>$1.2	& 8.35	& $\pm$0.95	\\
GOODS-S $P(D)$	& 100	& $>$0.2	& 9.32	& $^{+5.47}_{-3.66}$ \\ 
Power-law	& 100	& Total		& 12.61	& $^{+8.31}_{-1.74}$ \\ 
\hline
All fields	& 160	& 2.0-350	& 9.17	& $\pm$0.59	\\
All fields	& 160	& $>$2.0	& 9.49	& $\pm$0.59	\\
GOODS-S $P(D)$	& 160	& $>$0.3	& 11.31	& $^{+4.00}_{-2.43}$ \\ 
Power-law	& 160	& Total		& 13.63	& $^{+3.53}_{-0.85}$ \\ 
\hline
\end{tabular}
\caption{The cosmic infrared background inferred from PEP data.}
\label{tab:cib}
\end{table}

\begin{table*}[!ht]
\centering
\begin{tabular}{l c c | c c | c c | c c | c c}
\hline
\hline
Description & band   & Flux range & \multicolumn{2}{c|}{$0.0<z\le0.5$} & \multicolumn{2}{c|}{$0.5<z\le1.0$} & \multicolumn{2}{c|}{$1.0<z\le2.0$} & \multicolumn{2}{c}{$2.0<z\le5.0$} \\
            & $\mu$m & $[$mJy$]$  & $\nu I_\nu$ & Error & $\nu I_\nu$ & Error & $\nu I_\nu$ & Error & $\nu I_\nu$ & Error \\
	    &	     &            & \multicolumn{2}{c|}{$[$nW m$^{-2}$ sr$^{-1}]$} & \multicolumn{2}{c|}{$[$nW m$^{-2}$ sr$^{-1}]$} & \multicolumn{2}{c|}{$[$nW m$^{-2}$ sr$^{-1}]$} & \multicolumn{2}{c}{$[$nW m$^{-2}$ sr$^{-1}]$} \\
\hline
GOODS-S		& 70  & 1.2-36 & 1.57 & 0.97 & 1.70 & 1.10 & 0.47 & 0.23 & 0.08 & 0.06 \\
\hline
GOODS-S		& 100 & 1.2-45 & 2.02 & 1.19 & 2.38 & 1.14 & 1.44 & 0.46 & 0.51 & 0.22 \\
All fields	& 100 & 1.2-142& 3.35 & 0.77 & 2.80 & 0.63 & 2.03 & 0.37 & 0.61 & 0.22 \\
\hline
GOODS-S		& 160 & 2.0-55 & 1.64 & 0.85 & 2.59 & 0.94 & 1.91 & 0.70 & 1.16 & 0.50 \\
All fields	& 160 & 2.0-357& 3.21 & 0.75 & 3.02 & 0.71 & 2.43 & 0.62 & 1.16 & 0.45 \\
\hline
\end{tabular}
\caption{The contribution of galaxies at different epochs to the cosmic infrared background.}
\label{tab:cib_redshift}
\end{table*}


\begin{acknowledgements}
We wish to thank M.~Bethermin, A.~Franceschini, C.~Gruppioni, G.~Marsden, S.~Niemi,
M.~Rowan-Robinson, and E.~Valiante for providing new, extended model outputs for us, 
and the anonymous referee for useful comments.
PACS has been developed by a consortium of institutes led by MPE (Germany) and 
including UVIE (Austria); KU Leuven, CSL, IMEC (Belgium); CEA, LAM (France); 
MPIA (Germany); INAF-IFSI/OAA/OAP/OAT, LENS, SISSA (Italy); IAC (Spain). 
This development has been supported by the funding agencies BMVIT (Austria), 
ESA-PRODEX (Belgium), CEA/CNES (France), DLR (Germany), ASI/INAF (Italy), 
and CICYT/MCYT (Spain).
\end{acknowledgements}




\bibliographystyle{aa}
\bibliography{biblio_PEP_counts}


\begin{appendix} 

\section{A cheap $P(D)$ estimator}\label{sect:pdd_method}

In addition to directly detected sources and to stacking of 24 $\mu$m dense
catalogs, further information about the shape of PACS number counts comes from
the statistical properties of observations, probing the counts at even fainter
fluxes.
The ``probability of deflection'', or $P(D)$ distribution, is basically the
distribution  of pixel values in a map (see Sect. \ref{sect:pdd_counts}). 

For a large density of sources, the contribution to the
$P(D)$ is a Poisson distribution with a large mean value, convolved with the
instrumental noise (typically assumed to be Gaussian). 
Generally speaking, for sufficiently steep number counts at faint flux densities,
the depth reached with the $P(D)$ analysis is significantly higher than what
is provided by individually-detected sources, and is a powerful tool to probe counts
slopes in cases dominated by confusion.

Given differential number counts $dN/dS$, and a beam function 
(e.g. PSF) $f(\theta,\phi)$, describing the point-source spatial-response at
position $x$, the probability of deflection $P(D)$ can be written as
\citep[e.g.][]{patanchon2009}: 
\begin{equation}
P(D) = \mathcal{F}^{-1} \left[ exp \left( \int_0^\infty R(x) e^{i\omega x} dx -
\int_0^\infty R(x)dx \right) \right]
\end{equation}
\noindent or, alternatively, isolating the real part only
\citep[e.g.][]{franceschini2010}:
\begin{eqnarray}
\nonumber P(D) &=& 2 \int_0^\infty exp\left\{-\int_0^\infty R(x)\left[ 1-cos\left(2\pi\omega
x\right)\right]dx\right\} \ \times \\
 &\times& cos\left[ \int_0^\infty R(x)
sin\left(2\pi\omega x\right)dx-2\pi \omega D\right] d\omega
\end{eqnarray}
\noindent where the mean number of source responses of intensity $x$ in the beam is
given by:
\begin{equation}
R(x) = \int_0^\infty
\frac{dN\left[\frac{x}{f\left(\theta,\phi\right)}\right]}{dS} \
\frac{d\omega}{f\left(\theta,\phi\right)}\textrm{.}
\end{equation}

We defer to \citet{patanchon2009}, \citet{franceschini2010}, \citet{glenn2010}
for a full treatment and description of the $P(D)$ formalism.
Only in the case of very simple $dN/dS$ functional forms and trivial beams (e.g.
Gaussian), the above equations can be solved analytically. For an effective beam
that is not strictly positive, or is not azimuthally symmetric, it is necessary
to use the full 2-D beam map. 

For example, the Herschel/PACS beam is characterized by 3 prominent lobes, which
cannot be described analytically. In such a case, if one likes to account for the 
real beam function, a numerical treatment of $R(x)$, and possibly $P(D)$ is
necessary. 

Finally, in case of real observations, the instrumental noise contribution to
pixel flux densities must be included.

\subsection{The numerical $P(D)$ approach}

Given an instrumental PSF $f(\theta,\phi)$ and a number counts model $dN/dS$, 
it is possible to predict the pixel flux density distribution of a map with $M$
pixels, simply rolling random numbers. We describe here the
principles of the method we developed in the frame of the PEP survey. This
approach turns out to be of simple implementation, avoiding integrations of
oscillating functions or Fourier Transforms, and it is accessible to any user
with a basic knowledge of random number generators. Furthermore, it
avoids any analytic simplification of the beam function, thus allowing to employ
the real instrumental PSF, regardless of how complex its bi-dimensional profile
is.

For a given $dN/dS$, we describe how a synthetic $P(D)$ is computed. The adopted form of
number counts and fitting technique are described in the next Section.
Parameters are varied by means of a MCMC engine and for each realization a new
$P(D)$ is computed and compared to the observed data.

The input information needed by the $P(D)$ numerical algorithm are:

\begin{itemize}
\item description of number counts, $dN/dS$;
\item beam function, e.g. in form of an observed PSF 2-D map;
\item value of instrumental noise.
\end{itemize}

Further secondary inputs, derived from the three primaries above, are:

\begin{itemize} 
\item the maximum radius $r_{max}$ in pixels from a given source within which a pixel 
is affected by the source itself, e.g. the radius at which 99\% of the observed
PSF flux is enclosed; if the pixel scale of the map is $\alpha$ arcsec/pix, the
corresponding angular radius is $\alpha r_{max}$;
\item the total expected number, $N_0$, of sources within the solid angle $\Omega$ defined by
$r_{max}$, simply given by the integral of number counts:
$N_0=\Omega\int_{S_{min}}^{S_{max}} \frac{dN}{dS} \ dS$ 
\item the cumulative probability that a random source has a flux lower than $S$,
given by: $P(S)=\frac{\Omega}{N_0}\int_{S_{min}}^S \frac{dN}{dS^\prime} \
dS^\prime$.
\end{itemize}

As mentioned above, the procedure relies on several dice rolls, using random
numbers to determine the actual number of sources expected to contribute to each
pixel of the map, the distance of the given source to the pixel, the fluxes of
sources. 

\begin{description}
\item[{\bf Step 1.}] We assume that the distribution of sources in the sky is randomly
uniform, i.e. we ignore the clustering properties of sources in the sky.
Clustering would mainly affect the $P(D)$ over 
scales larger than the size of the GOODS-S field under exam here 
\citep[e.g.][]{glenn2010,viero2009,lagache2007}.

Given $N_0$, the number of sources affecting the given pixel in the current
realization is a random Poisson deviate for expected $N_0$. This number is
recomputed for each pixel and each $dN/dS$ realization.
\item[{\bf Step 2.}] For each source affecting the given pixel, it is necessary to
determine the fraction $f_i$ of flux affecting the pixel. 
Roll $N$ random, uniformly distributed positions, inside $r_{max}$ (defined
either by $r, \phi$ or by $x,y$ coordinates). Based on the position of the given
source, and the beam function, derive the fraction of flux affecting the pixel
for each source.
\item[{\bf Step 3.}] Finally assign a random flux $S_i$ to each of the $N$ sources affecting the given
pixel, according to the input $dN/dS$. To this aim, the probability $P(S)$
(defined above) is used to transform a flat random number distribution in the
needed flux values. 
\end{description}

Assuming that any further background has been subtracted from the map, the flux
of the given pixel is finally given by $D=D_0+\sum_{i=1}^N f_i S_i$, where $D_0$
is the term describing pure noise (i.e. random flux, in case of no sources at
all) and is given by a Gaussian deviate as wide as the noise measured on actual
PEP r.m.s. maps (see Tab. \ref{tab:fields}).

\subsection{Number counts model and minimization}

Number counts are modeled similarly to \citet{patanchon2009} and
\citet{glenn2010}. We adopt a simple, parametric model, defined by the
differential counts $dN/dS$ at a set of flux density knots, or nodes. The model is
defined within a $S_{min}$ and $S_{max}$ lower and upper flux boundaries;
outside of this flux range, we simply set $dN/dS$ to zero. Between nodes, number
counts are described as power-laws, connecting at knots.

We used a Markov-Chain Monte Carlo (MCMC) sampling of the parameter 
space, to explore the likelihood function of the model to reproduce the observed
$P(D)$. We use a Metropolis-Hastings algorithm as our MCMC sampler.
In order for this sampler to converge efficiently, we draw new steps from
univariate Gaussian distributions, whose widths were given by dummy MCMC runs.

The position of knots and the corresponding amplitude of counts are free
parameters, and are varied at each MCMC map realization. The upper boundary of
the covered flux range is fixed to 100 mJy, a value driven by the limits of
observed, resolved counts in GOODS-S.  
In order to minimize the number of free parameters, and further degeneracies, we
limit the description of number counts to a broken power-law with three
sections, for a total of seven free parameters (3 positions and 4 amplitudes).

In the past, a standard approach was to subtract bright individually detected
sources from the $P(D)$ analysis, and work only at the faint side, but it has
been shown that a more robust estimate of the source counts is obtained if
the $P(D)$ is fitted across the whole available flux density range
\citep[e.g.][]{patanchon2009}. Therefore, to derive the observed $P(D)$ to be
reproduced, we use the observed science maps without subtracting any object.

Section \ref{sect:pdd_counts}, Figs. \ref{fig:p_of_d}, \ref{fig:counts_not_norm}
and Tab. \ref{tab:pow_law_pdd} describe the results of the whole $P(D)$ analysis.

\subsection{Effects of map-making high-pass filtering}

It is necessary to recall that PEP maps have been obtained by
applying a high-pass filtering process along the data time-line, in order to get
rid of $1/f$ noise. This procedure consist in a 
running-box median filter, and induces two different side effects onto 
PACS maps: {\em 1)} the background is naturally removed from the maps; {\em 2)}
the PSF profile is eroded and wings are suppressed.

At very faint fluxes, and high source
densities, close to and beyond the confusion limit, the contribution of sources to
the model $P(D)$ basically generates a ``background'' sheet of sources, which
shifts the $P(D)$ histogram to a non-zero peak and median. In principle, the
source component of the $P(D)$ is always positive, and the addition of noise
generates the negative nearly-Gaussian behaviors at the faint side of $P(D)$.
The adopted high-pass filtering shifts the whole flux pixel distribution
function to a zero peak/median. In order to take this into account, with the current form of PEP maps, 
we shift the model $P(D)$ to the same zero peak, prior of fitting. 

The second consequence of applying an high-pass filter is 
a modification of object profiles. Detected objects are masked 
and excluded from the median filter derivation.
Tests were done adding simulated sources to the time-lines
before masking and before high-pass filtering (see Popesso et al. 
\citeyear{popesso2011} and Lutz et al. \citeyear{lutz2011}). 
The result is that the filtering modifies the fluxes of masked 
sources by less than 5\%, and those of unmasked point sources --- below the 
detection limit --- by a factor $\le16$\%.
In order to understand the impact of this on the $P(D)$ analysis, 
we built simulated maps 20 times deeper than the GOODS-S
3$\sigma$ threshold, once using an un-filtered (i.e. masked) PSF at 
all fluxes, and once using an un-filtered PSF above 3$\sigma$, but 
a filtered (unmasked) PSF below. The $P(D)$ obtained in the two cases
are well consistent to each other, with unit median ratio and 
a $<$5\% scatter across all fluxes; no systematic shift, nor skewness are observed.

\end{appendix}


\end{document}